\documentclass{aa}  
\usepackage{graphicx}
\usepackage{txfonts}
\usepackage{float}
\usepackage{mathtools}
\usepackage{subcaption}
\usepackage{array, blindtext}
\usepackage{epstopdf}
\usepackage{xcolor}
\usepackage{gensymb}
\usepackage[colorinlistoftodos]{todonotes}
\usepackage{indentfirst}
\usepackage{ulem}
\usepackage{mathtools}

\usepackage{hyperref}
\usepackage[version=4]{mhchem}
\usepackage{gensymb}
\usepackage{tikzsymbols}
\usepackage{hypcap}

\begin{document} 
   \title{
   Mineral cloud and hydrocarbon haze particles in the atmosphere of the hot Jupiter JWST target WASP-43b}
   \author{Ch. Helling \inst{1,2,3}
   \and 
   Y. Kawashima\inst{3}
   \and
   V. Graham\inst{4}
   \and
   D. Samra\inst{1,2}
   \and
   K. L. Chubb\inst{3}
   \and 
   M. Min\inst{3}
   \and 
   L.B.F.M. Waters\inst{3}
   \and
   V. Parmentier\inst{5}
   }
   
    \institute{Centre for Exoplanet Science, University of St Andrews, North Haugh, St Andrews, KY169SS, UK\\
             \email{ch80@st-andrews.ac.uk}
         \and
             SUPA, School of Physics \& Astronomy, University of St Andrews, North Haugh, St Andrews, KY169SS, UK
         \and
         SRON Netherlands Institute for Space Research, Sorbonnelaan 2, 3584 CA Utrecht, NL
         \and  
         School of Physics and Astronomy, University of Glasgow, University Avenue, Glasgow, G12 8QQ, UK
         \and
            Department of Physics, University of Oxford, Parks Rd, Oxford, OX1 3PU, UK}
   \date{\today}
     \date{Received September 15, 2996; accepted March 16, 2997}

\abstract
    {Having a short orbital period and being tidally locked makes WASP-43b an ideal candidate for JWST  phase curve measurements. Phase curve observations of an entire orbit will enable the mapping of the atmospheric structure across the planet, with different wavelengths of observation allowing different atmospheric depths to be seen.}
    {We provide insight into the details of the clouds that may form on WASP-43b, and their impact on the remaining gas phase, in order to prepare the forthcoming interpretation of the JWST and follow-up data.}
    {We follow a hierarchical modelling strategy. We utilize 3D GCM  results as input for a kinetic, non-equilibrium model for mineral cloud particles, and for a kinetic model to study a photochemicaly-driven hydrocarbon haze component.}
    {Mineral condensation seeds form throughout the atmosphere of WASP-43b.  This is in stark contrast to the ultra-hot Jupiters, like WASP-18b and HAT-P-7b. The dayside is not cloud free but is loaded  with few but large mineral cloud particles in addition to hydrocarbon haze particles of comparable abundance. Photochemically driven hydrocarbon haze appears on the dayside, but does not contribute to the cloud formation on the nightside. The geometrical cloud extension differs across the globe due to the changing thermodynamic conditions. Day and night differ by 6000km in pressure scale height. As reported for other planets, the C/O is not constant throughout the atmosphere and varies between 0.74 and 0.3. The mean molecular weight is approximately constant in a  \ce{H2}-dominated  WASP-43b atmosphere because of the moderate day/night-temperature differences compared to the super-hot Jupiters.}
    {WASP-43b is expected to be fully covered in clouds which are not homogeneously distributed throughout the atmosphere. The dayside and the terminator clouds will be a combination of mineral particles of locally varying size and composition, and of hydrocarbon hazes. The optical depth of  hydrocarbon hazes is considerably lower than that of mineral cloud particles such that a wavelength-dependent radius measurement of WASP-43b would be determined by the mineral cloud particles but not by hazes. }

  \maketitle

%Section 1 :Wasp43b individual plots from other PDF
%Section 2: All comparison plots.
\section{Introduction}
WASP-43b is a giant gas planet that has been selected as a JWST target because it orbits a near-solar abundance K7V star (T$_{\rm eff}= 4500$K) at an orbital distance of $a\approx 0.0153$au  in 0.813\,days (\citealt{2012A&A...542A...4G}). Its atmosphere can therefore be well characterised through transmission spectroscopy and thermal emission measurements.  The dayside of WASP-43b is suggested to have a black-body equivalent temperature of $\approx 1700\pm200$K and an equilibrium temperature of T$_{\rm eq}=1400$K (\citealt{2014ApJ...781..116B}). With a planetary mass of 1.78$\pm$0.01 M$_{\rm J}$ and a radius of R$_{\rm P}=0.93\pm 0.09$R$_{\rm Jup}$ (\citealt{2011A&A...535L...7H}), its bulk density is 2.744$\pm 0.4$ g cm$^{-3}$.  Applying retrieval approaches, \ce{H2O} (\citealt{2014ApJ...793L..27K,2017AJ....153...68S}) and CO (\citealt{2016ApJ...829...52F}) are suggested  to be present in the atmospheric gas that is accessible from thermal emission (but no \ce{CH4} as of yet). A comprehensive summary of observational results for WASP-43b can be found in \cite{2017AJ....153...68S} who present a full-orbit phase curve in the warm Spitzer bands 3.6$\mu$m and 4.5$\mu$m.
WASP-43b is classified as a giant gas planet similar to HD\,189733b and HD\,209458b but  appears to show less day-night heat distribution (\citealt{2017ApJ...835..198K}) despite orbiting closer to its host star compared to  HD\,189733b (a\,$\approx 0.0031$au) and HD\,209458b (a\,$\approx 0.0047$au). We note that both, HD\,189733b and HD\,209458b, have smaller mean bulk densities of 1.8 g cm$^{-3}$ and 0.616 g cm$^{-3}$, respectively (Table 1 in \cite{2016MNRAS.460..855H}).  \cite{2015ApJ...801...86K}  and \cite{2014Sci...346..838S,2017AJ....153...68S}  point out that the day-side phase curve can be fitted with a cloud-free GCM but such a GCM predicts too high flux on the nightside.   \cite{2018ApJ...869..107M} discuss non-equilibrium chemistry caused by advection for CO, \ce{CO2}, \ce{H2O}, and \ce{CH4}, suggesting a general advection driven overabundance of CO and  \ce{H2O} for a high C/O=2.0, and for \ce{CH4} on the night side if C/O=0.5. \cite{2019AJ....157..205M} reanalysed the WASP-43b data suggesting a nightside brightness between the results of \cite{2017AJ....153...68S}  and \cite{2018ApJ...869..107M}. {  \cite{kchubb2019} performed a re-analysis of the HST/WFC3 and Spitzer/IRAC  transmission data for WASP-43b using all available opacity data,  and  report the detection of AlO and H$_2$O in transmission with a high statistical significance. No evidence for the presence of \ce{CO}, \ce{CO2} nor \ce{CH4} in transmission could be reported, however the study notes the small wavelength region analysed. The presence of AlO points to a rich atmosphere chemistry in WASP-43b, possibly leading up to complex metal-oxide clusters similar to AGB star envelopes \citep{2017A&A...608A..55D}.}
 \cite{20VePaBl.wasp43b} present the JWST WASP-43b atmosphere consortium paper with  predictions for atmospheric structures, the most important kinetic gas-phase species and cloud properties. 
 
% [somehow I cannot compile if I include the following sentence. Is there any page limit?]
%Recently, 
%\cite{2020arXiv200104759V} simulated the radiative transfer, chemical kinetics, cloud microphysics, and global circulation in the atmosphere of WASP-43b to predict JWST/MIRI observations by also using the spectral retrieval models.
 
In this paper, we apply a hierarchical modelling approach where we study kinetic cloud formation,  kinetic gas and hydrocarbon haze chemistry by utilizing the output from a 3D GCM  for WASP-43b. We demonstrate that WASP-43b is not cloud free.  We model the formation and explore the properties of two chemically different cloud particle ensembles, mineral cloud particles and hydrocarbon hazes, that form by distinctly different processes. One of the most remarkable differences is that mineral cloud particles form from an oxygen-rich gas and that hydrocarbon hazes require the presence of the hydrocarbon precursors such as $\mathrm{CH_4}$ in the same oxygen-rich, atmospheric environment.  We make suggestions about their composition, particle sizes and derive implications for the gas phase composition to be expected in chemical equilibrium, but also for the C/H/O/N chemistry calculated from photochemical modelling. 

Our results suggest that WASP-43b is covered in clouds,  a finding similar to HD\,189733b and HD\,209458b, and hazes that mainly occur on the dayside and in the terminator regions. We suggest that spectra and phase-curves of irradiated exoplanets, and giant gas planet like WASP-43b in particular, may be confused by the co-existence of hydrocarbon haze particles and mineral seed particles in the upper most atmosphere that is accessible by transmission spectroscopy. These are atmospheric regions of low gas density where photochemical processes  enable the formation of hydrocarbon haze particles but where surface growth processes are still far too inefficient to cause a substantial growth of the mineral particles (in fact, of any particle).

Section~\ref{s:ap} summarises our approach and provides necessary details to allow a comparison to our previous modelling method. Section~\ref{s:Tp} gives an account of the temperature structure which is a major driver of the cloud formation.  Sect.~\ref{s:clouds} presents our analysis of the kinetically forming mineral clouds in WASP-43b. Section~\ref{s:cheq} summarises our general findings for the gas-phase composition assuming chemical equilibrium. Here we discuss the validity of assuming a constant mean molecular weight, the carbon-to-oxygen ratio and the degree of ionsation. Section~\ref{s:ncheq} addresses the effect of mixing processes on the abundance of C/H/O/N-binding molecules and the formation of hydrocarbon hazes under the effect of the host star's radiation. Section~\ref{s:twocloud} presents the comparison between the models applied here leading to mineral particles and hydrocarbon haze, and Sect.~\ref{aA} addresses the challenge of systematic errors within cloud property comparisons from different model approaches.    Section 8 concludes the paper.

    \begin{figure}
\includegraphics[width=1\linewidth]{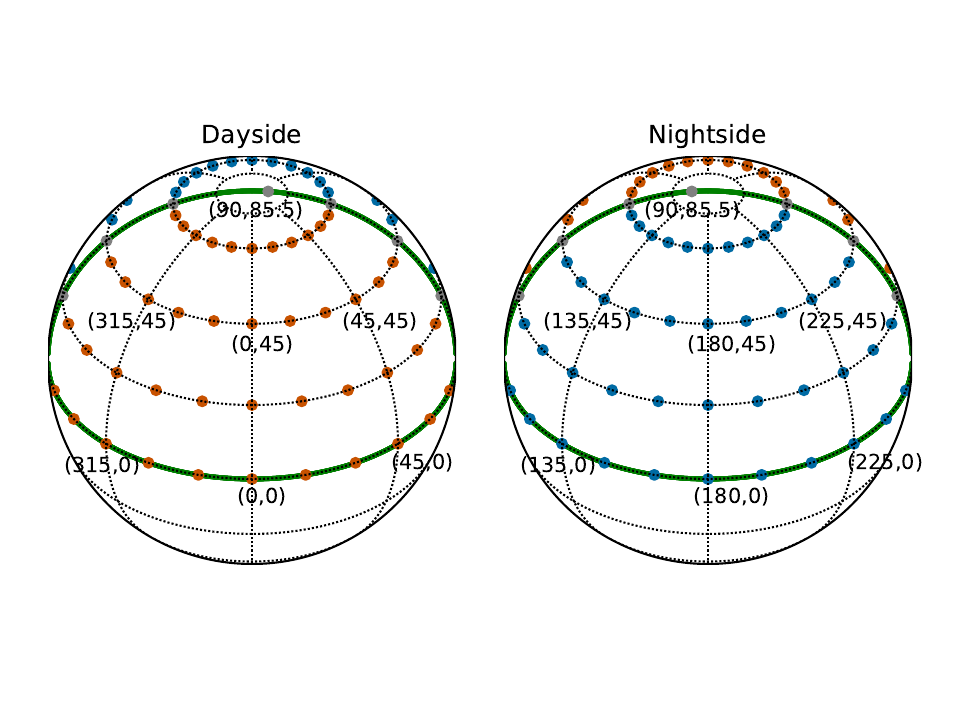}\\*[-2cm]
\caption{Positions of the 1D profiles taken from the GCM (left: dayside; right:  nightside), with the points $(\phi, \theta)=(0\degree,0\degree)$ and $(\phi, \theta)=(180\degree,0\degree)$ being the sub-stellar (phase=0.5) and anti-stellar points, respectively ($(\phi, \theta)$ = (longitudes, latitudes)). 1D profiles that sample the dayside (red), nightside (blue), morning  and evening terminator (both grey) are indicated. The green lines show the locations of slices used in subsequent  maps along the equator ($\theta = 0\degree$) and the terminator (longitudes $\phi = 90\degree$ --  evening terminator,  $\phi = 270\degree$ --  morning terminator)}%
\label{ThetaPhi}
\end{figure}

\section{Approach}\label{s:ap}
We adopt a hierarchical three-model approach in order to examine the cloud structure on WASP-43b  similar to works on the hot Jupiters HD\,189733b and HD\,209458b (\citealt{Lee2015,2016MNRAS.460..855H}), and the ultra-hot Jupiters WASP-18b (\citealt{2019arXiv190108640H}) and HAT-P-7b (\citealt{2019arXiv190608127H,Mol2019}): The first modelling step produces pre-calculated, cloud-free 3D GCM results. These results are used as input for the second modelling step which is a kinetic cloud formation model consistently combined with equilibrium gas-chemistry calculations. We utilise 97 1D (T$_{\rm
  gas}$(z), p$_{\rm gas}$(z), v$_{\rm z}(x,y,z)$)-profiles for WASP-43b similar to our work on HAT-P-7b as visualised in Fig.~1: T$_{\rm gas}$(z) is the local gas temperature [K], p$_{\rm gas}$(z) is the local gas pressure [bar], and v$_{\rm z}$(x,y,z) is the local vertical velocity component [cm s$^{-1}$].  The same 1D profiles are used as input for our third modelling step, the kinetic haze simulation  (\citealt{2018ApJ...853....7K, 2019ApJ...876L...5K, 2019ApJ...877..109K}). The same, height-dependent mixing coefficient $K_{\rm zz}$, which is assumed as $H v_{\rm z}$ with $H$ being the atmospheric scale height, is used in both the  kinetic cloud formation and the kinetic haze formation model. We use the solar element abundances from  \cite{2009ARA&A..47..481A} for the undepleted element abundances.

\begin{figure*}
\begin{tabular}{p{8cm}p{8cm}}
\includegraphics[width=1.0\linewidth]{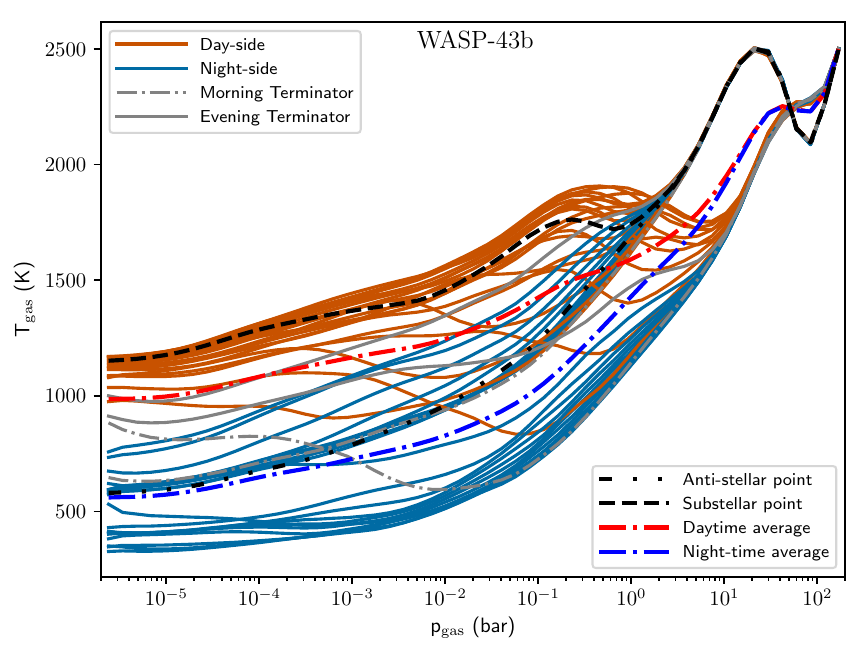}&
 \includegraphics[width=1.1\linewidth]{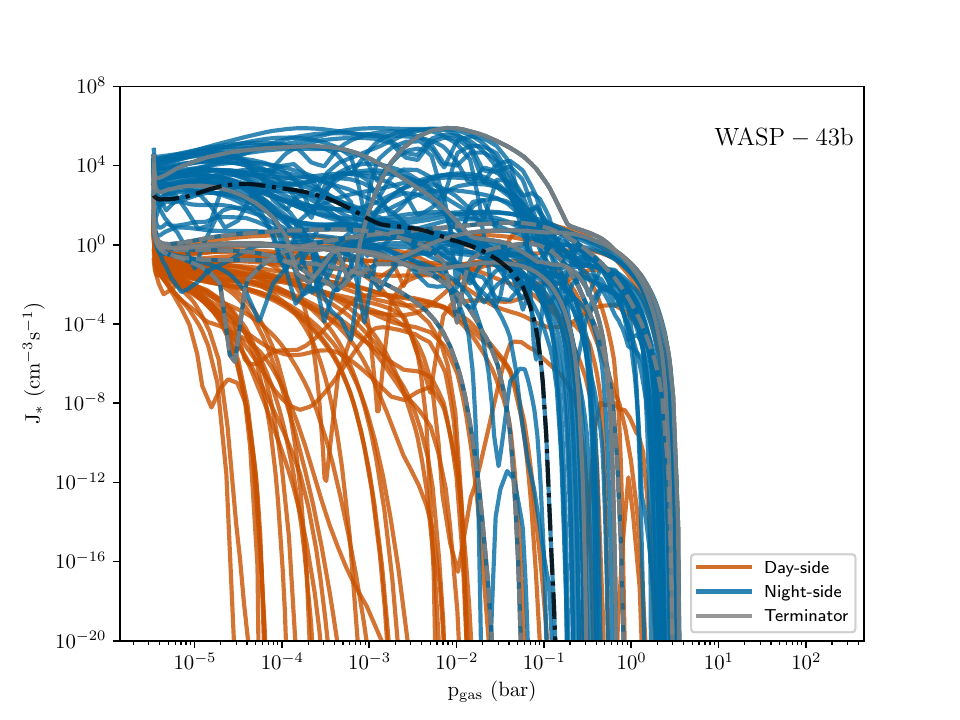}\\
\includegraphics[page=1,width=0.55\textwidth]{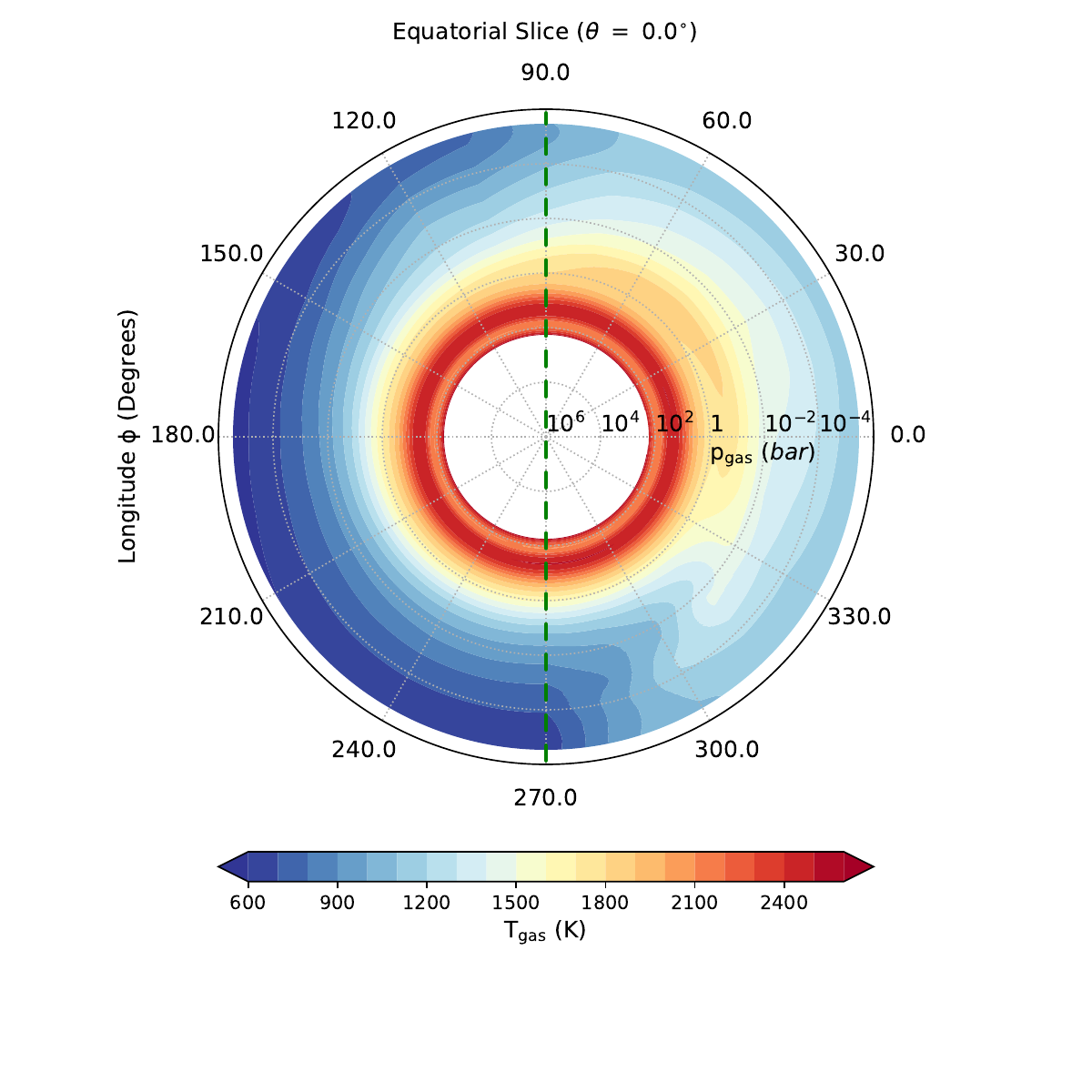}&
\includegraphics[page=7,width=0.55\textwidth]{images/Slice_Plots.pdf}\\*[-2.2cm]
\includegraphics[page=2,width=0.55\textwidth]{images/Slice_Plots.pdf}&
 \includegraphics[page=8,width=0.55\textwidth]{images/Slice_Plots.pdf}\\*[-1.5cm]
\end{tabular}\caption{{\bf Left:} Local gas temperature and local gas pressure (T$_{\rm gas}$, p$_{\rm gas}$) extracted from the 3D GCM \cite{Parmentier2018}, {\bf Right:} Total mineral seed formation (nucleation) rate, $J_*$ [cm$^{-3}$ s$^{-1}$]. 
{\bf Top} 1D profiles for the probed 97 profiles {  (red – dayside, blue – nightside, grey dashed – morning terminator ($\phi= 270^o$), grey-solid – evening terminator ($\phi= 90^o$), black dashed – substellar point($\phi= 0^o$) black dot-dashed – antistellar point($\phi= 180^o$)), } {\bf  Middle} 2D cut  visualising the temperature distribution (left) and the distribution of the formed condensation seeds (right)  in the equatorial plane, {\bf  Bottom:} 2D cut along the terminator  for the a tidally locked WASP-43b. }
  \label{TpNuc}
\end{figure*}

\begin{figure}
\includegraphics[width=0.9\linewidth]{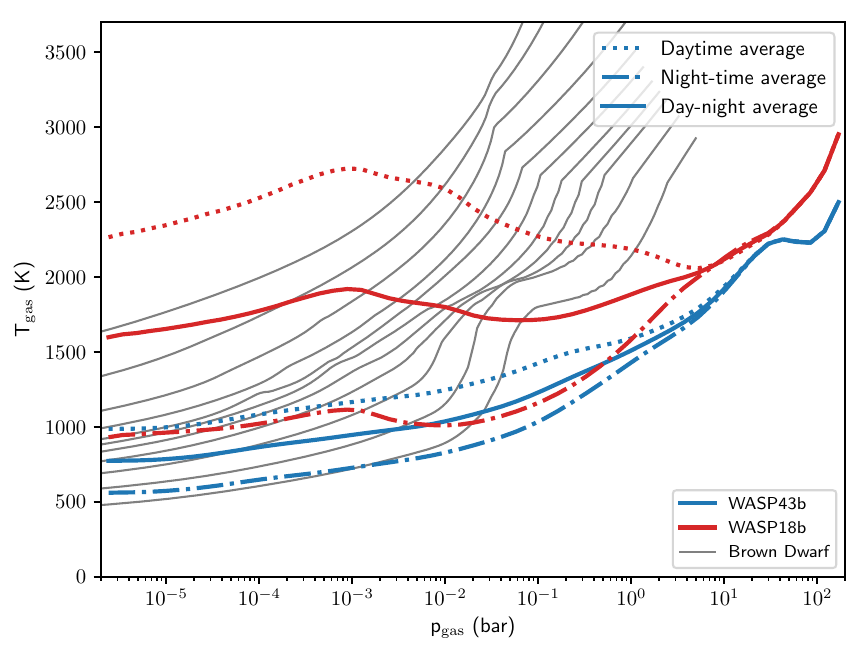}
\caption{Comparison of WASP-43b to the ultra-hot Jupiter WASP-18b (red) and {\sc Drift-Phoenix} young planets / brown dwarf atmosphere structures for log(g)=3.0  and T$_{\rm eff}=1000\,\ldots\,3000$K {  by steps of 200K}  (\citealt{2009A&A...506.1367W}). }
  \label{TpDP}
\end{figure}
%\begin{figure}
%\includegraphics[width=1\linewidth]{images/Wasp43b.pdf}\\
%\includegraphics[width=1.4\linewidth, angle=-90]{images/Katy_PTlimits.pdf}\\
%\caption{Katy's retrieved (T$_{\rm gas}$, p$_{\rm gas}$) profile from ... data.}%
%\label{OldpT}
%\end{figure}

The hierarchical approach has the
  limitation of not explicitly taking into account the potential effect of
  horizontal winds on cloud formation. However, processes governing the formation of mineral clouds are determined by local thermodynamic properties which are the result of 3D dynamic atmosphere  simulations. Horizontal transport can affect both the cloud formation and the cloud properties. Figure 4 in \cite{Lee2015} shows for the gas giant HD\,189733b (orbiting a similar host star to that of WASP-43b but at a larger orbital distance of 0.0314 AU) that the (vertical) settling of the cloud particles occurs at time scales longer then a typical (horizontal) advection time scale on the nightside but that the settling and advection are of comparable magnitudes on the dayside.  Cloud particle properties such as particle size or particle composition should be smeared out in longitude compared to the results shown here. We note that comparing~\citet{Lee2015} (without horizontal advection) and ~\citet{Lee2016} (including horizontal advection), the non-coupled problem is both more computationally feasible, easier to interpret and provides reasonable first order insights into the expected mean cloud particle size and composition, cloud distribution and gas phase depletion in exoplanet atmospheres. The situation is somewhat different for photochemically triggered cloud formation. Photochemical hydrocarbon-haze production, for example, is determined by the external radiation field. A certain smearing out of the hydrocarbon haze properties should also be expected for this cloud particle population in particular in transient regimes like at the day-night terminators.  \cite{2014A&A...564A..73A} and \cite{Drummond2018a} demonstrate from HD\,189733b and HD\,209458b simulations that longitudinal quenching of C/H/O/N species occurs  inside the jet regions and on the nightside.  \cite{Mol2019} point out for HAT-P-7b, however, that the atmospheric optical depth is little effected, in particular if cloud particles form through other paths than C/H/O/N chemistry.

\smallskip\noindent
{\it 3D GCM input:}We use the SPARC/MITgcm~\citep{Showman2009} to determined the 3D thermal structure of WASP-43b. The model solves the primitive equations coupled to non-grey radiative transfer. We assume solar composition abundances including rainout of condensates but do not consider any cloud radiative feedback. Local chemical equilibrium is assumed, {  since}, as shown by~\citet{Steinrueck2019}, the effect of disequilibrium chemistry on the thermal structure is small compared to the horizontal temperature contrasts. More details of the specific model run can be found in %~\citet{Irwin2019} and ~\citet{Venot2020}. 
 \cite{Parmentier2018,2019arXiv190903233I}. For more details on the GCM approach, we refer the reader to the description in \cite{2019arXiv190608127H}. The outputs of our WASP-43b models are very similar to the ones of \cite{2015ApJ...801...86K} since both are utilising the same base 3D GCM code.

{  We run the simulation for 300 Earth days and average all physical quantities over the last 100 days of simulation. The deeper atmosphere would need a much longer, impractical integration time to converge. Our results therefore rely on the assumption that the deep atmospheric properties do not significantly alter the photospheric layers, an often made and  reasonable assumption as shown by~\citet{2016A&A...595A..36A}. Although this has recently been called into question in the context of grey radiative transfer schemes~\citep{2019arXiv190413334C,Wang2020}, the timescales needed for equilibriation (50,000 to 100,000 days) are impractical when using non-grey radiative transfer. 

\cite{20VePaBl.wasp43b} show in their Fig. 9 that the cloudless GCM spectra fits very well to the dayside atmosphere but is a poor fit to the
nightside atmosphere. In Fig. 11 they show that post-processed clouds allow a much better fit to the nightside atmospheres. When active clouds are added in the GCM the nightside spectra does not change significantly compared to the post-processed case, showing that the nightside temperature structure is not extremely affected. This experiment shows that the cloudless GCM is a reasonable approximation to the temperature map of WASP-43b.}

\smallskip
\noindent
{\it Kinetic cloud formation:} To preserve consistency, we apply the same set-up of our kinetic cloud formation model (nucleation, growth, evaporation, gravitational settling, element consumption and replenishment)  and equilibrium gas-phase calculations like for HAT-P-7b (Sect 2.1 in \cite{2019arXiv190608127H}).  In addition we include KCl as  nucleation and growth species. The seed forming species need to be considered as surface growth material, too, since both processes (nucleation and growth) compete for the participating elements (Ti, Si, O, C, K, and Cl in this work). We consider the formation of 16 bulk materials (s=\ce{TiO2}[s], \ce{Mg2SiO4}[s], \ce{MgSiO3}[s], MgO[s], SiO[s], \ce{SiO2}[s], Fe[s], FeO[s], FeS[s], \ce{Fe2O3}[s], \ce{Fe2SiO4}[s], \ce{Al2O3}[s], \ce{CaTiO3}[s], \ce{CaSiO3}[s], C[s], KCl[s]; {  s indicating that this is a condensate material}) that form from 11 elements (Mg, Si, Ti, O, Fe, Al, Ca, S, C, K and Cl) by 128 surface reactions. The abundance of these 11 elements will decrease if cloud particles are forming (nucleation, growth) and increase if cloud particles evaporate. Sulfur has not been included in our present mineral cloud model.  Sulfuric materials in form of S[s], FeS[s],  MgS[s] would contribute by less than 10\% in volume fraction in a solar element abundance gas \citep[see Fig. 6 in][]{2019AREPS..47..583H}.

\smallskip\noindent
{\it Kinetic haze formation}: We perform the simulations of photochemistry and haze formation using the models of \cite{2018ApJ...853....7K}. The procedure is the same as our previous studies \citep{2019ApJ...876L...5K, 2019ApJ...877..109K}: we first perform photochemical simulations to derive the steady-state distribution of gaseous species. Then, assuming the production rate of haze monomers  
%({\bf utilzing a mass conversion rate based on experiences with Titan}) 
at each altitude as the sum of the photodissociation rates of the major hydrocarbons in our photochemical model, $\mathrm{CH_4}$, $\mathrm{HCN}$, and $\mathrm{C_2H_2}$, we derive the steady-state distribution of haze particles by the particle growth simulations. In the particle growth simulation, distribution of haze particles is derived considering coagulation of  homo-disperse monomers, gravitational sedimentation, and diffusion. Neither evaporation nor condensation is considered. Thermal decomposition such as considered in \cite{2017ApJ...847...32L} can be important in such a hot atmosphere of WASP-43b, but is ignored due to the uncertainty of the thermal stability of haze. We adopt the haze monomer radius of 1~nm and its density of 1~$\mathrm{g}$~$\mathrm{cm^{-3}}$.  The haze formation feedback on the atmospheric chemistry such as the elemental depletion is not considered due to the high computational cost.  For the details of the photochemical and particle growth models, see \cite{2018ApJ...853....7K}.

 We choose four representative points along the equator ($\theta = 0\degree$) to perform the haze formation simulations because of the high computational cost; sub-stellar point ($\phi = 0\degree$), evening terminator ($\phi = 90\degree$), anti-stellar point ($\phi = 180\degree$), and morning terminator ($\phi = 270\degree$). For the anti-stellar point ($\phi = 180\degree$), we only perform the chemical simulation and do not perform the haze formation calculation because of the absence of a sufficient radiation field and hence, no photochemistry.

As for the elemental abundance ratios used in the photochemical calculation, we adopt the original solar ones. Though we have also simulated the photochemistry using the reduced oxygen abundance as a result of mineral cloud formation, we have confirmed that  the depletion of the oxygen abundance in the cloud-forming region is replenished by the vertical eddy diffusion transport from the lower atmosphere with the solar abundance. However, if we are able to couple mineral cloud formation with photochemical simulation, the oxygen abundance in the cloud-forming region should be kept depleted because of the much shorter timescale of cloud formation compared to that of diffusion. Coupling those two simulations is our future work.
For the UV spectrum of WASP-43, since it has not been observed, we use that of HD~85512 constructed by the MUSCLES Treasury Survey \citep{2016ApJ...820...89F, Youngblood:2016ib, 2016ApJ...824..102L} because of its similar stellar type to WASP-43. The attenuation of the UV spectrum due to the wavelength-dependent molecular absorption at each altitude is taken into account.
For the simulation of the sub-stellar point, we ignore the diurnal variation factor of 1/2 in the expression of the photodissociation rate \citep[see][Eq.~(3)]{2018ApJ...853....7K}, which we adopt for the terminator points, and for that of the anti-stellar point, we ignore photochemistry.
Finally, as for the zenith angle of the stellar irradiation, which should be $0.0^{\circ}$ for the sub-stellar point while the value somewhat smaller than $90.0^{\circ}$ for the terminator if the host star is spatially resolved, we use the globally-averaged value of $57.3^{\circ}$ \citep{2012ApJ...761..166H} for all the points for simplicity.

\section{The warm atmosphere of WASP-43b}\label{s:Tp}

Figure~\ref{TpNuc} (left) visualises the thermodynamic input that we utilize for our following study of cloud and haze formation and the gas-phase composition in the atmosphere of WASP-43b. WASP-43b is a hot Jupiter that is assumed to be tidally locked with its host star which results in a dayside that is hotter than the nightside. The resulting  gas temperature difference of $\approx 700$K at a gas pressure of $\approx 10^{-4}$~bar drives a global circulation from the day to the nightside (\citealt{Parmentier2018,2018AJ....155..150M}).  
%This global circulation patter will be affected by the inner boundary (\citealt{2019arXiv190413334C}). 
A smaller temperature difference is suggested between the substellar and anti-stellar point (top left) and the terminator profiles. We also show the dayside (red dash-dot) and the nightside  (blue dash-dot) averages which are poor representations of the day- and the nightside temperature distributions, respectively.

The inspection of the equatorial temperature distribution (Fig.~\ref{TpNuc} bottom left) suggests a west-east asymmetry  with the east ($\phi\rightarrow 90^o$ ) being warmer than the westerly ($\phi\rightarrow 270^o$)  dayside. The day-night temperature difference can clearly be seen from this 2D plots, and the changing temperature throughout the whole terminator regions that is probed by transmission spectroscopy is best seen in  Fig.~\ref{TpNuc} (middle left).

Setting the JWST target WASP-43b in perspective, we further demonstrate that, within the uncertainty of the GCM temperature structures, WASP-43b's day- and  nightside are on average cooler than those of a ultra-hot Jupiter like WASP-18b (Fig.~\ref{TpDP}). We also note that the vertical extension of the GCM profiles fall short of what 1D atmosphere simulations can provide by comparing to our 1D {\sc Drift-Phoenix}  model (Fig.~\ref{TpDP}, solid gray lines). \cite{2019arXiv190413334C} have demonstrated the effect of the inner boundary on the flow regimes in the atmosphere of WASP-43b. This has substantial impact also on the local thermodynamic conditions inside the atmosphere and we will need to return to this challenge in due time.

\section{A theoretical analysis of kinetically forming, mineral clouds on WASP-43b}\label{s:clouds}

In this section, we discuss the principle results for mineral cloud formation in the hot Jupiter WASP-43b to enable a detailed insight into the cloud properties that will shape the JWST spectrum of WASP-43b. In Sect.~\ref{s:ncheq}, we show in how far photochemical hydrocarbon haze formation does complete this picture of cloud formation in the atmosphere of WASP-43b leading to the idea of the possible presence of two chemically distinct cloud ensembles if photochemical processes affect the chemical carbon reservoir of the atmosphere (Sect.~\ref{s:twocloud}).

\begin{figure}
    \includegraphics[width=1.0\linewidth]{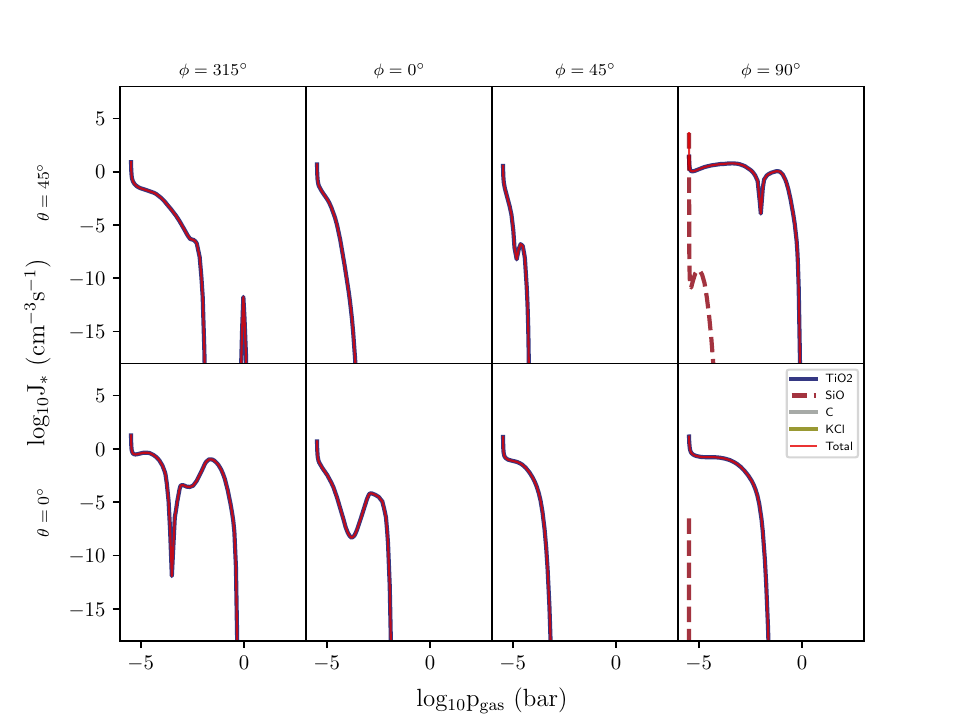}\\
    \includegraphics[width=1.0\linewidth]{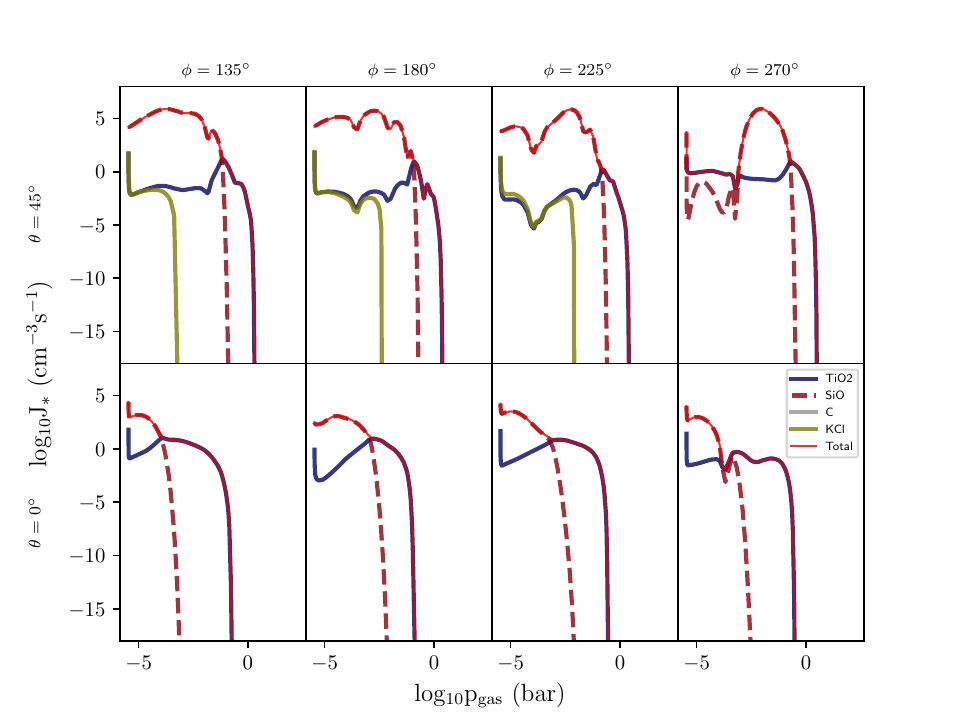}
    \caption{The total seed formation (nucleation) rate, $J_*=\Sigma J_{\rm s}$ [cm$^{-3}$ s$^{-1}$], and the individual nucleation rates, $J_{\rm s}$ [cm$^{-3}$ s$^{-1}$], for s=\ce{TiO2}, SiO, KCl, C on four selected (longitude, latitude) 1D profiles from the 3D GCM (\citealt{Parmentier2018}) for WASP-43b. {\bf Top:} dayside, {\bf Bottom:} nightside.  }
    \label{fig:NucRateDayNight}
\end{figure}

{\it Firstly,} we observe that nucleation seeds form for all the 1D profiles used here to probe cloud formation in the 3D atmosphere of WASP-43b (Fig.~\ref{TpNuc}, right, top). This is in stark contrast to the ultra-hot Jupiters, like WASP-18b and HAT-P7b, but in line with other hot Jupiters like HD\,189733b and HD\,209458\,b. {\it Secondly,} the rate of seed formation differs strongly between the day and the nightside because of the different thermodynamic conditions. The seed formation is more efficient on the nightside where SiO seeds dominate in the low-pressure part of the atmosphere. Once SiO[s] surface growth kicks in efficiently, SiO nucleation is terminated and \ce{TiO2} seeds form more efficiently in the middle atmosphere at $\approx 1$bar. KCl seeds do form on the night side in the low-pressure atmosphere but with a rate that is too low to be significant (Fig.~\ref{fig:NucRateDayNight})

\subsection{Asymmetric cloud coverage and non-homogeneous cloud properties}
The equatorial cut  (Fig.~\ref{TpNuc}, middle right) shows the day-night  and west-east ($\phi=270^o\rightarrow 90^o$) asymmetry in the formation rate of condensation seeds in the atmosphere of WASP-43b. The equatorial asymmetry in the seed formation rate is less strong (Fig.~\ref{TpNuc}, bottom). Based on the present 1D profiles, the eastern dayside of WASP-43b ($\phi=0^o\rightarrow 90^o$) forms condensation seeds in a narrow range at p$_{\rm gas}<10^{-4}$bar. The western dayside  ($\phi=0^o\rightarrow 270^o$) features a wider seed formation zone compared to the eastern dayside which deepens with increasing longitude $\phi$ in the equatorial region of WASP-43b.

\begin{figure}[h]
%    \centering
   \includegraphics[width=\linewidth]{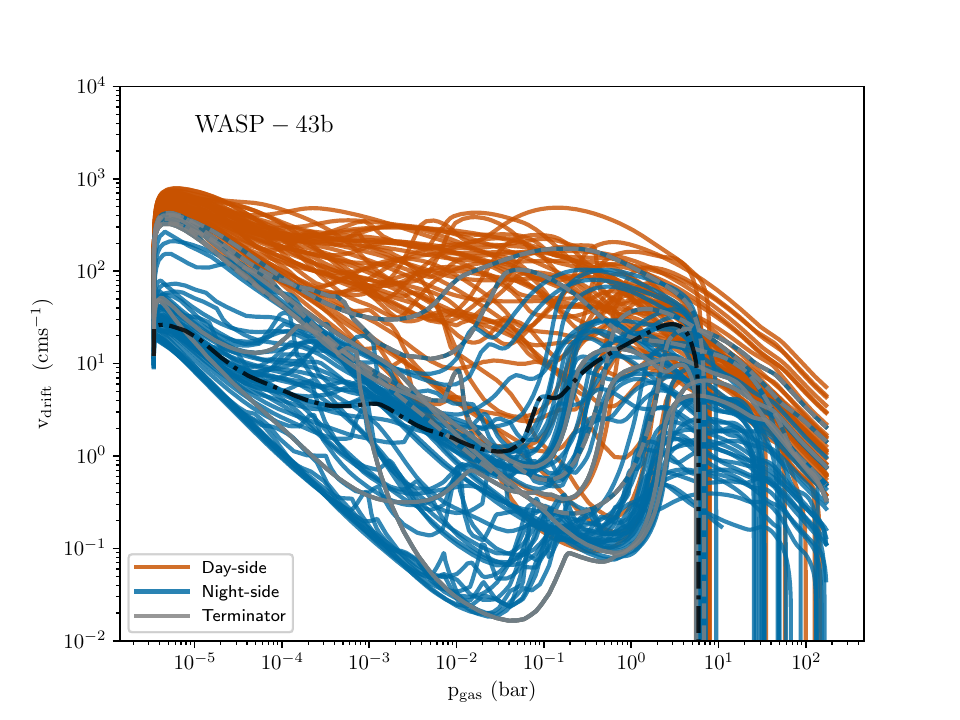}
    \caption{Drift velocity for all 1D trajectories of WASP-43b. The nighside profiles are shown in blue, the dayside profiles in orange, the terminator profiles in grey.}
    \label{fig:vdr}
\end{figure}

%Mean particle size <a>

\begin{figure*}
\begin{tabular}{p{8cm}p{8cm}}
    \includegraphics[width=1.1\linewidth]{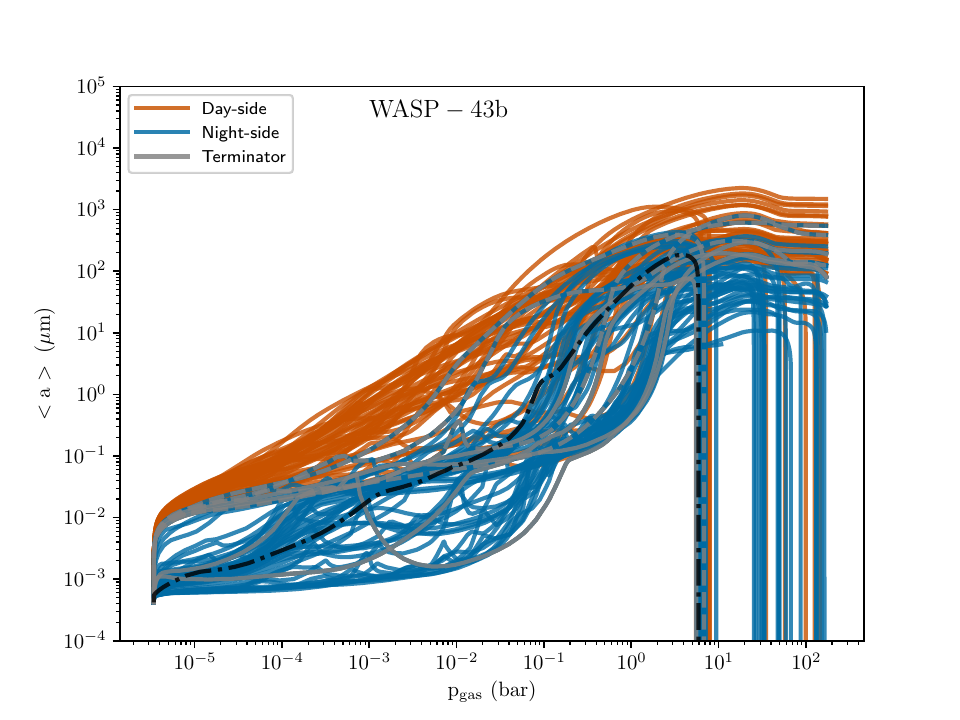} & 
    \includegraphics[width=1.1\linewidth]{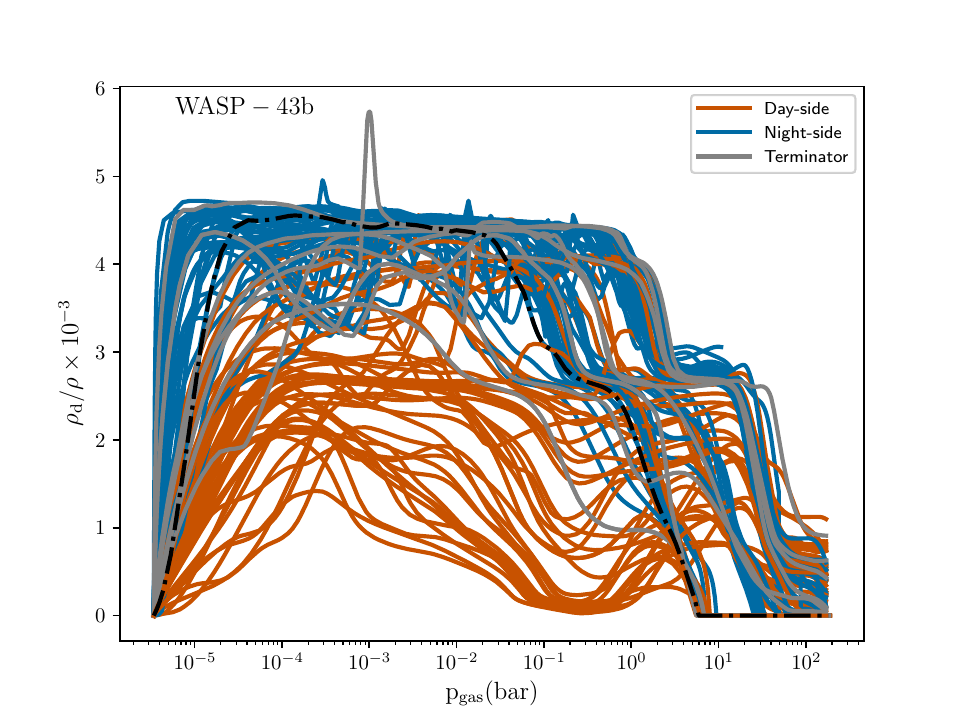}\\
 \includegraphics[page=5,width=0.55\textwidth]{images/Slice_Plots.pdf} &
  \includegraphics[page=3,width=0.55\textwidth]{images/Slice_Plots.pdf}\\*[-3cm]
\includegraphics[page=6,width=0.55\textwidth]{images/Slice_Plots.pdf}&
 \includegraphics[page=4,width=0.55\textwidth]{images/Slice_Plots.pdf}\\*[-1.5cm]
 \end{tabular}
    \caption{{\bf Left:} Mean cloud particle size $\langle a\rangle$ [$\mu$m], {\bf Right:} cloud particle load in terms of dust-to-gas mass ratios, $\rho_{\rm d}/\rho_{\rm gas}$ [$10^{-3}$]. {\bf Top:} for all 1D trajectories. The nightside profiles are shown in blue, the dayside profiles in orange, the terminator profiles in grey. {\bf Middle:} 2D cut  visualising the spatial location in the equatorial plane, {\bf  Bottom:} 2D cut along the terminator for a tidally locked WASP-43b.}
    \label{fig:DusttoGas}
\end{figure*}

\begin{figure*}
     \includegraphics[width=0.45\linewidth]{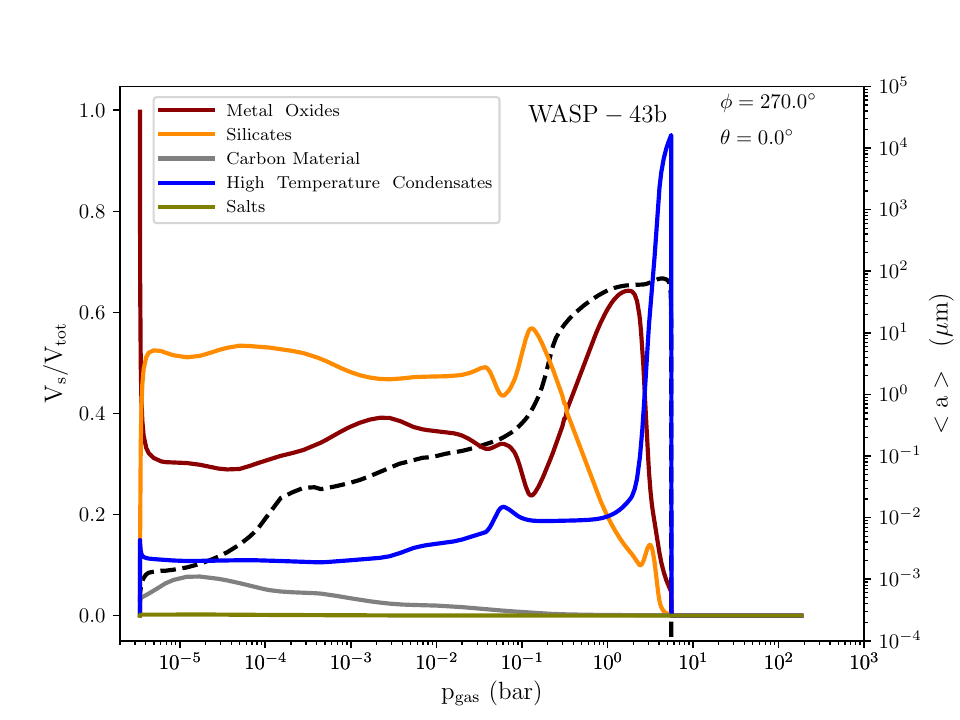}    \includegraphics[width=0.45\linewidth]{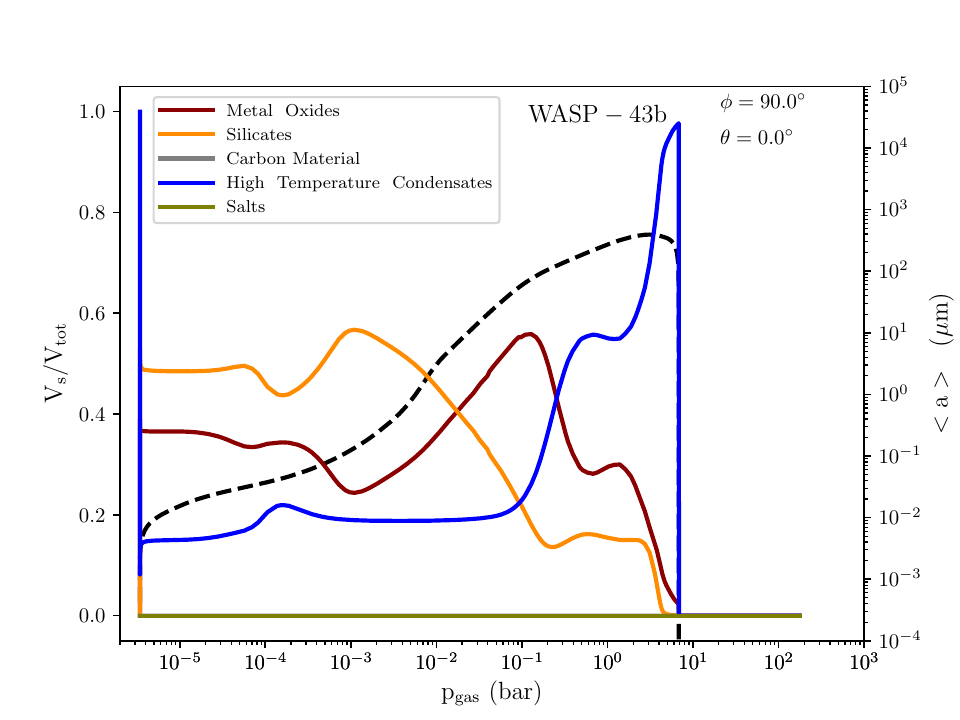}\\*[-0.0cm]
   \includegraphics[width=0.45\linewidth]{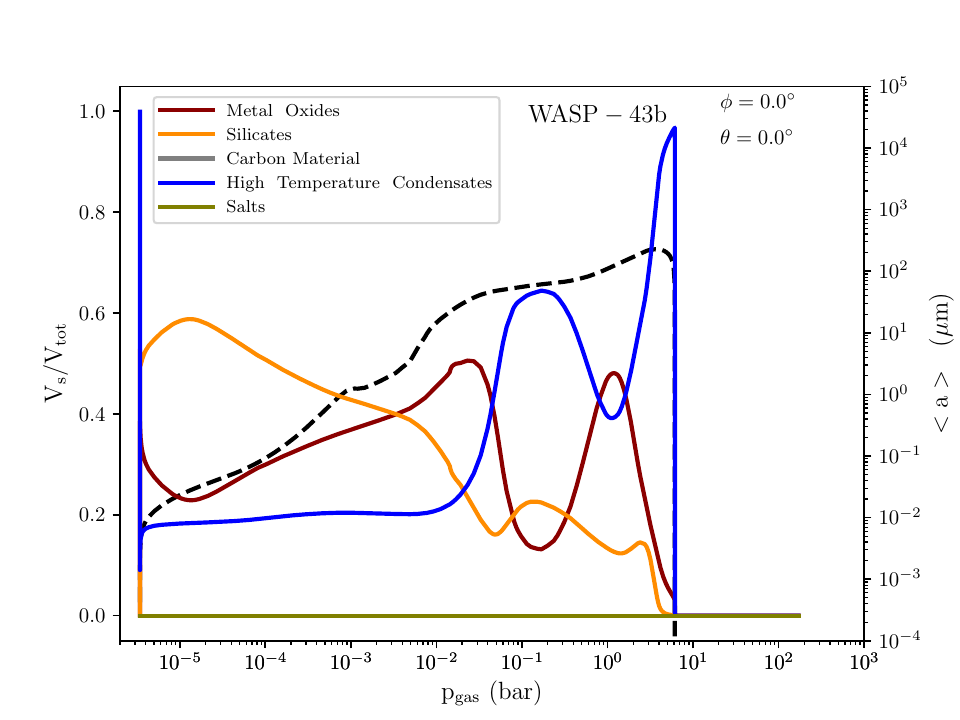}
 \includegraphics[width=0.45\linewidth]{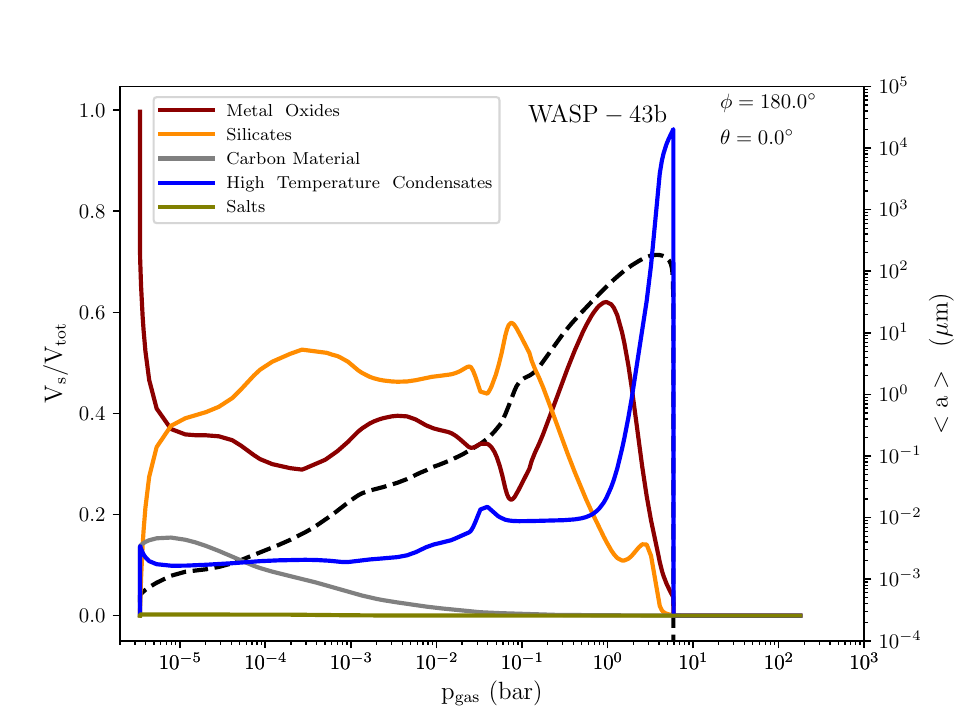}
    \caption{The cloud particle material composition grouped into 5 categories for the  morning terminator (top left), evening terminator (top right), sub-stellar point (bottom left), anti-stellar point (bottom right). }     
    \label{fig:5line_MT}
\end{figure*}

%\begin{figure}
%    \includegraphics[width=\linewidth]{images/mat_vol_frac_CombinedTerminators.pdf}
%    \caption{Material volume fractions compared for both terminators.}
%    \label{fig:matvolfrac_combined}
%\end{figure}

The lower boundary of the seed formation region (see Fig.~\ref{TpNuc}, right) does not coincide with the lower boundary of the cloud (compare Fig.~\ref{fig:DusttoGas}, left) because the particles grow and fall (gravitational settle) through the atmosphere into deeper, warmer atmospheric regions of increasing pressure. The growth in these warm regions is supported by the increasing gas pressure which increases the collision rates. The higher pressures further stabilises materials that already formed at high temperatures. The inner cloud boundary is set by the thermal stability of the particles of a particular size; big particles survive longer as they have more material to evaporate. 

Figure~\ref{fig:DusttoGas} (left) demonstrates that the cloud particles sizes are not homogeneous but change for all 1D profiles probed in our study. Figure~\ref{fig:DusttoGas} further demonstrates that cloud particles are present thought almost the entire atmospheric volume sampled here, in particular also on both the day- and the nightside. The characteristic cloud properties, however, are different between the day- and the nightside. As consequence of the seed formation discussed with regard to Fig.~\ref{TpNuc} (right), the mean cloud particle sizes are larger over a larger pressure range on the easterly dayside ($\phi=0^o\rightarrow 90^o$) compared to the westerly dayside ($\phi=0^o\rightarrow 270^o$)  and the nightside (Fig.\ref{fig:DusttoGas}, middle left). A similar but less pronounced asymmetry occurs for the terminator regions (Fig.\ref{fig:DusttoGas}, bottom left). 

The cloud opacity is determined by the cloud particles sizes, their abundance and their material properties. The dust-to-gas ratio, $\rho_{\rm d}/\rho_{\rm gas}$ (Fig.\ref{fig:DusttoGas}, right) combines these properties and visualises where the highest cloud mass load occurs in an atmosphere. $\rho_{\rm d}/\rho_{\rm gas}$ shows that only the inner part of the WASP-43b atmosphere is cloud-free. The largest $\rho_{\rm d}/\rho_{\rm gas}$ occurs on the nightside and reaches well across the morning terminator at $\phi=270^o$. The dayside is not cloud free but has a considerable lower cloud load, hence, it might be mistakenly interpreted as cloud-free.

Simple time scale estimate shows that cloud particle of sizes $\langle a\rangle\approx 0.001\,\ldots\,0.1\mu$m (Fig.~\ref{fig:DusttoGas}) {  occurring in the terminator and dayside regions} remain in the atmospheric regions accessible to transmission spectroscopy of p$_{\rm gas}\approx 10^{-5}\ldots\,10^{-4}$ bar for  about 2h, using $H\approx 8\cdot10^6$cm (Fig.~\ref{fig:meanmolweight}) and $v_{\rm dr}=10^3$cm/s (Fig~\ref{fig:vdr}). With a slightly smaller nightside scale height of  $H\approx 3\cdot 10^6$cm and $v_{\rm dr}=5$cm/s, the respective cloud particles that formed in the low-pressure regions of p$_{\rm gas}\approx 10^{-5}\ldots\,10^{-4}$ bar   and grew to a mean particle size of only $\langle a\rangle\approx 10^{-3}\mu$m remain $\approx 170$h (or $\approx 7$ d with a day of 24h). Once the nightside particles have sank into the deeper atmosphere, the time scale decrease considerably as their size increases by orders of magnitudes and, although time scale vary throughout the atmosphere, the settling time scale become comparable to the dayside example stated above.

\subsection{General material composition of cloud particles}
Figure~\ref{fig:5line_MT} demonstrates that the material composition of the cloud particles on WASP-43b does not change dramatically between the day- and the nightside in the low pressure regions where  p$_{\rm gas}\approx 10^{-6}\ldots\,10^{-4}$ bar for the 1D profiles inspected in this study: The dominating materials are silicates (\ce{MgSiO3}[s], \ce{Mg2SiO4}[s], \ce{Fe2SiO4}[s], \ce{CaSiO3}[s]) and metal oxides (MgO[s], SiO[s], \ce{SiO2}[s], FeO[s], \ce{Fe2O3}[s], \ce{CaTiO3}[s]). Silicates  dominate the cloud particle volume by far with values of 45\%$\,\ldots\,60\%$. High temperature condensates (\ce{TiO2}[s],  Fe[s], \ce{Al2O3}[s]) dominate the cloud particle volume at higher temperatures where silicates have evaporated. Carbon contributes a little bit, and salts (here KCl[s]) play no significant role for the material composition in the temperature ranges of the WASP-43b atmosphere investigated here.  KCl did occur as a minor nucleation species in the northern hemisphere of the nightside of WASP-43b. A detailed account of the individual contributions of the 16 materials considered in our study is given in Fig.~\ref{fig:matvolfrac_ASP} for the two terminators, the substellar and the anti-stellar point.
We note that the terminator structures are rather comparable in the frame of the models applied here.

\begin{figure}
    \includegraphics[width=0.5\textwidth]{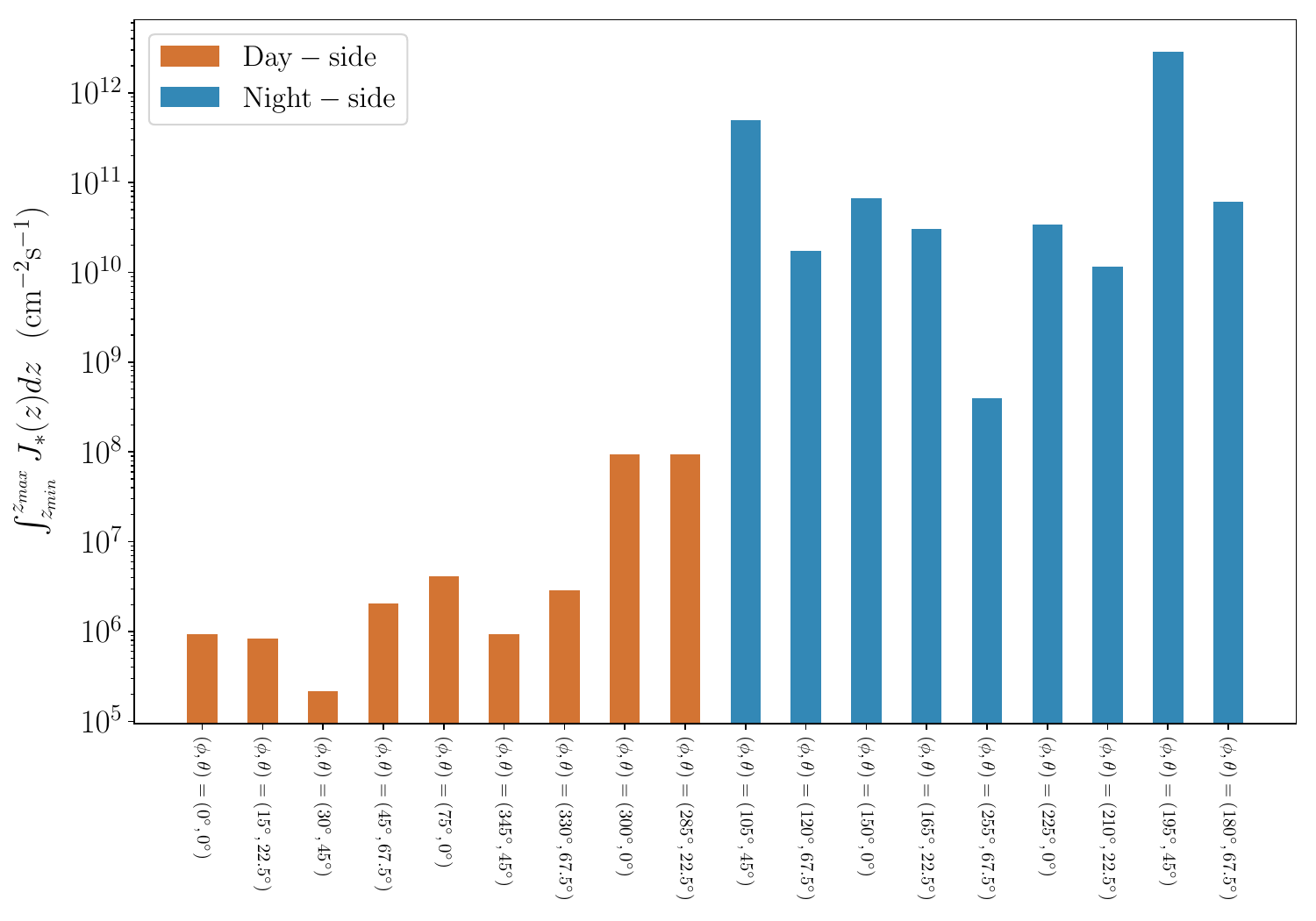}
    \includegraphics[width=0.5\textwidth]{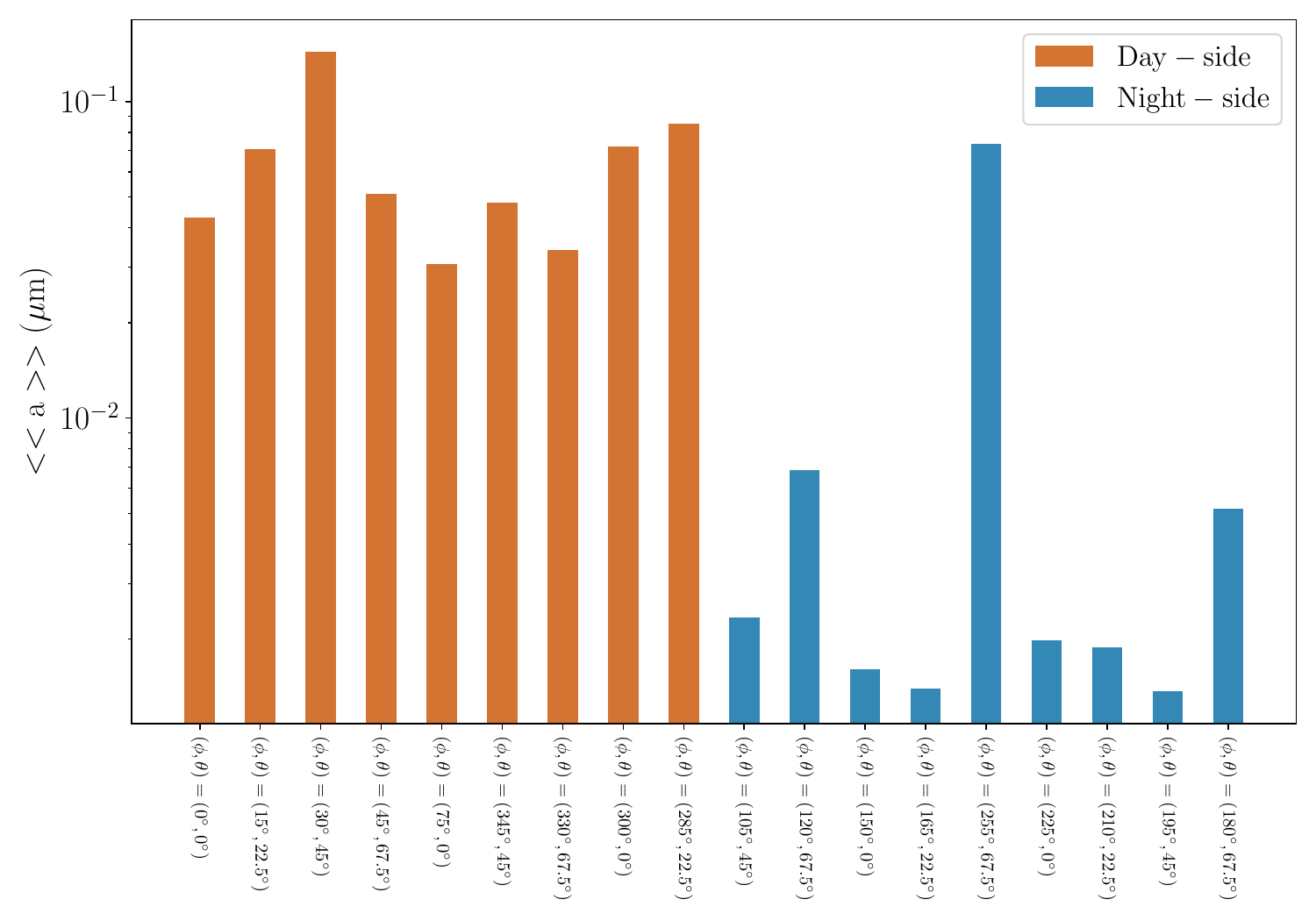}\\
    \includegraphics[width=0.9\linewidth]{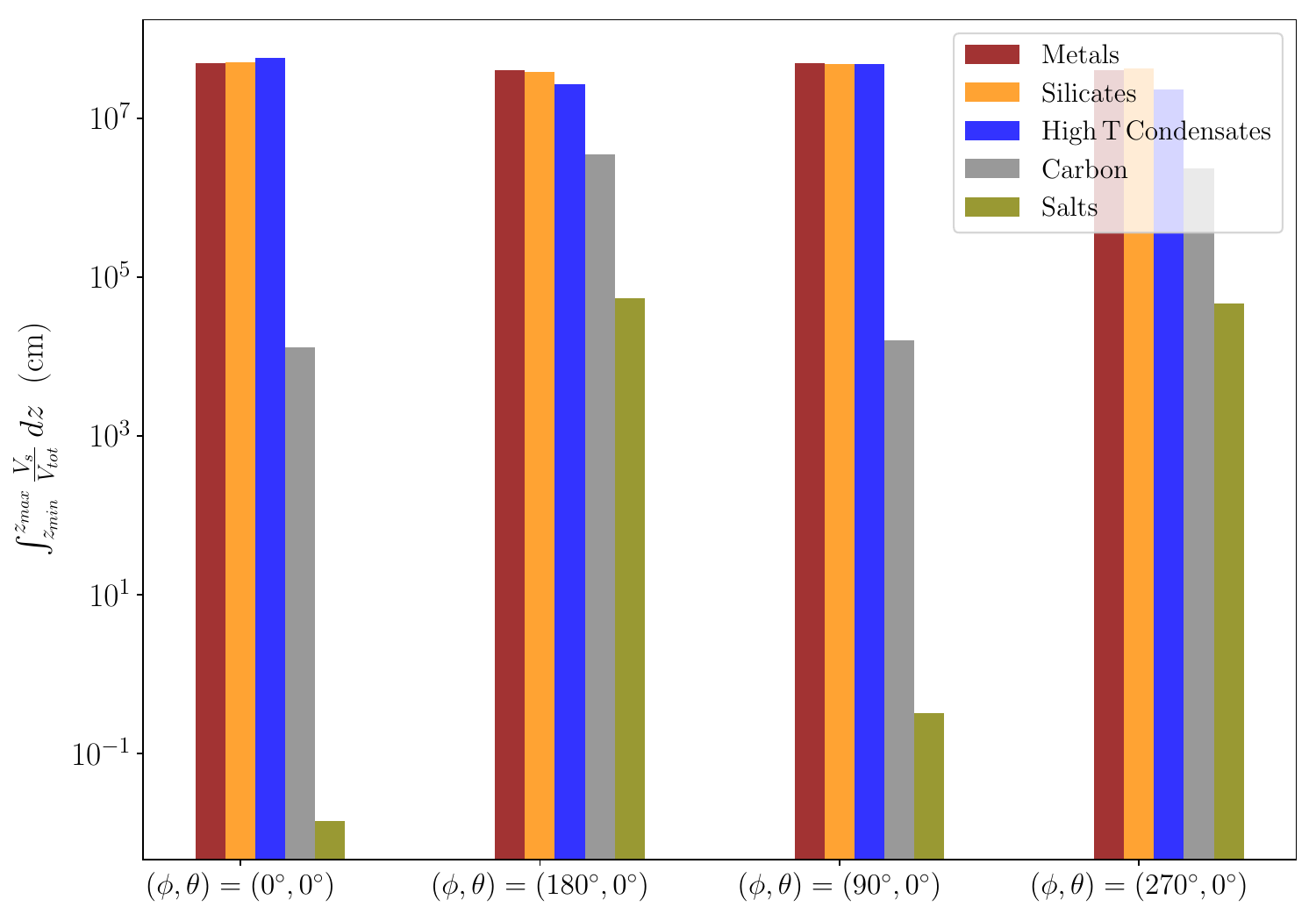}%
    \caption{Column integrated cloud properties for WASP-43b for 18 selected 1D profiles. {\bf Top:} Nucleation rate shown  $\int_{\rm zmin}^{\rm zmax} J_*(z) dz$ [cm$^{-2}$ s$^{-1}$] with $J_*=\sum_{\rm s} J_{\rm s}$ ($s=\,$\ce{TiO2}, SiO, C, KCl) {\bf Middle:}  
    $\langle\langle a \rangle\rangle = (\int_{\rm zmin}^{\rm zmax} n_{\rm d}(z)\langle a(z) \rangle \,dz)/( \int_{\rm zmin}^{\rm zmax} n_{\rm d}(z) \,dz)$ [$\mu$m] with $n_{\rm d}(z)$ [cm$^{-3}$] the local cloud particle number density.
    {\bf Bottom:}  Material volume fraction: $\int_{\rm zmin}^{\rm zmax} (V_{\rm s}(z)/V_{\rm tot}(z))\,dz$.
    %Column integrated mean particle size normalised by $\text{z}_{max}-\text{z}_{min}$, shown for 18 profiles. Normalised integrated <a> = $\frac{\Sigma_{i} \: \text{<a>}_{i} \: (\text{z}_{i+1}-\text{z}_{i})}{\text{z}_{max}-\text{z}_{min}}$
    }
    \label{fig:<<a>>_Parm}
\end{figure}

%Mean Molecular Weight
\begin{figure}
    \includegraphics[width=1.\linewidth]{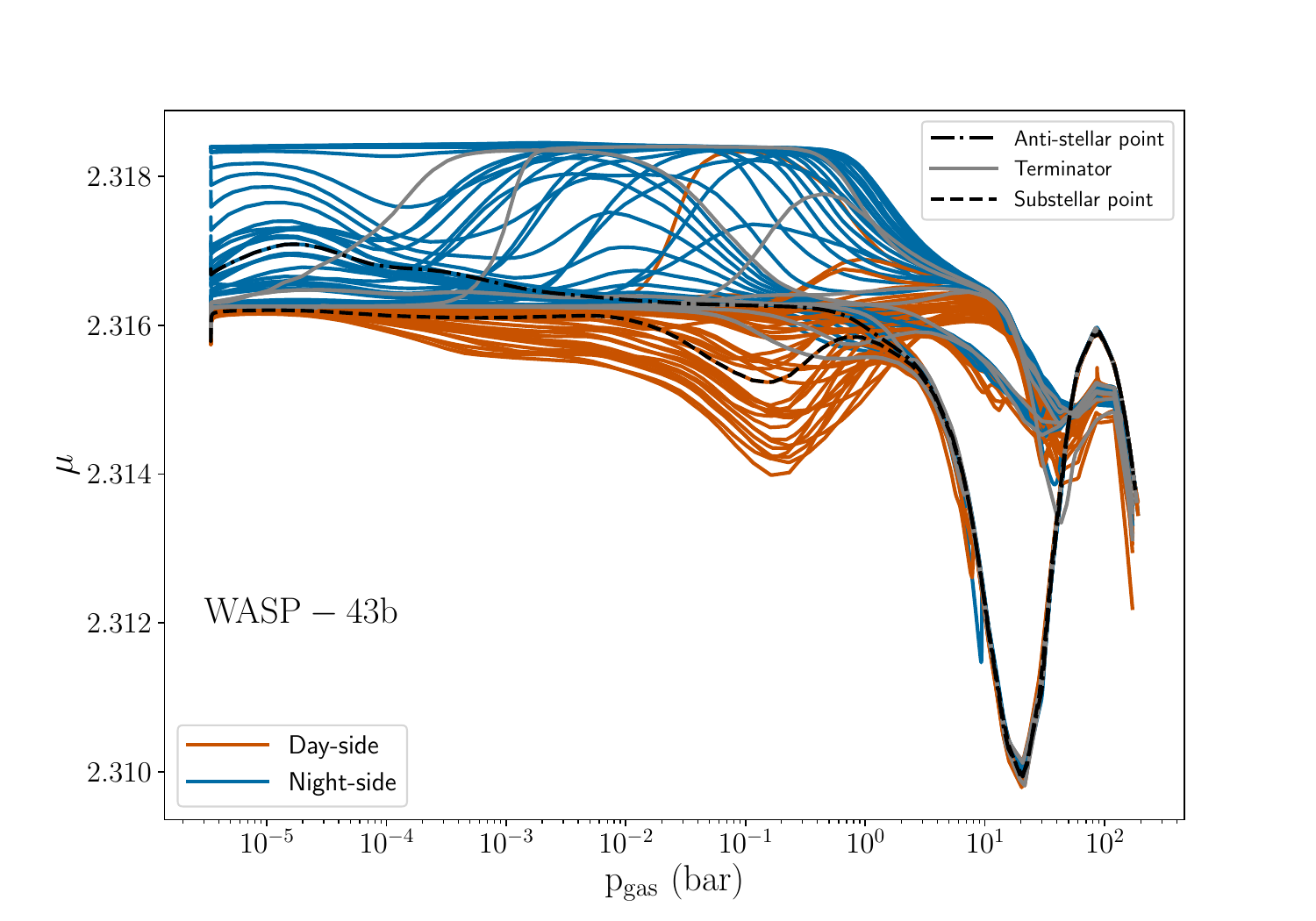}\\*[-0.47cm]
\includegraphics[width=1.0\linewidth]{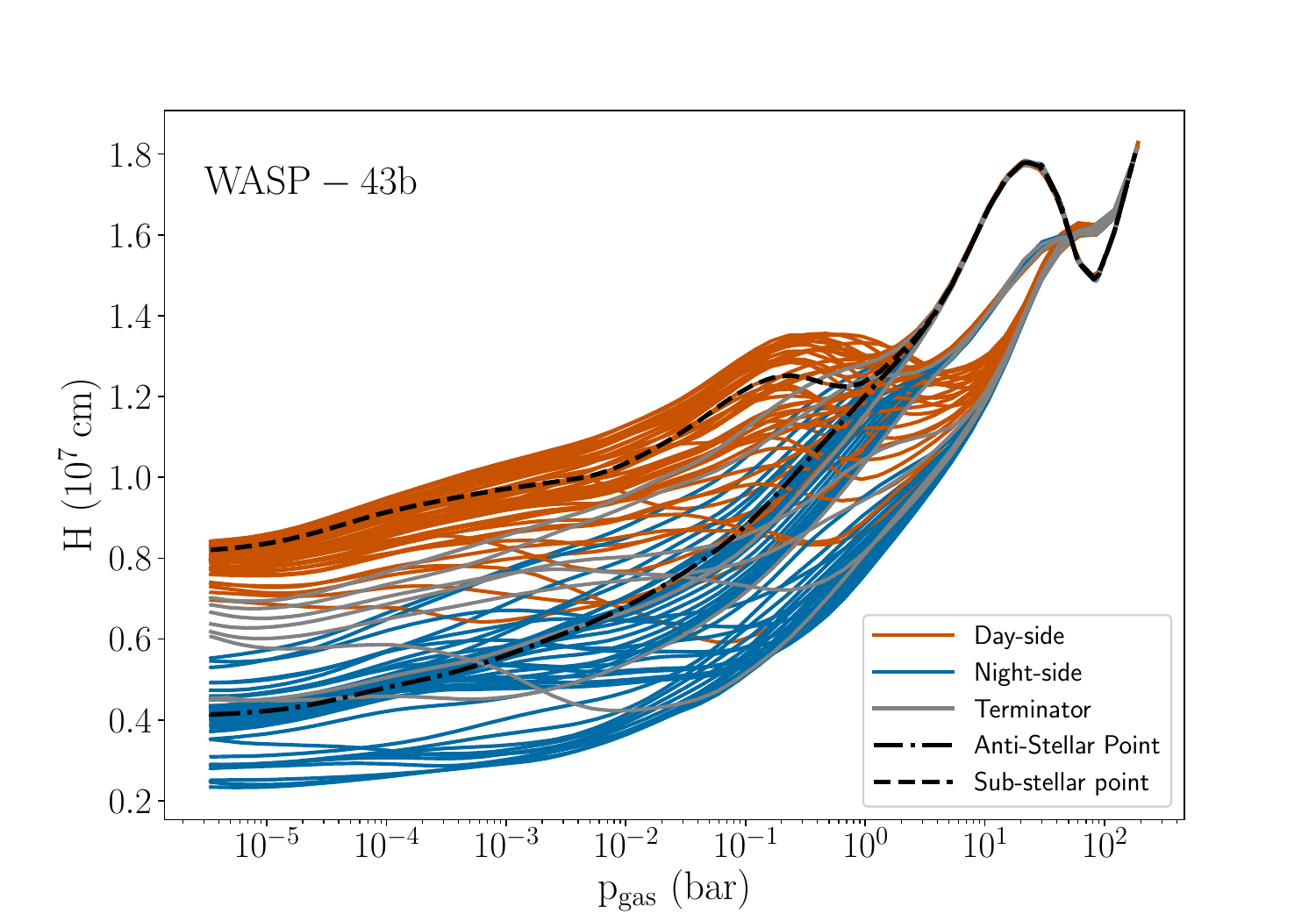}
    \caption{{\bf Top:} {  Variations in} the mean molecular weight, $\mu$ [amu], around the canonical value for a \ce{H2}/He gas of $\mu=2.316$. {\bf Bottom:} The resulting change in the atmospheric pressure scale height, $H$ [$10^7$cm], changes by $\Delta H\approx 6\cdot10^6$cm from the day- to the nightside at p$_{\rm gas}<0.5$bar. }
    \label{fig:meanmolweight}
\end{figure}

%C/O Ratio
\begin{figure}[h]
    \centering
    \includegraphics[width=\linewidth]{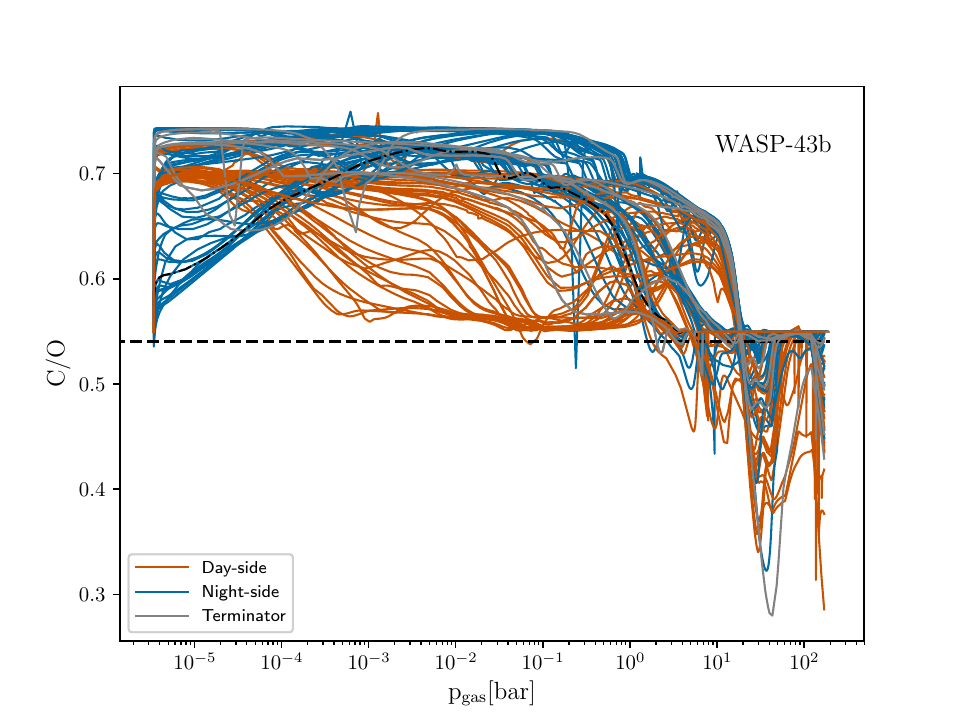}\\*[-0.6cm]
    \includegraphics[width=\linewidth]{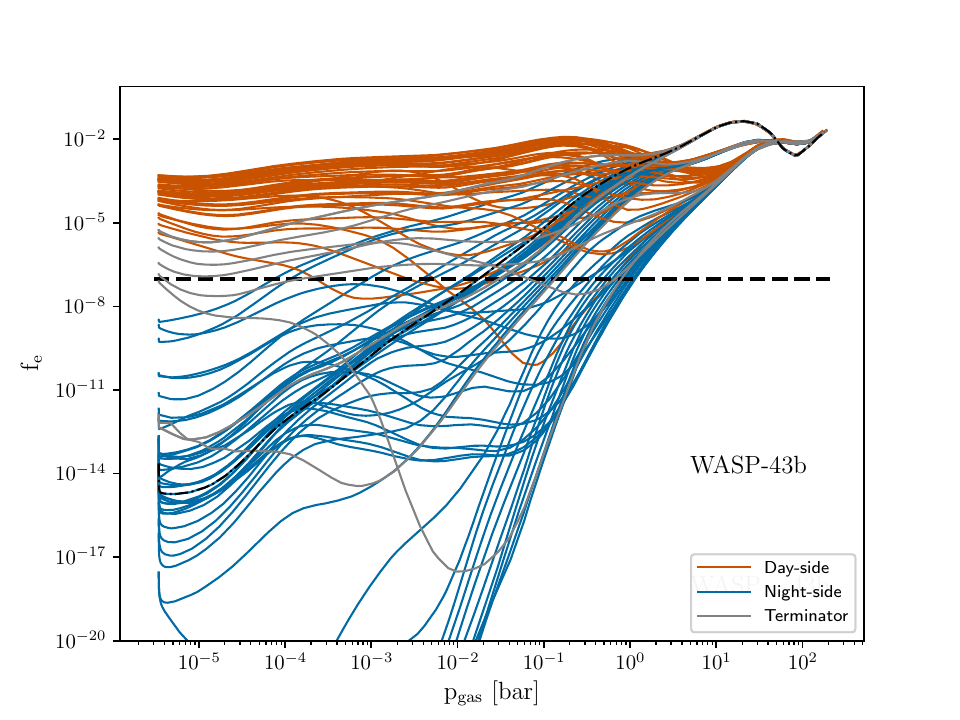}
    \caption{{\bf Top:} Carbon-to-oxygen ratio, C/O {(dashed horizontal line: solar C/O)}. {\bf Bottom:} Degree of ionisation, $f_{\rm e}=p_{\rm gas, tot}/p_{\rm e}$ (dashed horizontal line:  $f_{\rm e}=10^7$ as a threshold value for plasma behaviour). C/O and $f_{\rm e}$ are shown  for all 1D profiles studied here for the atmosphere of WASP-43b. The nighside profiles are shown in blue, the dayside profiles in orange, the terminator profiles in grey.}
    \label{fig:COratio}
\end{figure}

\subsection{Column integrated cloud properties}

Column integrated cloud properties may enable us to compare complex information without compromising the information content too much. All three characteristic cloud properties, nucleation rate, mean particles size and material composition, can change throughout a planetary atmosphere because the thermodynamic conditions change vertically and horizontally. This has been demonstrated for WASP-43b in the previous sections. The vertical structures have been shown to be very inhomogeneous such that not one value suffices to describe exoplanet clouds, and the same holds for the horizontal cloud distributions for which we have not only demonstrated a large day-night difference but our model suggests also a west-east change in cloud properties on the dayside of WASP-43b. Despite this complexity, column integrated mean values of selected cloud properties do recover our major findings from the previous, more detailed discussions as follows.

Figure~\ref{fig:<<a>>_Parm} (top) shows the column integrated total nucleation rate  $\int_{\rm zmin}^{\rm zmax} J_*(z) dz$ [cm$^{-2}$ s$^{-1}$] with $J_*=\sum_{\rm s} J_{\rm s}$ ($s=\,$\ce{TiO2}, SiO, C, KCl),  $z_{\rm min}$ and $z_{\rm max}$ spanning the geometric extension of the atmosphere at each of 18 (out of 97) selected  (longitudes, latitude)-profiles. Figure~\ref{fig:<<a>>_Parm} (middle) provides the height-averaged, number density weighted cloud mean particle size $\langle\langle a \rangle\rangle = (\int_{\rm zmin}^{\rm zmax} n_{\rm d}(z)\langle a(z) \rangle \,dz)/( \int_{\rm zmin}^{\rm zmax} n_{\rm d}(z) \,dz)$ [$\mu$m] with $n_{\rm d}(z)$ [cm$^{-3}$] the local cloud particle number density, and  $\langle a(z) \rangle$ the mean particle size as defined in Eq.~\ref{eq:amean2}. Figure~\ref{fig:<<a>>_Parm}  very clearly shows the anti-correlation between the nucleation rate (top) and the mean cloud particle size (middle) also in these column integrated values: The dayside (orange) shows a distinctly lower column integrated  seed formation rate than the nightside which corresponds to distinctly larger average mean particle sizes on the dayside compared to the nightside. The terminator profiles are not included in Fig.~\ref{fig:<<a>>_Parm}. 

Figure~\ref{fig:<<a>>_Parm} (bottom) shows the column integrated values for the material composition ($\int_{\rm zmin}^{\rm zmax} (V_{\rm s}(z)/V_{\rm tot}(z))\,dz$) for the substellar and the anti-stellar point, and the two equatorial terminator points.  These values demonstrate the contribution of the individual material groups to the composition of the whole cloud at this longitudes and latitudes. It suggests that the mineral clouds that form on WASP-43b are mainly made of silicates, metal oxides and high-temperature condensates. Salts like KCl[s] play a less important role in the frame of the model used here.

These integrated values do, however, not provide much insight into cloud formation in  particular pressure regimes like for example the low-pressure regime that is tested by transmission spectroscopy.

%_______________________________________________________________________________
\section{Gas phase composition in chemical equilibrium}\label{s:cheq}

The atmospheric composition on WASP-43b is expected to be that of a warm, dense planet where cloud formation has affected the abundance of  11 elements (Mg, Si, Ti, O, Fe, Al, Ca, S, C K, Cl) in a hydrogen-dominated atmosphere. Consequently, \ce{H2} is the most abundant gas-phase species inside the collision dominated part of the atmosphere of WASP-43b.  This can change for the outermost, more diluted atmospheric layers where external high-energy radiation may cause \ce{H2} to photoionise similar to brown dwarfs (\citealt{2018A&A...618A.107R}). Brown dwarfs do form a shell of ionised gas (an ionosphere) in these photo-dissociation regions.  CO appears as the most abundant molecule with \ce{H2O} becoming more abundant only in the uppermost, low-pressure atmosphere p$_{\rm gas}<10^{-4}$bar on the nightside.
The abundance hierarchy of the 3rd and 4th most abundant gas species depends considerably on the local thermodynamic conditions (Fig.~\ref{fig:Mol_ASP}). The equatorial evening terminator (longitude $\phi= 90^o$) shows \ce{CO2} and \ce{SiO}  as the next most abundant gas species in chemical equilibrium.  The equatorial morning terminator (longitude $\phi= 270^o$) has considerably more \ce{CH4} than \ce{CO2} throughout. The substellar and the anti-stellar point show this shift in SiO/\ce{CH4}/CO abundances more clearly. On the dayside, \ce{H2O} appears with a number density  comparable to CO for p$_{\rm gas}<0.1$bar. Small carbon (C, \ce{C2}) and hydrocarbon (CH, CN, \ce{C2H2}, HCN) molecules pertain their abundance hierarchy  in contrast to the changing abundance of \ce{CH4} for the four profiles studied in more detail in Fig.~\ref{fig:Mol_ASP}. Section~\ref{s:ncheq} will assess their abundances under the effect of an external radiation field and vertical advection.

\subsection{Mean molecular weight and atmospheric scale height}

Complex models require physically justified assumptions. One of them is the mean molecular weight which allows an easy conversion of the total gas pressure into the total gas density applying the ideal gas law. Being able to assume a constant mean molecular weight when running an 3D GCM is computationally advantageous. We demonstrate here that this assumption is justified for the collision dominated part of the atmosphere of WASP-43b, and we demonstrate its effect on the hydrostatic atmospheric scale height. This will change where processes occur that e.g. photionise \ce{H2}, hence in the maybe molecular-dynamics dominated upper atmosphere where mass-loss mechanisms may take effect.

Based on our chemical equilibrium calculations that are affected by the depletion and enrichment of elements due to cloud formation (Fig.~\ref{fig:Mol_ASP}, top) and based on the 1D (T$_{\rm gas}$, p$_{\rm gas}$)-profiles extracted from the Parmentier-3D GCM, the mean molecular weight is approximately constant ($\mu = 2.31\,\ldots\,2.3185$) around the solar value of an \ce{H2}-dominated atmosphere of $\mu=2.316$ (Fig.~\ref{fig:meanmolweight}, top). 
Our photochemical simulations also show that the mean molecular weight is almost constant below the pressure range of $10^{-6}$~bar (i.e., GCM calculation range), above which it decreases except for the nightside as H becomes the dominant species due to the photodissociation of $\mathrm{H_2}$ (see Fig.~\ref{fig:photo}).
The nightside tends towards the higher values while the dayside tends towards the lower value in this range. Casting this in terms of the hydrostatic pressure scale height, $H=kT_{\rm gas}/(\mu m_{\rm H} g)$ (Fig.~\ref{fig:meanmolweight}, bottom), the large temperature differences result in a change of the scale height of $H=2.5\,\ldots\,8.5\cdot 10^6$cm in the upper, low pressure part of the atmosphere. 
%\yui{[Please modify this sentence because the temperature difference is larger than that of the mean molecular weight]} 
The day-night-difference of $\Delta H \approx 6\cdot10^6$cm (6000km) prevails throughout the atmospheric profiles for pressures p$_{\rm gas}<0.5$bar.

\subsection{The carbon-to-oxygen ratio}

Mineral ratios like Si/O,  Mg/O, and the carbon-to-oxygen ratio, C/O, may enable linking observations to evolutionary stages of exoplanets. The challenge herewith is that observations have so far only provides snippets of the atmosphere spectrum, i.e. information about limited wavelength intervals. This limitation in combination with a limited range of molecular absorbers included in retrieval codes has led to suggestions of C/O>1, i.e. carbon-rich exoplanet atmospheres. The carbon abundance can only exceed the oxygen abundance if either carbon is produced (like in AGB stars) or inserted otherwise into the system (e.g. meteoritic influx), or if the oxygen is simply reduced below the original carbon-abundance level. The latter might occur if the planet already starts out with a primordial atmosphere that is oxygen-depleted due to the formation of \ce{H2O}/CO\, ices inside the planet-forming disk.  The accretion of the gaseous envelope would then occurs  from this oxygen-depleted gas and further processing of the atmospheric gas into clouds can then lead to a C/O>1 (\citealt{2014Life....4..142H}).

Figure ~\ref{fig:COratio} details C/O for the 97 1D profiles that we have investigated for the atmosphere of WASP-43b. C/O traces the location of the cloud in pressure space: A sharp increase defines the cloud top which is where oxygen is effectively stored in cloud particles. The detailed shape of the C/O curves depend on the details of the atmospheric temperature and pressure but it is clear that depletion pushes C/O to values of $\approx 0.74$. The trend reverses in deeper atmospheric layers where high temperature cause oxygen-carrying solids to evaporate. An enrichment of oxygen results and, hence, C/O decreases to values as low as 0.3. C/O values for WASP-43b do vary throughout the atmosphere and between the day- and the nightside.

\subsection{The ionised fraction of the atmospheric gas on WASP43b}

WASP-43b is a giant gas planet that orbits its host star rather closely at a distance of $a\approx 00153$au in 0.813 days, but its host star is comparably cool compared to ultra-hot Jupiters like WASP-18b or HAT-P-7b. The degree of ionisation, $f_{\rm e}=p_{\rm e}/(p_{\rm gas}$ ($p_{\rm e}$ -- electron pressure, $p_{\rm gas}$ -- the total gas pressure), is low on the day- and on the nightside (Fig.~\ref{fig:COratio}, bottom). The atmosphere on the dayside can be considered as partially ionised as $f_{\rm e}\approx 10^{-3.5} > 10^{-7}$. While $f_{\rm e}\approx 10^{-7}$ has been postulated as threshold for possible plasma behavior in a gas, follow-up studies are required to quantify this for WASP-43b as outlined in \cite{2015MNRAS.454.3977R}.

The dominating electron donors are \ce{K+} and  \ce{Na+}, followed by \ce{Ca+}. This is very much in line with results obtained for brown dwarfs (see Fig. 3 in \citealt{2015MNRAS.454.3977R}).  \ce{H-} appears as the third most abundant ion for WASP-43b (Fig.~\ref{fig:Ions_ASP}).

%_________________________________________________________________________

\begin{figure*}[h]
\begin{minipage}[b]{0.5\linewidth}
  \centering
  \includegraphics[width=\linewidth]{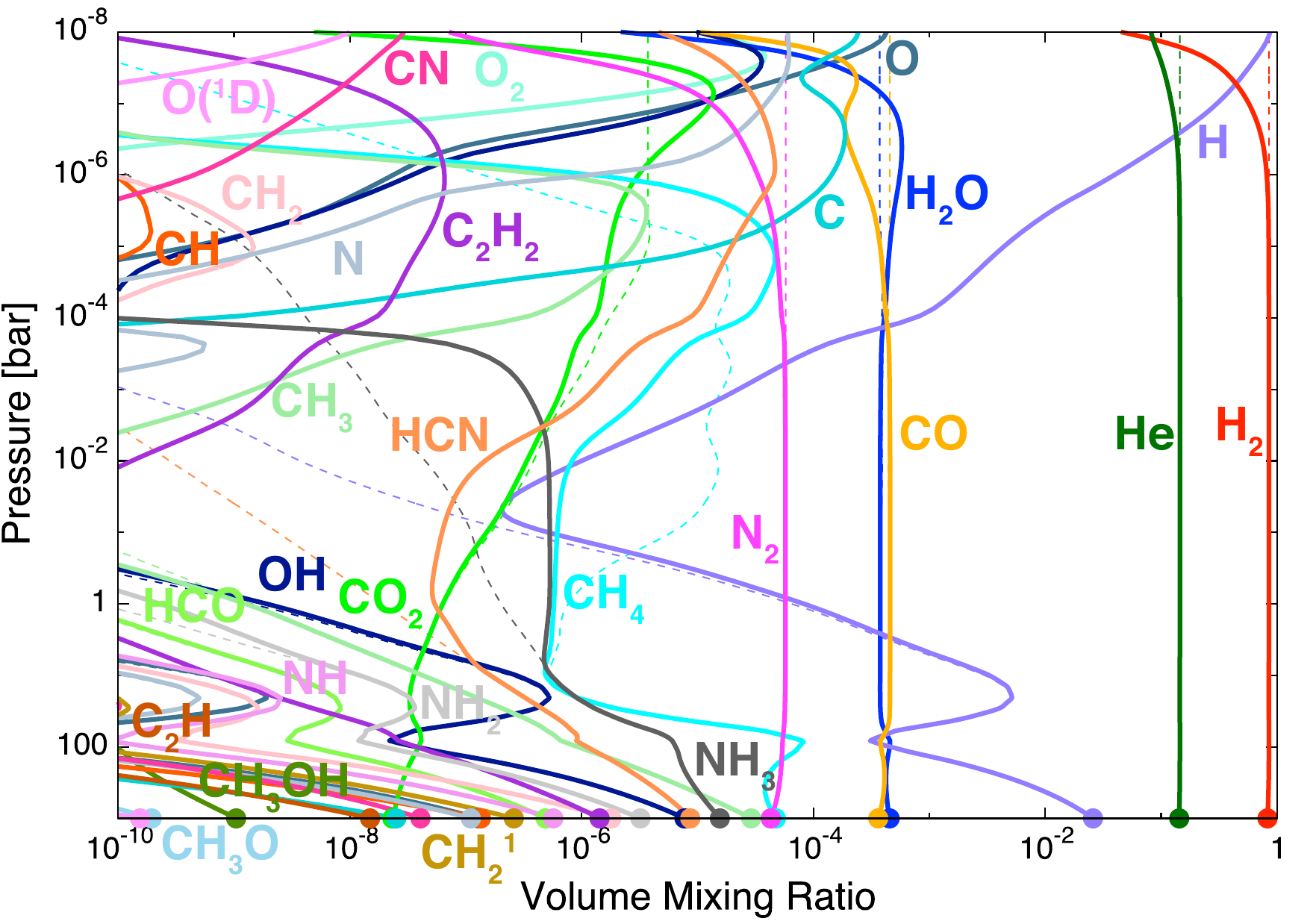}
  \subcaption{Morning terminator ($\phi = 270\degree$)}
 \end{minipage}
 \begin{minipage}[b]{0.5\linewidth}
  \centering
  \includegraphics[width=\linewidth]{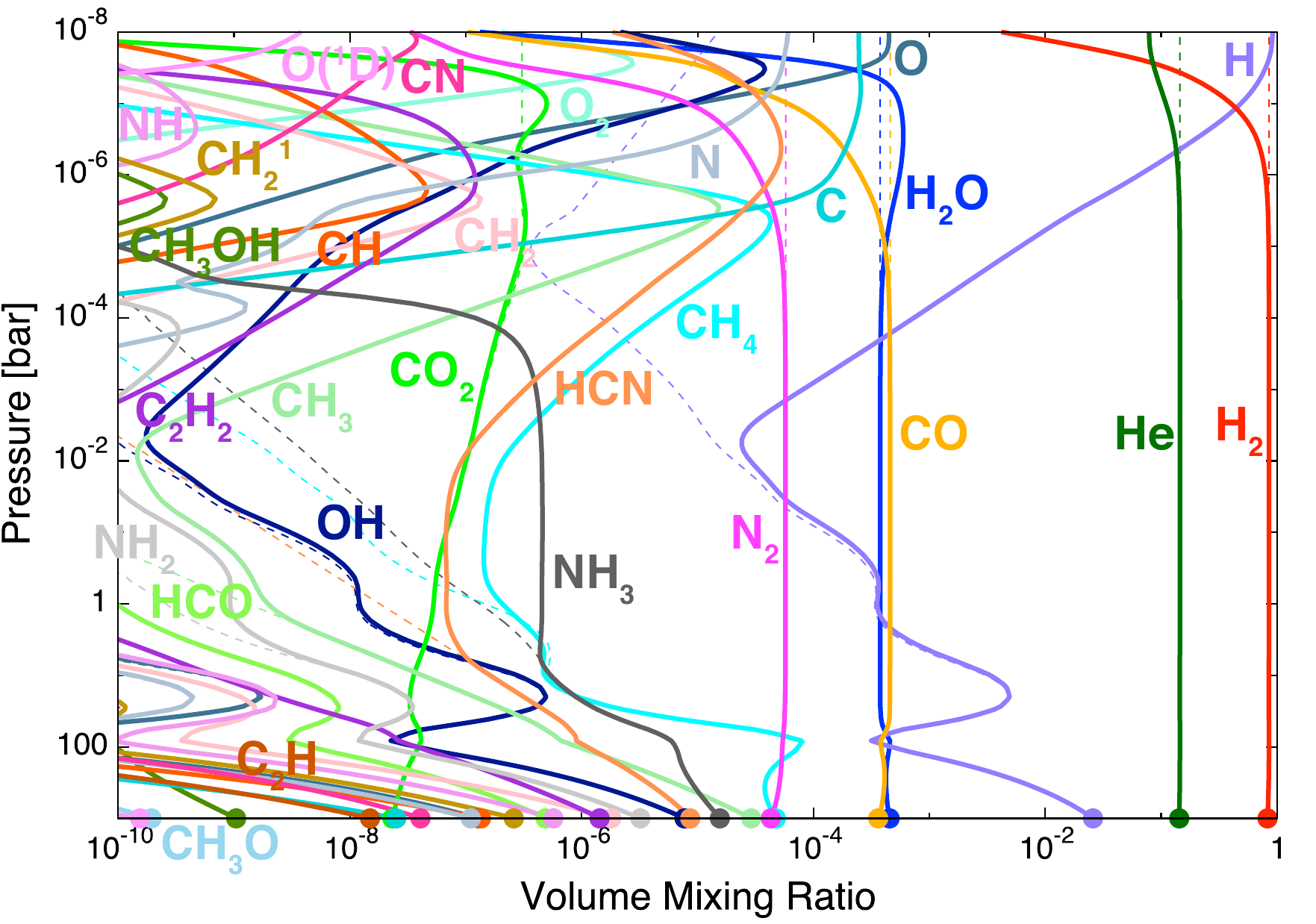}
  \subcaption{Evening terminator ($\phi = 90\degree$)}
 \end{minipage}\\
 \begin{minipage}[b]{0.5\linewidth}
  \centering
  \includegraphics[width=\linewidth]{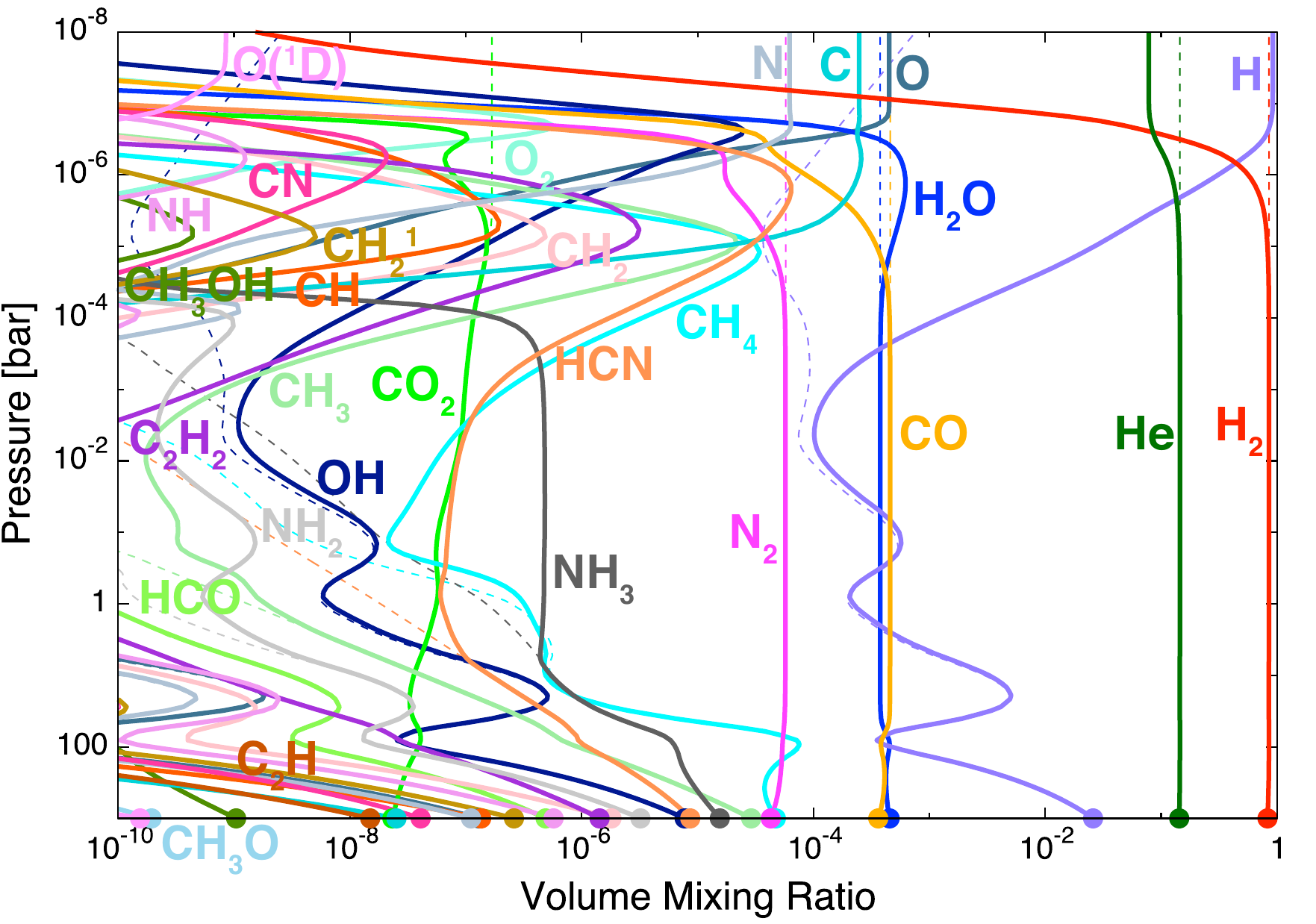}
  \subcaption{Sub-stellar point ($\phi = 0\degree$)}
 \end{minipage}
 \begin{minipage}[b]{0.5\linewidth}
  \centering
  \includegraphics[width=\linewidth]{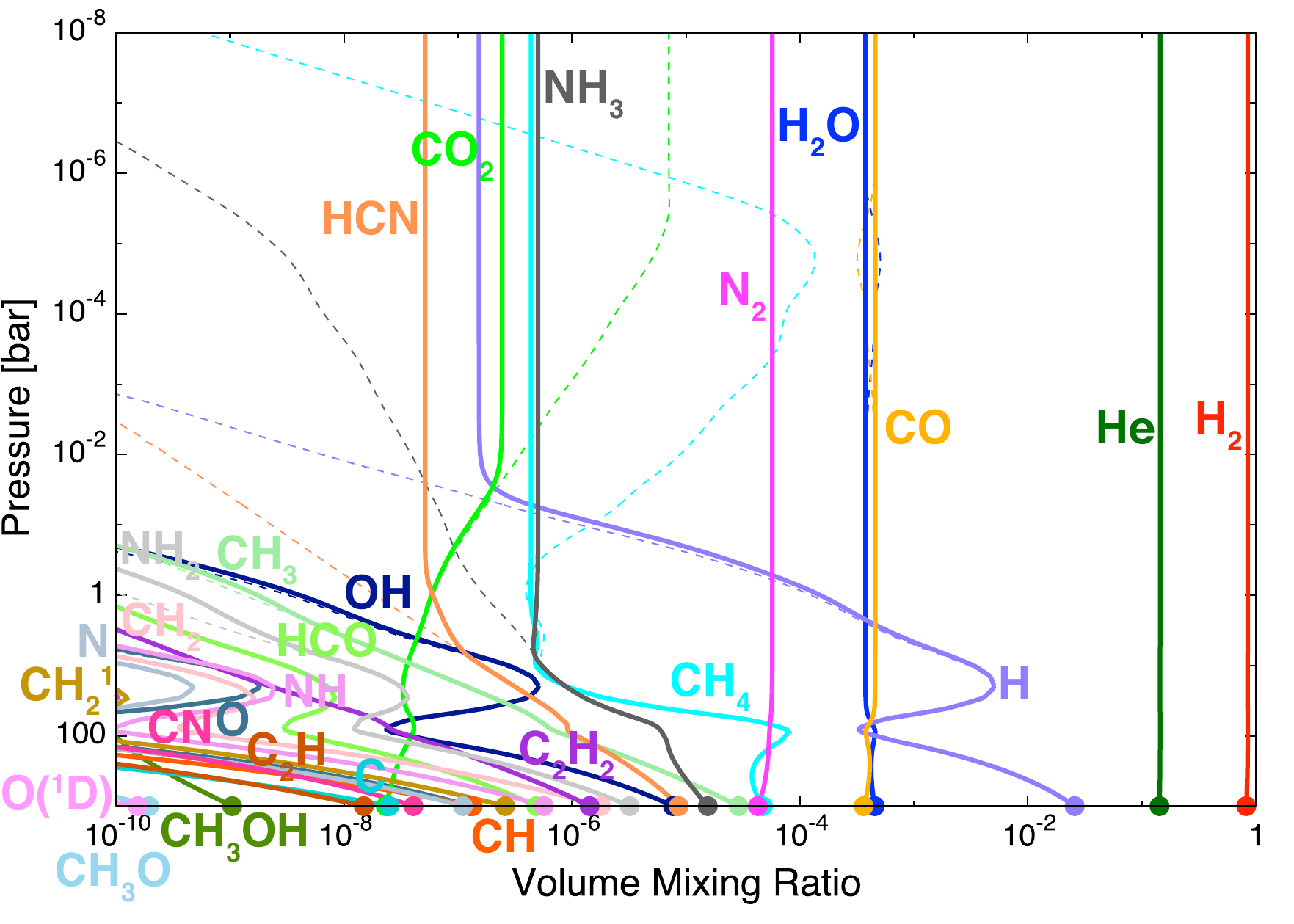}
  \subcaption{Anti-stellar point ($\phi = 180\degree$)}
 \end{minipage}
 \caption{Kinetic  photochemical equilibrium  C/H/O/N gas-phase results  (solid line) for the 1D WASP-43b profiles
 at (a)~the morning terminator ($\phi = 270\degree$), (b)~the evening terminator ($\phi = 90\degree$), (c)~the sub-stellar point ($\phi = 0\degree$),   and (d)~the anti-stellar point ($\phi = 180\degree$). The filled circles represent the thermochemical equilibrium values at the lower boundary. For reference, the dashed lines show the abundances in thermochemical equilibrium, where the eddy diffusion transport is also ignored.}\label{fig:photo}
\end{figure*}

\begin{figure}%[h]
 \begin{minipage}[b]{0.85\linewidth}
  \centering
  \includegraphics[width=\linewidth]{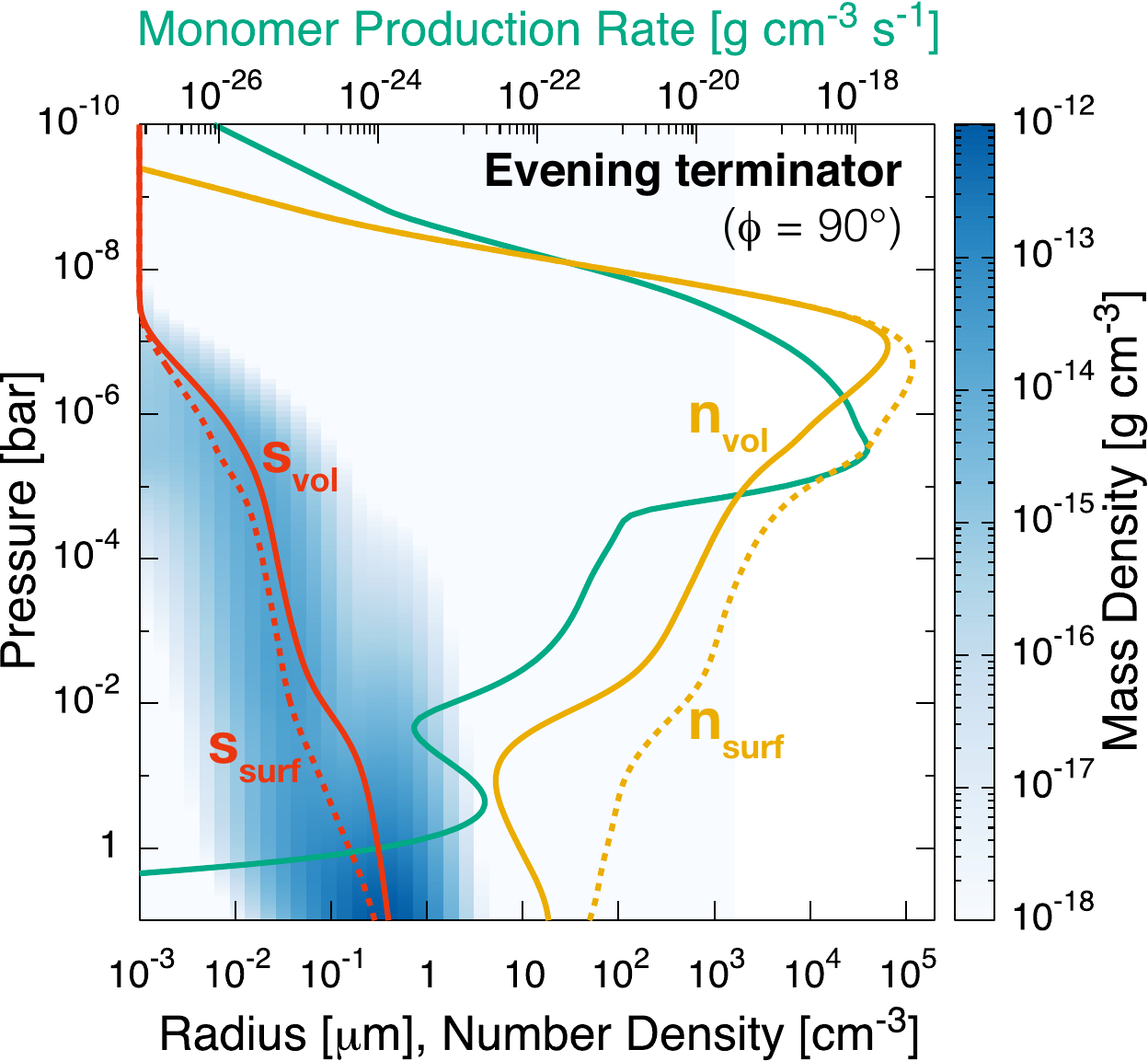}
 \end{minipage} \begin{minipage}[b]{0.85\linewidth}
  \centering
  \includegraphics[width=\linewidth]{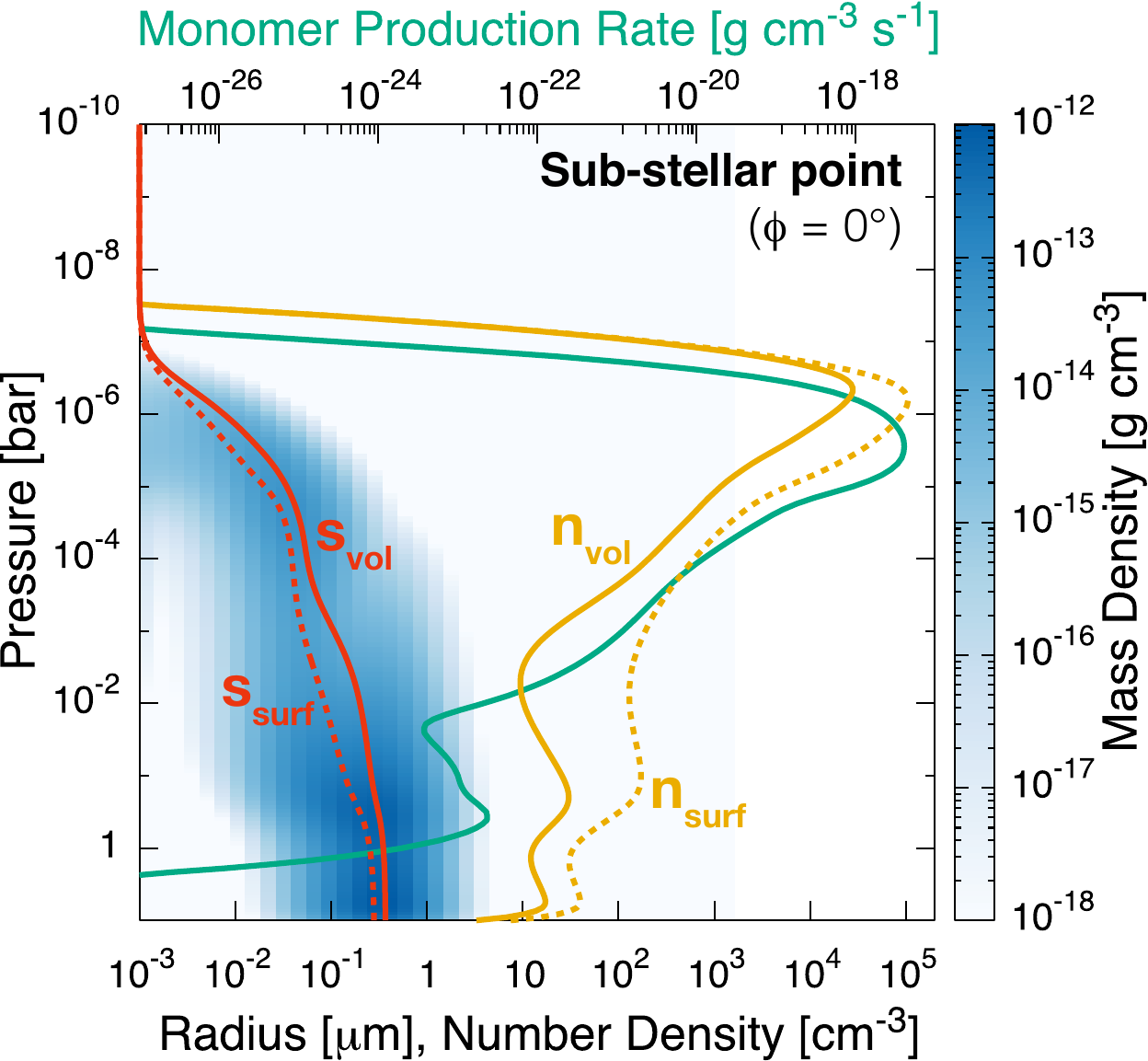}
\end{minipage}
 \begin{minipage}[b]{0.85\linewidth}
  \centering
  \includegraphics[width=\linewidth]{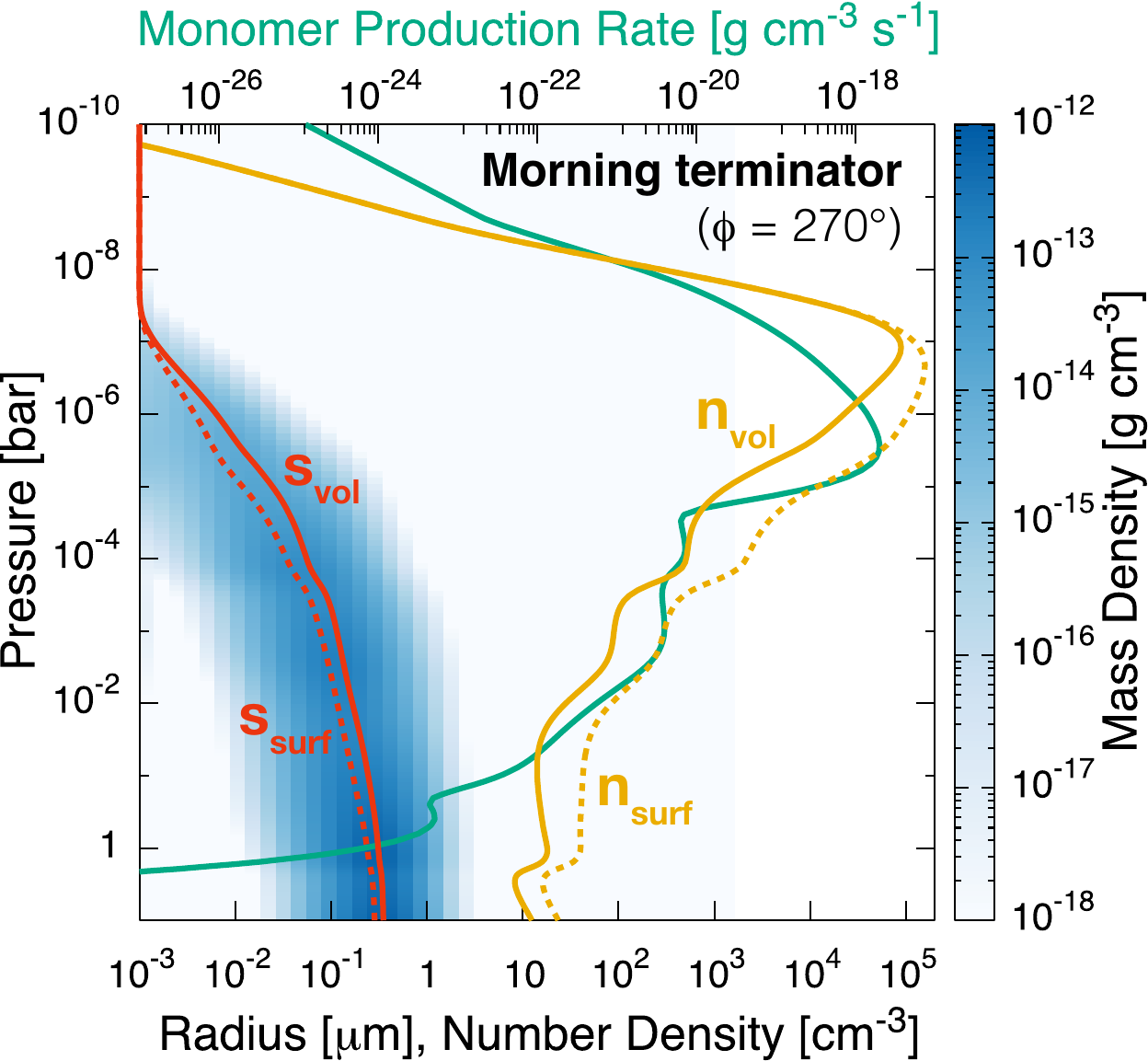}
 \end{minipage}
%  \begin{minipage}[b]{0.85\linewidth}
%  \centering
%  \includegraphics[width=\linewidth]{images/haze_Phi0.eps}\\[-6.0cm]
%  \subcaption{Sub-stellar point ($\phi = 0\degree$)}
%{\small Sub-stellar point ($\phi = 0\degree$)}
%\end{minipage}\\[5.5cm]
% \begin{minipage}[b]{0.85\linewidth}
%  \centering
%  \includegraphics[width=\linewidth]{images/haze_Phip90.eps}\\[-5.5cm]
%  \subcaption{Evening terminator ($\phi = 90\degree$)}
%{\small Evening terminator ($\phi = 90\degree$)}
% \end{minipage}\\[5cm]
% \begin{minipage}[b]{0.85\linewidth}
%  \centering
%  \includegraphics[width=\linewidth]{images/haze_Phim90.eps}\\[-6cm]
%  \subcaption{Morning terminator ($\phi = 270\degree$)}
%{\small Morning terminator ($\phi = 270\degree$)}
% \end{minipage} \\[5cm]
 \caption{
 %Vertical profiles of 
 The haze particles profiles for the 1D WASP-43b profiles at the evening terminator ($\phi = 90\degree$, top), the sub-stellar point ($\phi = 0\degree$, middle), and the morning terminator ($\phi = 270\degree$, bottom); volume average radius $s_\mathrm{vol}$ (solid red) and number density $n_\mathrm{vol}$ (solid orange), and the surface average radius $s_\mathrm{surf}$ (dotted red) and number density $n_\mathrm{surf}$ (dotted orange), along with that of the monomer mass production rate (solid green). See \cite{2018ApJ...853....7K} for the definition of each quantity. %The mass densities for all size bins at each pressure level are also plotted with the blue color contour.}
 }\label{fig:haze}
\end{figure}

%----------------------------------------------------

\section{The C/H/O/N molecule abundances  under the effect of mixing and UV radiation, and the formation of hydrocarbon hazes on WASP-43b}\label{s:ncheq}
%\textcolor{blue}{(>Christiane As there are not almost any difference between the results from the original and cloud-depleted elemental abundance ratios, I decided to present the results from the original elemental abundance ratios.)}

We wish to explore the effect of photochemistry on the cloud particle formation in the atmosphere of WASP-43b. We explore the formation of hydrocarbon hazes as result of hydrocarbon monomers coagulating to larger hydrocarbon aggregate structures from an oxygen-rich (C/O<1) atmospheric gas in addition to the formation of mineral cloud particles. In a non-irradiated oxygen-rich atmosphere (C/O<1), most of the carbon would be locked up in CO in such a high-temperature atmosphere  unless the strong CO bond is destroyed. $\mathrm{CH_4}$ becomes the most dominant one in low-temperature atmospheres. We perform kinetic C/H/O/N chemistry calculations for the two terminators ($\phi=90^o, 270^o$), the substellar ($\phi=0^o$) and the anti-stellar point ($\phi=180^o$) at the equator  ($\theta=0^o$). First, we discuss the  the photochemical C/H/O/N gas-phase results (Sect.~\ref{ss:kph}). Section~\ref{ss:haze} discusses our results for the hydrocarbon haze at the three selected atmospheric profiles. We note, however, that the profiles of temperature and eddy diffusion coefficient have been extrapolated outsided the GCM calculation domains in order to capture the full potential of hydrocarbon haze formation in  the upper and lower atmospheres (Fig.~\ref{fig:Kzz}). The photochemical calculation domain extends from 1000~bar to the pressure at the highest altitude grid, which is different for each calculation and also varies during the calculation due to the change in mean molecular weight in the upper atmosphere because of the molecular hydrogen decomposition. 
The calculation region to the lower atmosphere is very important to simulate the atmospheric region where the pressure is high enough for the thermochemistry to dominate because the chemistry in the lower atmosphere greatly affects that in the upper atmosphere through the vertical mixing. We note here that a deeper inner boundary of the 3D GCM simulation may effect the atmosphere structure \citep{2019arXiv190413334C} but this has not been considered in our paper. While the temperatures in the upper atmosphere are extrapolated isothermally, those in the lower atmosphere are extrapolated with the same temperature-pressure gradient. Eddy diffusion coefficients are extrapolated with the values at the upper and lower boundaries of the mineral cloud calculation region. The complete (T$_{\rm gas}$, p$_{\rm gas}$, K$_{\rm zz}$) information for these profiles are provided in Fig.~\ref{fig:Kzz}.

\subsection{Kinetic, photochemical gas-phase results for C/H/O/N chemistry}\label{ss:kph}
%\subsubsection{Photochemistry}\label{ss;pc}
In Figure~\ref{fig:photo}, we show the results of our photochemical calculations for (a)~morning terminator ($\phi = 270\degree$), (b)~evening terminator ($\phi = 90\degree$), 
(c)~sub-stellar point ($\phi = 0\degree$), and (d)~anti-stellar point ($\phi = 180\degree$).
Among the four atmospheric 1D profiles probed here, the anti-stellar point ($\phi = 180\degree$, (d)) is the simplest due to the absence of photochemistry. In the lower atmosphere ($p_{\rm gas} \gtrsim 1$~bar), the volume mixing ratio of each species (solid line) is consistent with its thermochemical equilibrium abundance (dashed line). This is because the timescales of thermochemical reactions are shorter than eddy diffusion transport timescales due to the high temperatures and pressures in the lower atmosphere. Many species such as HCN, $\mathrm{C_2H_2}$, and OH are created because of their stability at high temperatures. On the other hand, the upper atmosphere is dominated by eddy diffusion mixing as can be 
%clearly 
seen for some species that their abundances are quenched at certain pressure levels; $\sim 10^{-2}$~bar for $\mathrm{CO_2}$ and $\mathrm{H}$, $\sim 1$~bar for $\mathrm{CH_4}$ and HCN, and $\sim 10$~bar for $\mathrm{NH_3}$.

The other three cases with C/H/O/N photochemistry are more complex, but the chemistry in the lower atmosphere is similar since the (T$_{\rm gas}$, p$_{\rm gas}$)-profiles at such a high pressure range is almost the same (see Figure~\ref{fig:Kzz}). Also, it is the same as the case of (d)~anti-stellar point ($\phi = 180\degree$) that the middle atmosphere ($1$~bar $\lesssim p_{\rm gas} \lesssim 10^{-4}$~bar) is dominated by eddy diffusion mixing. The upper atmosphere ($p_{\rm gas} \lesssim 10^{-4}$~bar) is, however, governed by photochemistry. Many species, the abundances of which in thermochemical equilibrium are quite low, such as H, O, C, HCN, N, OH, $\mathrm{CH_3}$, $\mathrm{C_2H_2}$, $\mathrm{CH_2}$, $\mathrm{O_2}$, $\mathrm{CH}$, $\mathrm{CN}$, $\mathrm{CH_2^1}$, $\mathrm{NH}$, $\mathrm{O (^1D)}$, and $\mathrm{CH_3OH}$ are produced by photochemical reactions. As for the haze precursors (i.e., $\mathrm{CH_4}$, HCN, and $\mathrm{C_2H_2}$), they are all created through the photochemistry in the upper atmosphere while HCN and $\mathrm{C_2H_2}$ in the lower atmosphere are created thorough the thermochemical reactions.

Compared to the cases of the two terminators, where only the one side facing to the host star is irradiated, sub-stellar point ($\phi = 0\degree$, (c)) receives twice 
%\ch{(Why twice?)} 
the amount of the incoming stellar photon flux (see Sec.~\ref{s:ap}), resulting in more efficient photochemistry. This effect can be seen at, for example, the slightly higher pressure levels (lower altitudes) of the dissociation of $\mathrm{H_2}$, $\mathrm{H_2O}$, CO, $\mathrm{CH_4}$, HCN, and $\mathrm{C_2H_2}$, and the existence of photochemically-produced species mentioned above at somewhat higher pressure levels (lower altitudes). As for the comparison between the two terminators, the only differences in the calculation settings are the assumed profiles of temperature-pressure and eddy diffusion (see Figure~\ref{fig:Kzz}), and the amount of the incoming photon flux is the same. The effect of the higher temperatures for the evening terminator can be seen at the larger number of the photochemically-created species in the upper atmosphere (because high temperatures also favour these high-energy species).

The resultant sum of the photodissociation rates of the three haze precursors, which is assumed to be the monomer production rate in the particle growth simulations, integrated throughout the atmospheres are $2.61 \times 10^{-11}$, $8.19 \times 10^{-11}$, and $2.48 \times 10^{-11}$~[g~$\mathrm{cm^{-2}}$~$\mathrm{s}^{-1}$] ($6.23 \times 10^{9}$, $1.95 \times 10^{10}$, and $5.91 \times 10^{9}$~cm$^{-2}$ s$^{-1}$) for (a)~morning terminator ($\phi = 270\degree$), (b)~evening terminator ($\phi = 90\degree$), 
(c)~sub-stellar point ($\phi = 0\degree$), and (d)~anti-stellar point ($\phi = 180\degree$), respectively. Unlike the case of the cool ($\sim 500$~K) planets \citep{2019ApJ...877..109K}, we find that HCN rather than $\mathrm{C_2H_2}$ contributes most to the haze formation among the three precursors because of its stability at high temperatures for all the three longtitude points. Here we note the link to a possible lighting-induced boost of HCN in cloudy exoplanets (\citealt{2016ApJS..224....9R,2016MNRAS.461.1222H}). The monomer production rate for the sub-stellar point is more than twice the values of the terminators. We find that this is because of the decreased abundances of the major-photon absorbers of $\mathrm{H_2}$ and CO in the upper atmosphere such that HCN can use higher ratio of the incoming photons for its photodissociation for the case of the sub-stellar point. The production rates of the two terminators are almost similar, which is reasonable since the incoming stellar photon flux is exactly the same.

The atmospheric chemistry of WASP-43b was also explored by \cite{2018ApJ...869..107M} and \cite{20VePaBl.wasp43b}.
\cite{2018ApJ...869..107M} simulated the 3D profile of the atmospheric composition considering the horizontal and vertical transport with the use of the global circulation model THOR \citep{2016ApJ...829..115M}, but without photochemistry.
Our temperature-profiles of the sub- and anti-stellar points, and morning and evening terminators are similar to their cases of $G = 0.5$, where $G$ is their greenhouse parameter.
Comparing the vertical profiles of the volume mixing ratios of the four molecules they presented, namely $\mathrm{H_2O}$, CO, $\mathrm{CH_4}$, and $\mathrm{CO_2}$ at the four longtitude points with ours, the chemistry of the lower atmospheres is almost similar although we ignore horizontal transport. This is because the temperature profiles in the lower atmosphere are basically the same along the latitude and thus the volume mixing ratios at thermochemical equilibrium state are similar anyway. Note that they found that the timescale of the horizontal transport in the longtitude direction is always shorter than that of the vertical transport while that of the horizontal transport in the latitude direction is comparable to the vertical one. However, since they did not include photochemistry while we have ignored horizontal transport, the profiles in the upper atmosphere are different. The abundance of $\mathrm{CH_4}$ is much larger due to the molecular photodissociation and all the four molecules are dissociated in the very upper atmosphere in our case.
%\sout{Due to the horizontal transport, chemical profiles in the upper atmosphere should be smeared out in longitude compared to the results presented here.}

\cite{20VePaBl.wasp43b} studied the chemical structure of WASP-43b considering the effect of horizontal circulation with the use of their pseudo-2D model in addition to the 1D model. Our VMR profiles of the major species such as $\mathrm{H_2O}$, $\mathrm{CO}$, $\mathrm{NH_3}$, $\mathrm{N_2}$, $\mathrm{HCN}$, and $\mathrm{CO_2}$, are basically similar to their 1D results while the dayside profile of $\mathrm{CH_4}$ is different probably due to the differences of the assumed stellar UV spectrum and chemical network in addition to the different assumed profiles of temperature and eddy diffusion coefficient.
They demonstrated that the horizontal circulation reduces the gradient of chemical composition along the longitude especially in the pressure range of $10^{-4}-1$~bar. The abundance profiles for the two terminators and the anti-stellar point presented here should be somewhat homogenized to that of the dayside due to the horizontal quenching effect we have ignored.

\subsection{Hydro-carbon haze formation}\label{ss:haze}
%\textcolor{brown}{This section seems to repeat what has been said in Sect.~\ref{ss;pc}.}
%\textcolor{blue}{(This section has not been finished yet.)}
Applying the monomer production rate derived in Sect.~\ref{ss:kph}, the hydrocarbon cloud particles are now calculated from a coagulation model simulating the growth by particle-particle coagulation processes under the effect of gravity and vertical mixing due to eddy diffusion.

Figure~\ref{fig:haze} shows the calculated vertical profiles of haze properties for evening terminator ($\phi = 90\degree$, top), sub-stellar point ($\phi = 0\degree$, middle), and morning terminator ($\phi = 270\degree$, bottom). The resultant profiles of the haze particle are basically similar because of the only a factor of differences in the monomer production rate \citep[see][for the dependence on monomer production rate]{2018ApJ...853....7K}. The particles start to grow at $p_{\rm gas} \sim 10^{-6}$~bar and result in the average sizes of $\sim 3-4$~$\mu$m at the lower boundary of $p_{\rm gas} \sim 10$~bar.

The monomer production rate profiles (green lines) have the peaks at certain pressures since the photodissociation rates are determined by both the number of available photons, which decreases with increasing pressures, and the number of the molecules, which basically increases with increasing pressures. The second peaks can be seen at $p_{\rm gas} \sim 1$~bar come from the increase of the $\mathrm{C_2H_2}$ number density due to its stability at high temperatures. Compared to the two terminators, the case of the sub-stellar point has more concentrated monomer production profile because the photodissociation of the haze precursors begins to occur at higher pressure (lower altitude) but with the higher rate due to the intense incoming photon flux. This result in the more broader size distribution in this case, which can be seen as the the slightly larger difference between the volume and surfaced average radii.
Here we note that we have ignored the thermal decomposition due to the uncertainty of the thermal stability of haze, which can affect the particles in the high-temperature lower atmosphere.
\section{Two ensembles of chemically distinct cloud particles in oxygen-rich exoplanet atmospheres}\label{s:twocloud}

Mineral cloud particles may form by distinctly different processes than hydro-carbon cloud particles due to the presence of an radiation field in an (atmosphere) environment of changing density. The mineral cloud particles discussed here (Sect.~\ref{s:clouds}) to form in the atmosphere of WASP-43b originate from an oxygen-rich gas (C/O<1) through gas-gas and gas-particle reactions from a highly supersaturated gas. The hydro-carbon cloud particles, also called 'hydro-carbon haze', form under the influence of a UV radiation field by photo-chemical processes from the small hydro-carbon molecules like \ce{CH4}, \ce{C2H2} and \ce{HCN} (Sect.~\ref{ss:haze}). The resulting monomers grow through particle-particle collisions (coagulation) to larger cloud particles in form of aggregates. The processes involved in the formation of the two cloud particle ensembles,  mineral particles and hydrocarbon hazes,  are distinctly different and we present here a study how this two cloud particle ensembles  differ in their properties like nucleation rates, cloud particles sizes and cloud particle number densities.

\subsection{Mineral seed and hydro-carbon haze formation rates}

\begin{figure}
    \centering
    \includegraphics[width=\linewidth]{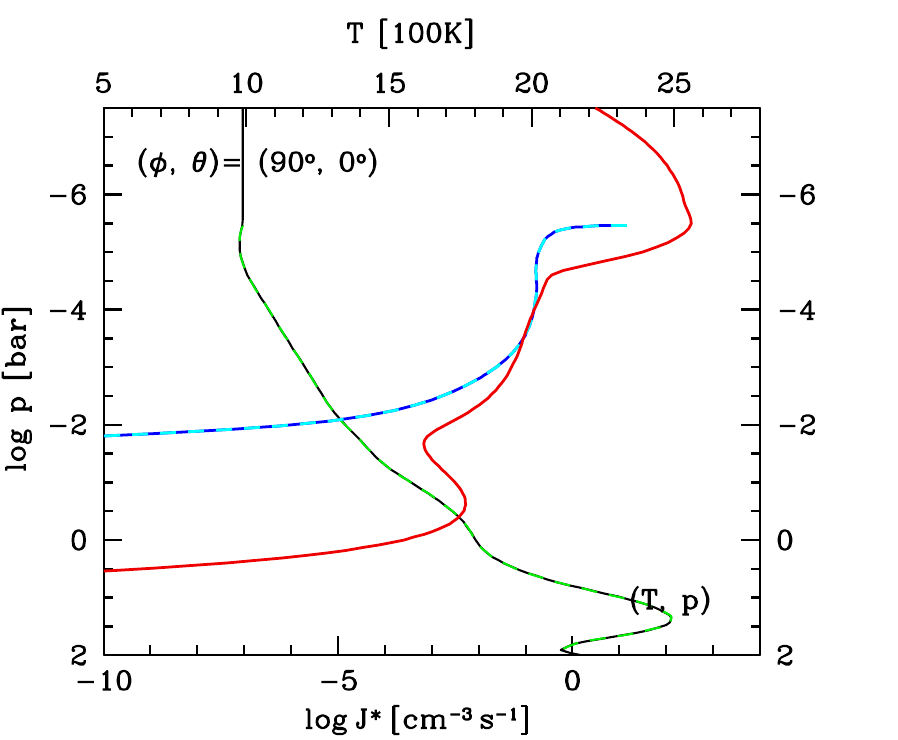}\\
    \includegraphics[width=\linewidth]{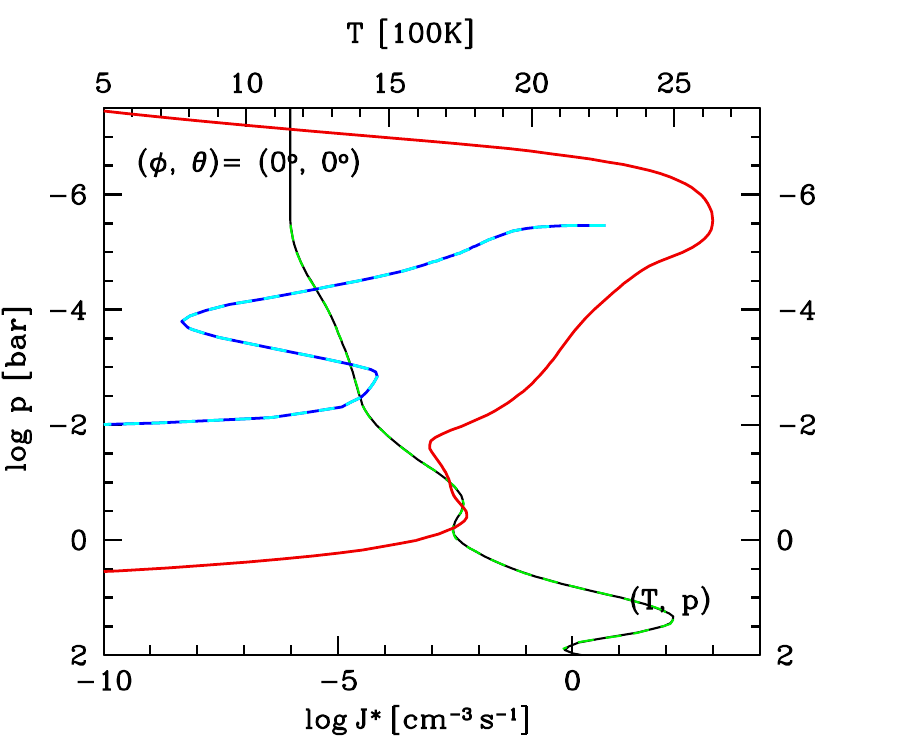}\\
    \includegraphics[width=\linewidth]{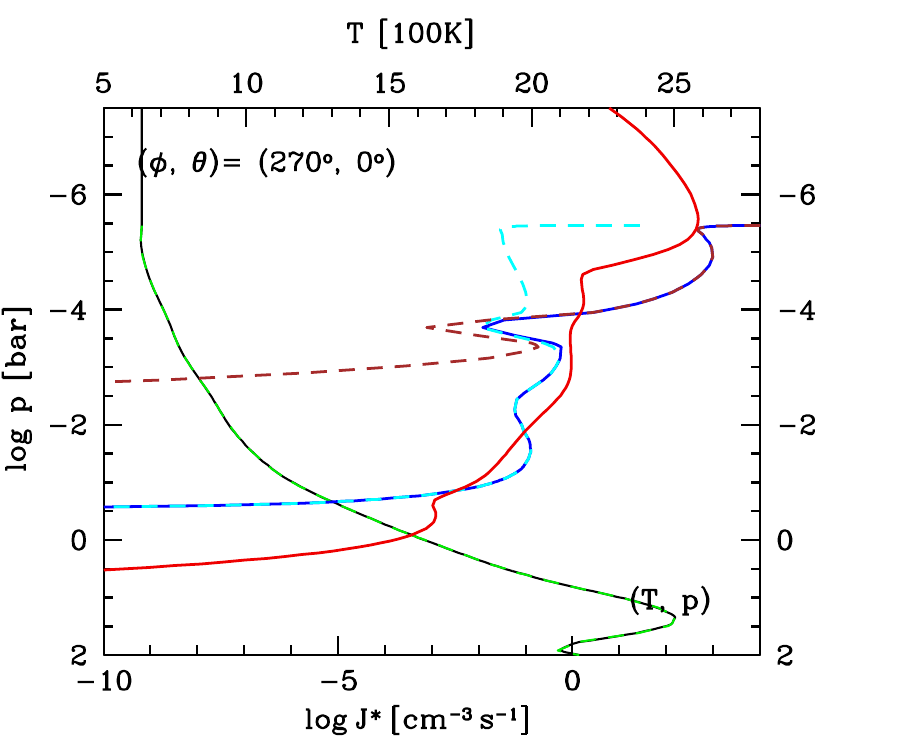}
    \caption{Two cloud particle ensembles: The mineral seed formation rate (blue: total value {  of the mineral seed particles} TiO$_2$ (cyan dashed), SiO (brown dashed), and KCl (green, not occurring at equator)) and the hydrocarbon haze particles formation rate (red). The (T$_{\rm gas}, p_{\rm gas}$) structure is shown in black (extrapolated) and green dashed (original 3D GCM).}
    \label{fig:J*comp}
\end{figure}

We present both cloud ensembles in  Figure~\ref{fig:J*comp} by comparing our results for mineral seed formation applying modified classical nucleation theory as described in Sect~\ref{s:ap} for the atmosphere of  WASP\,43b with a haze monomer production rate, which is defined as the sum of the photodissociation rates of the three haze precursor, $\mathrm{CH_4}$, HCN, and $\mathrm{C_2H_2}$, (note that this is the upper limit of the haze monomer production rate) by assuming a monomer size of 1~nm and its density of 1 g cm$^{-3}$ giving a mass of $4.19\cdot 10^{-21}$g per C/H/O/N haze monomer.
The mineral seed particles have a material density of 4.23 g cm$^{-3}$ for \ce{TiO2}, 2.64 g cm$^{-3}$ for SiO and 1.98 g cm$^{-3}$ for KCl.

For photochemically driven gas-phase reactions, the hydrocarbon haze formation rate, which is defined as the sum of the photodissociation rates of the haze precursors, appears to be more efficient than mineral seed particles  on the highly irradiated dayside of WASP-43b. 
%\yui{[Haze monomer production rate is just defined as the sum of the photodissociation rates of the haze precursors and any growth is not considered]}
The photodissociation rates of the haze precursors are 
%\sout{smaller for higher temperatures because of their less stability at higher temperatures.
%The effect is less prominent in terminator regions probed where the amount of available stellar UV photons is smaller than that in the dayside.}
almost the same for the three longitude points studied here because the sub-stellar point receives only two times higher UV irradiation compared to the two terminator points.  Negligible amount of photochemical hazes will form on the nightside as no or only very small amounts of stellar photons are available. Horizontal circulation may be able to transport particles from the dayside to the nightside.%, the effect of which will depend on 
%is of our future study.

\begin{figure}
    \centering
    \includegraphics[width=\linewidth]{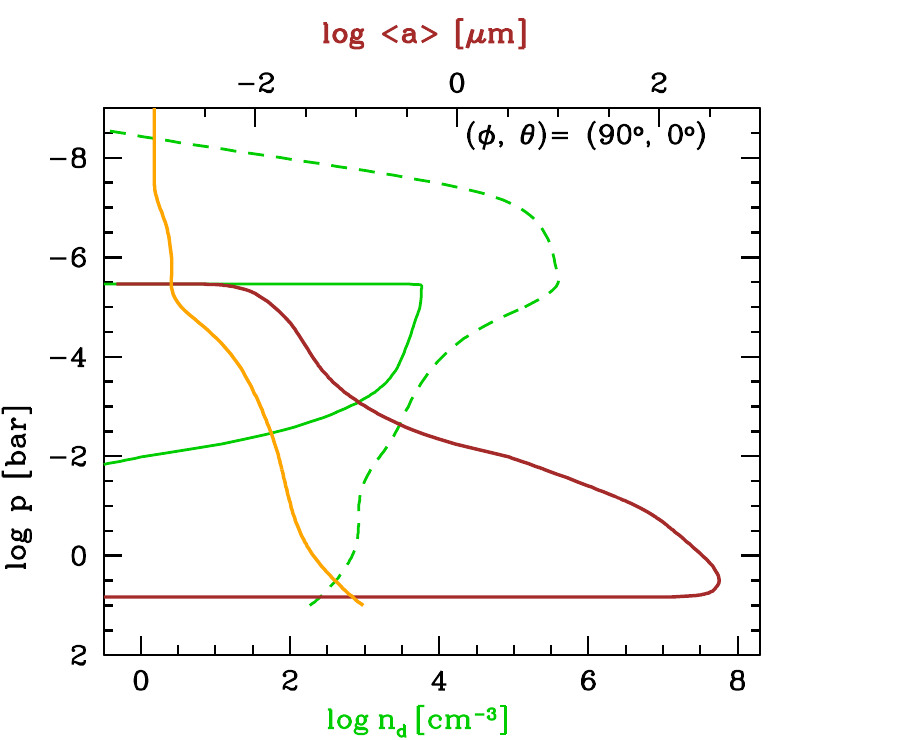}\\
    \includegraphics[width=\linewidth]{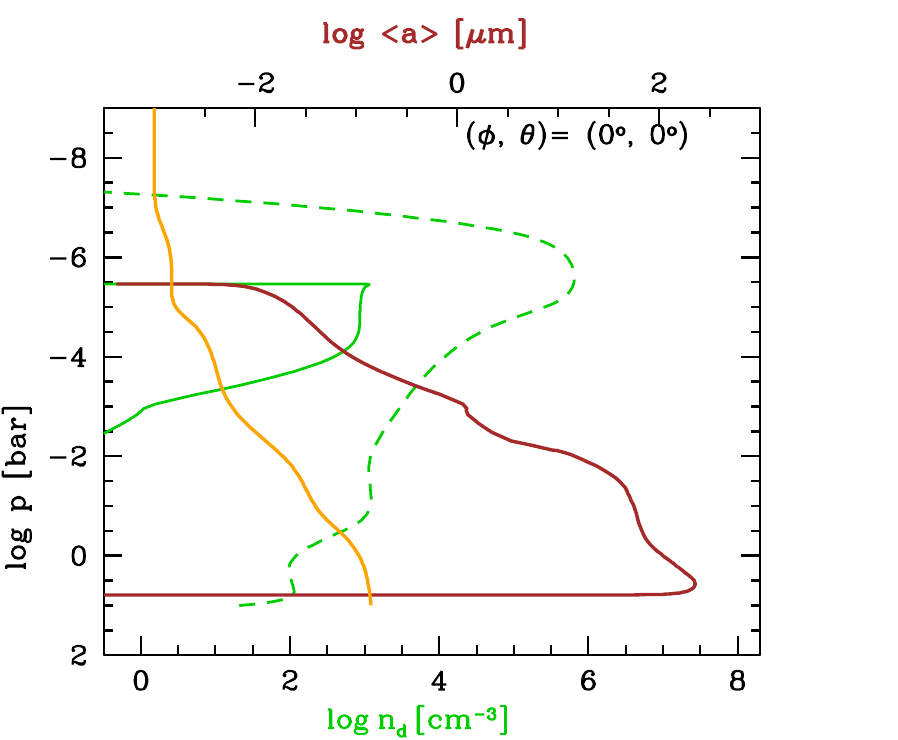}\\
   \includegraphics[width=\linewidth]{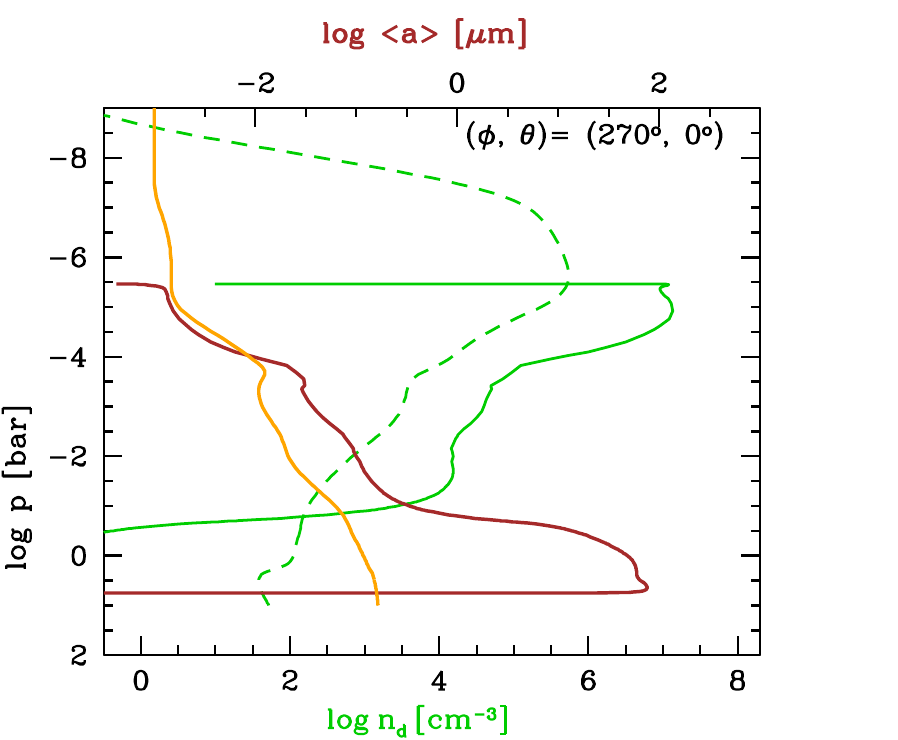}
    \caption{Characteristic properties of the two cloud particle ensembles: particle number density (n$_{\rm d}$ [cm$^{-3}$]):  mineral particles -- green solid, hydrocarbon haze -- green dashed.  Mean particles size $\langle a\rangle$ [$\mu$m] (Eq.~\ref{eq:amean2}): mineral particles -- brown; hydrocarbon hazes -- orange.  }
    \label{fig:a_nd_comp}
\end{figure}

The whole cloud extend throughout the whole computational domain of the utilized 1D trajectories (Figs.~\ref{TpNuc},~\ref{fig:DusttoGas}) because gravitational settling transports the cloud particles inwards. Gravitational settling, and thermal stability in the case of mineral cloud particle, determine the geometrical cloud extension, and hence, the inner cloud boundary. This is not yet been taken into account for the photochemical haze comparison as shown in  Fig.~\ref{fig:J*comp},~\ref{fig:a_nd_comp} due to the uncertainty of the thermal stability of haze.


\subsection{Mineral particle and hydrocarbon haze number density and mean particle sizes}

As a consequence of the differences between the mineral and the hydrocarbon haze nucleation rate, the hydrocarbon number densities (green dashed line in Fig~\ref{fig:a_nd_comp}) are larger than the mineral particles densities (green solid line) at the substellar point ($(\phi, \theta)= (0^o, 0^o)$) but also at the evening terminator ($(\phi, \theta)= (90^o, 0^o)$). With the local temperature decreasing by $\sim 500$K at the morning terminator ($(\phi, \theta)= (270^o, 0^o)$) compared to the substellar point, the mineral seed formation rate creates a larger abundance of mineral cloud particles 
at the morning terminator. The number of cloud particles has direct implications for the resulting (mean) sizes as a larger number provides more surface to condense on, and hence, mineral cloud particles will remain smaller. This is why the morning terminator ($(\phi, \theta)= (270^o, 0^o)$) has smaller mineral cloud particles than the evening terminator  ($(\phi, \theta)= (90^o, 0^o)$). This logic does not hold for the hydrocarbon haze particles which appear 
almost the similar sizes for all the three points because of the slight differences in the monomer production rates. This is because the mineral particle growth significantly depends on the surrounding bulk gas phase (and accelerates with increasing density) but for the case of the hydrocarbon haze, its formation is more governed by additional photochemical process. %\sout{(photoprocesses, dynamics/mixing) to create the needed C/H species}.}

Comparing model results by number hinges on an exact definition of the compared quantities.  Figure~\ref{fig:a_nd_comp2} (orange and brown lines) demonstrates  
the differences that result in the values of  cloud particle sizes when derived as weighted average ($\langle a\rangle$, Eq.~\ref{eq:amean2}, solid lines), as 
 surface weighted average ($\langle a\rangle_{\rm A}$, Eq.~\ref{eq:ameanA2}, dashed line) and as $\sqrt[3]{\langle a^3\rangle}$ (Eq.~\ref{eq:amean32}, dotted line) from our two different methods (moment equation vs. binning method). Fig~\ref{fig:a_nd_comp} demonstrates that differences  of 0.5 order of mag may occur.

 Our comparison  suggests that surface growth processes lead to larger particles in particular in the inner, denser atmospheric regions and that hydrocarbon cloud particles may remain smaller, more haze-like.  The upper, low-pressure regime where $p_{\rm gas}<10^{-4}$bar is populated by an ensemble of small hydrocarbon haze particles of $<10^{-2.5}\mu$m and mineral particles of $>10^{-2}\mu$m. Therefore, transmission spectra  may be affected by two cloud particle populations in low density atmospheres regions where photochemistry is efficient but surface growth processes are not yet efficient because of the low gas-surface collision rates. However, a simple retrieval of particles sizes will not suffice to distinguish between the two cloud particle ensembles, {  in particular as it is degenerate with the particle number density of the opacity species.}

 \section{Discussion of observational implications}
 
\begin{figure*}
     \includegraphics[width=0.5\linewidth]{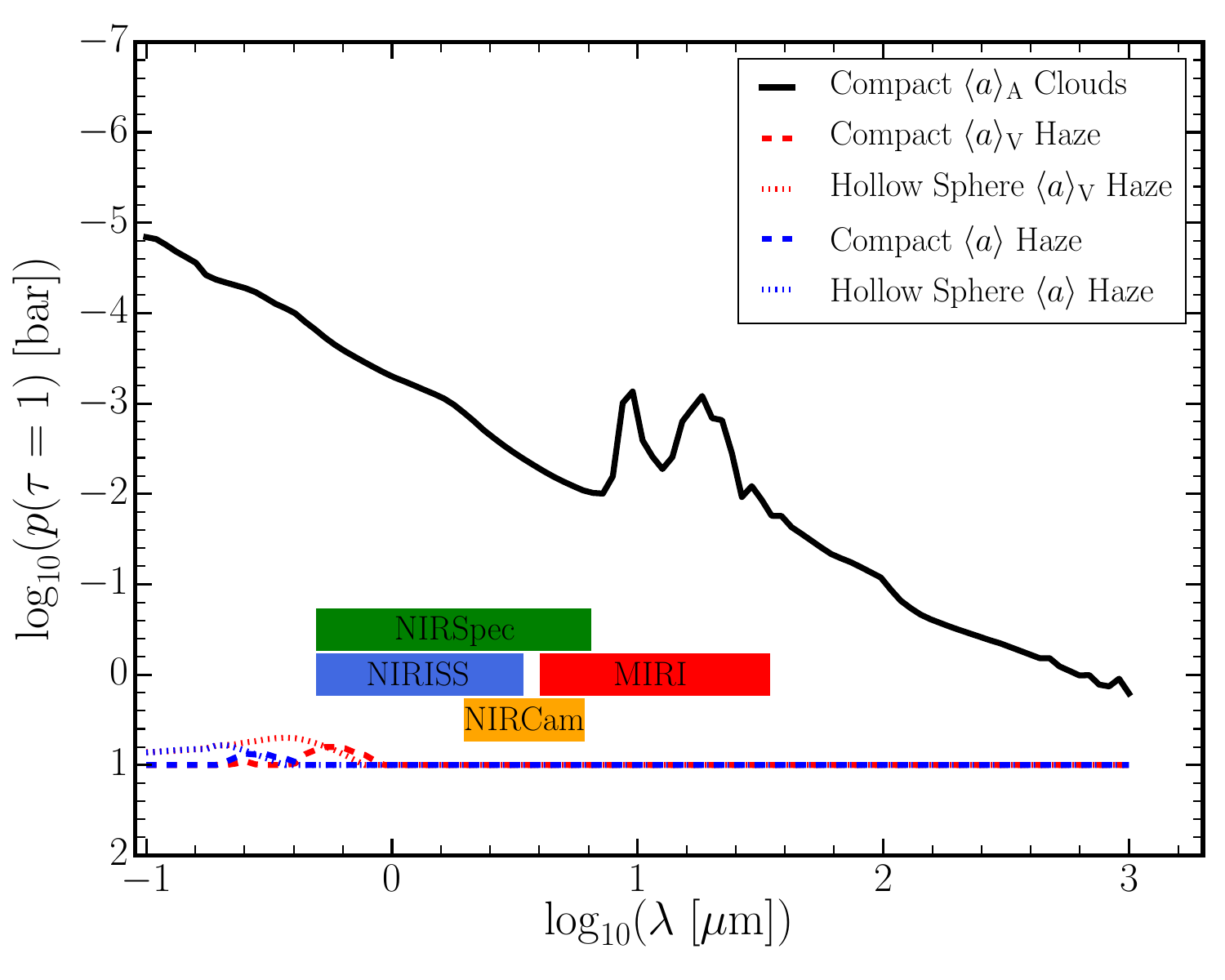}    \includegraphics[width=0.5\linewidth]{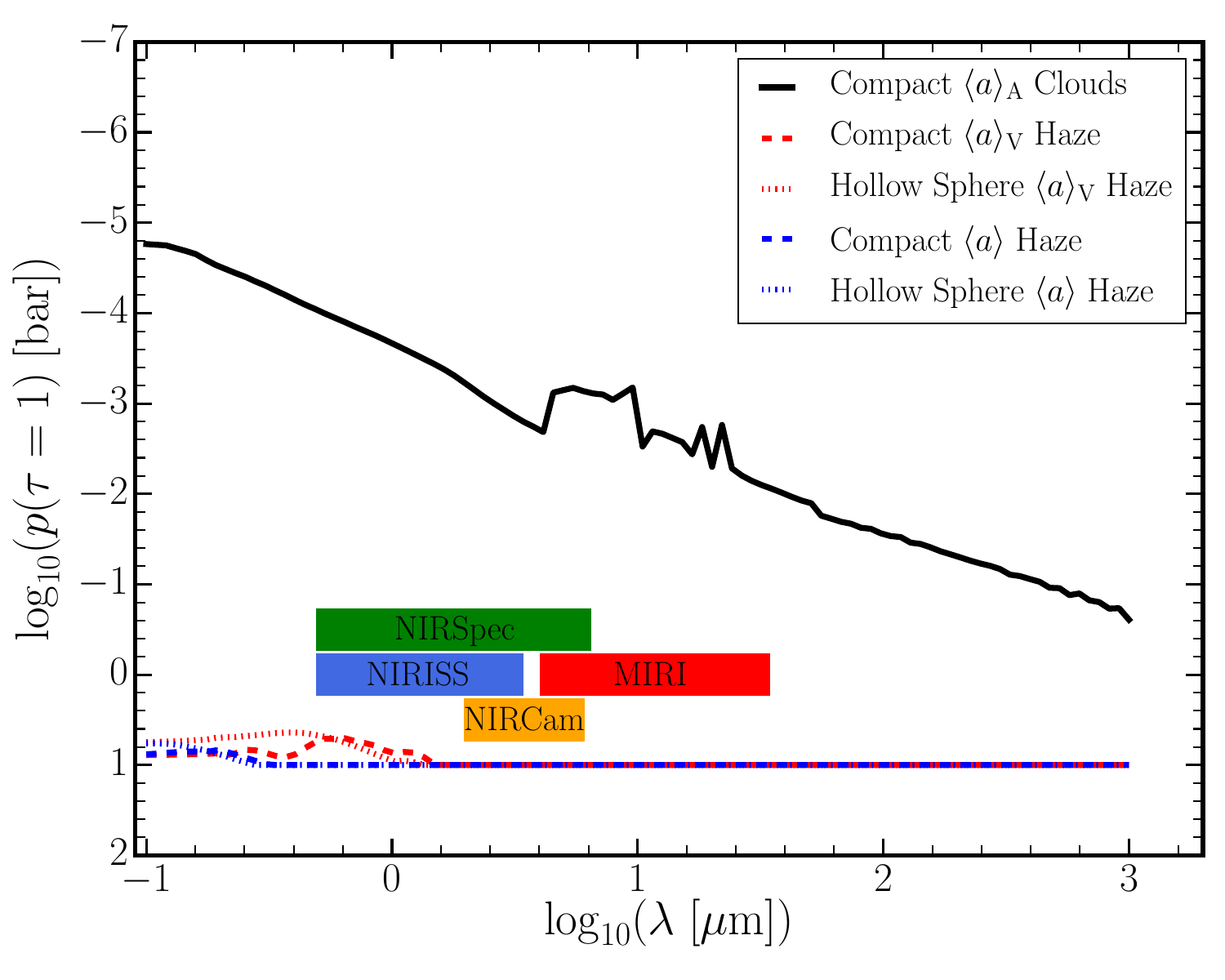}\\*[-0.0cm]
   \includegraphics[width=0.5\linewidth]{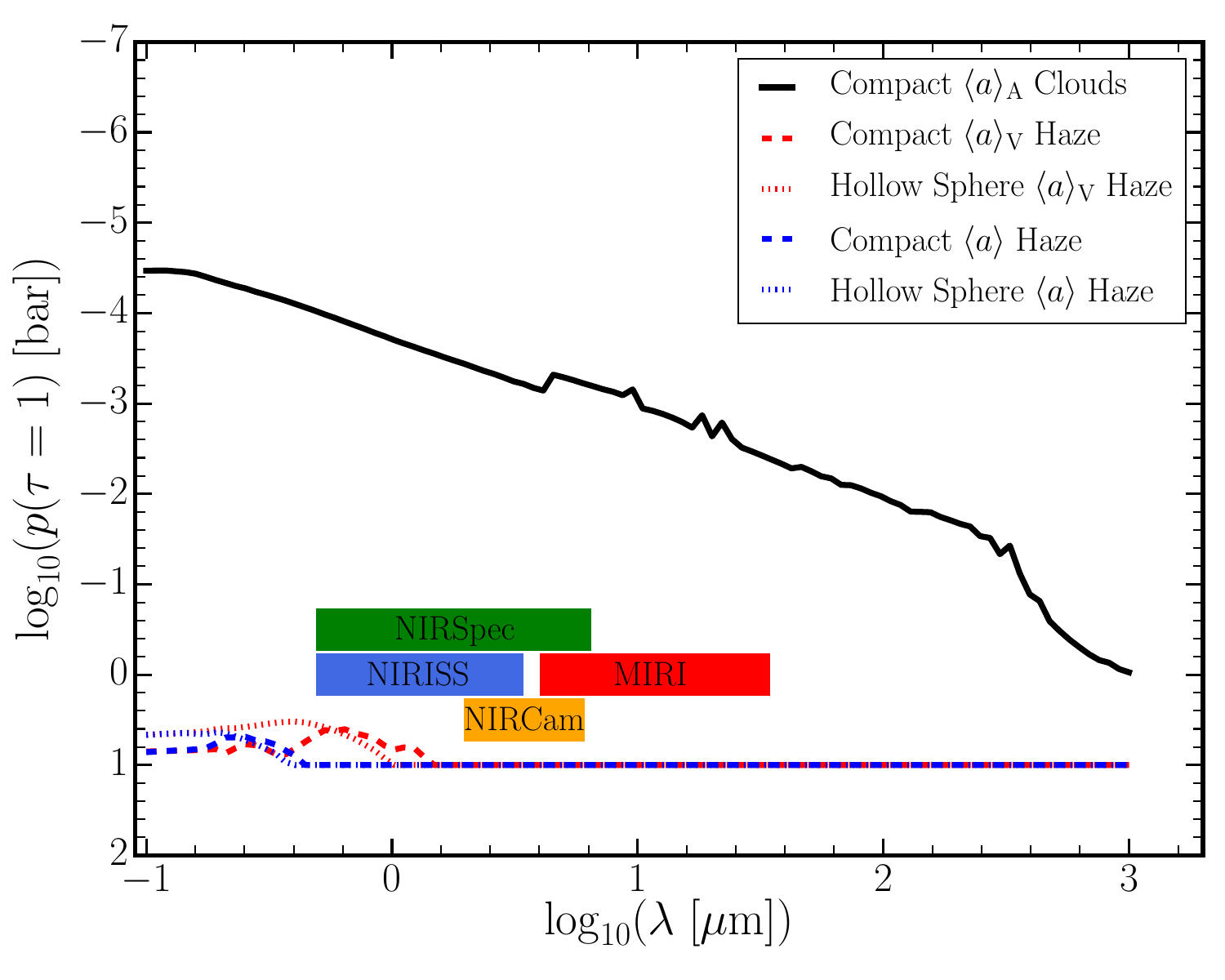}
 \includegraphics[width=0.5\linewidth]{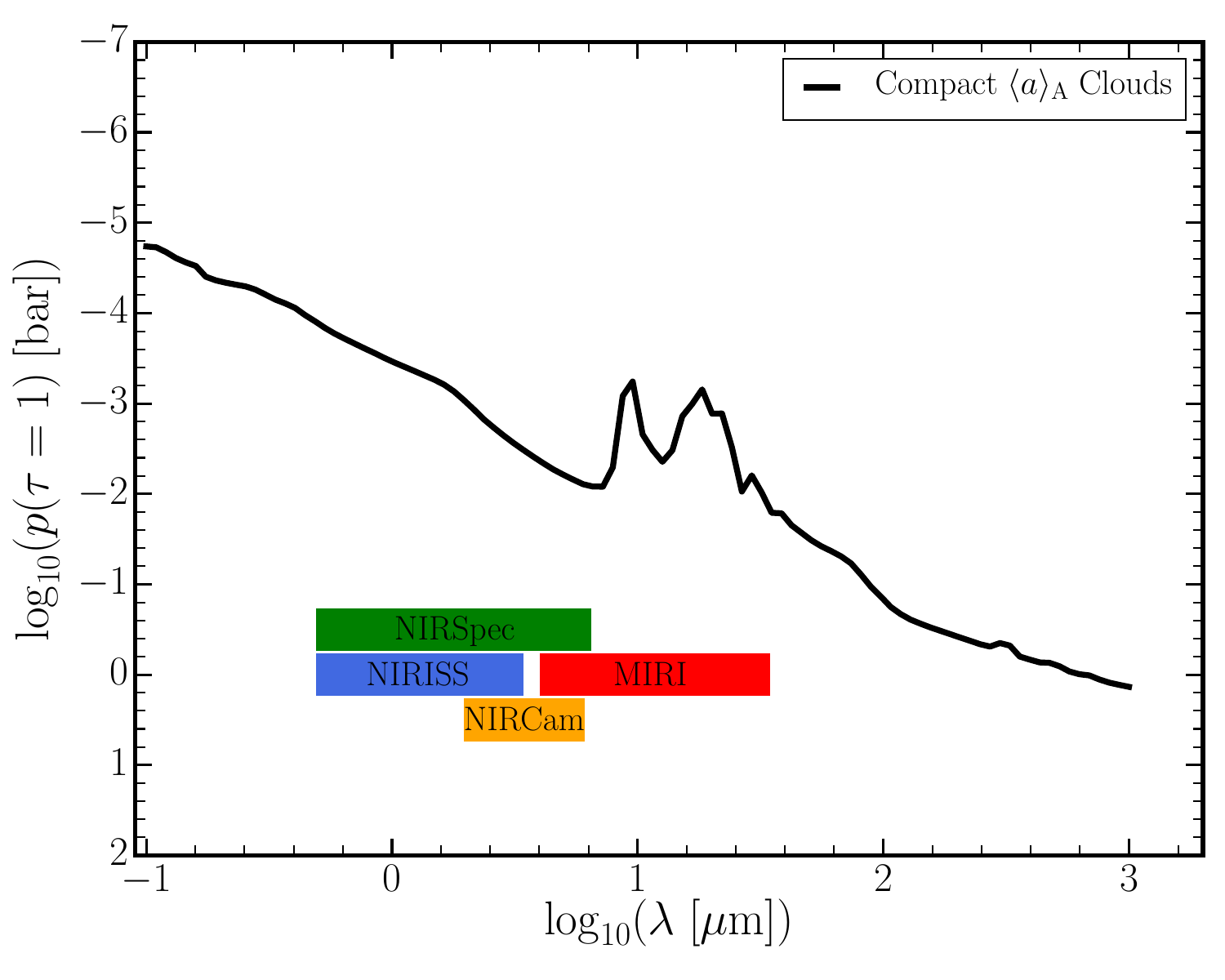}
    \caption{The atmospheric gas pressure, $p_{\rm gas}$ [bar], at which the wavelength-dependent vertical optical depth, $\tau(\lambda)$, of the mineral cloud particles (solid black lines) and the hydrocarbon haze particles (red and blue, dashed and dotted lines) reaches unity. Four equatorial profiles are considered: the morning terminator ($\phi=270^o$, top left), the evening terminator ($\phi=90^o$, top right), the substellar point ($\phi=0^o$, bottom left), the anti-stellar point ($\phi=180^o$, bottom right); similar to Figs.~\ref{fig:5line_MT},~\ref{fig:photo}. 
    The hydrocarbon hazes are considered to have a tholin opacity (Fig~\ref{fig:pgastau1zoomed}) for compact particles  and a distribution of hollow spheres for two averages of particle size. The mineral cloud particles (solid black line)  reach $\tau(\lambda)=1$ already at relatively low pressure in the optical but at $p_{\rm gas}\approx 1$ bar for longer wavelengths, hence, WASP-43b's atmosphere is more transparent in the far-IR. The hydrocarbon hazes remain optically thin. The coloured bars show wavelength ranges covered by the four {\sc JWST} instruments, based on modes suggested for transmission spectroscopy of exoplanets \citep{Stevenson2016}: {\sc MIRI} (red) --  based on Medium Resolution Spectroscopy modes, {\sc NIRCam} (orange) --  based on Grism Time Series mode, {\sc NIRSpec} (green) -- based on Fixed Slit Spectroscopy mode available filters, {\sc NIRISS} (blue) -- based on Single Object Slitless Spectroscopy mode.}    
    \label{fig:pgastau1}
\end{figure*}

\begin{figure}
     \includegraphics[width=\linewidth]{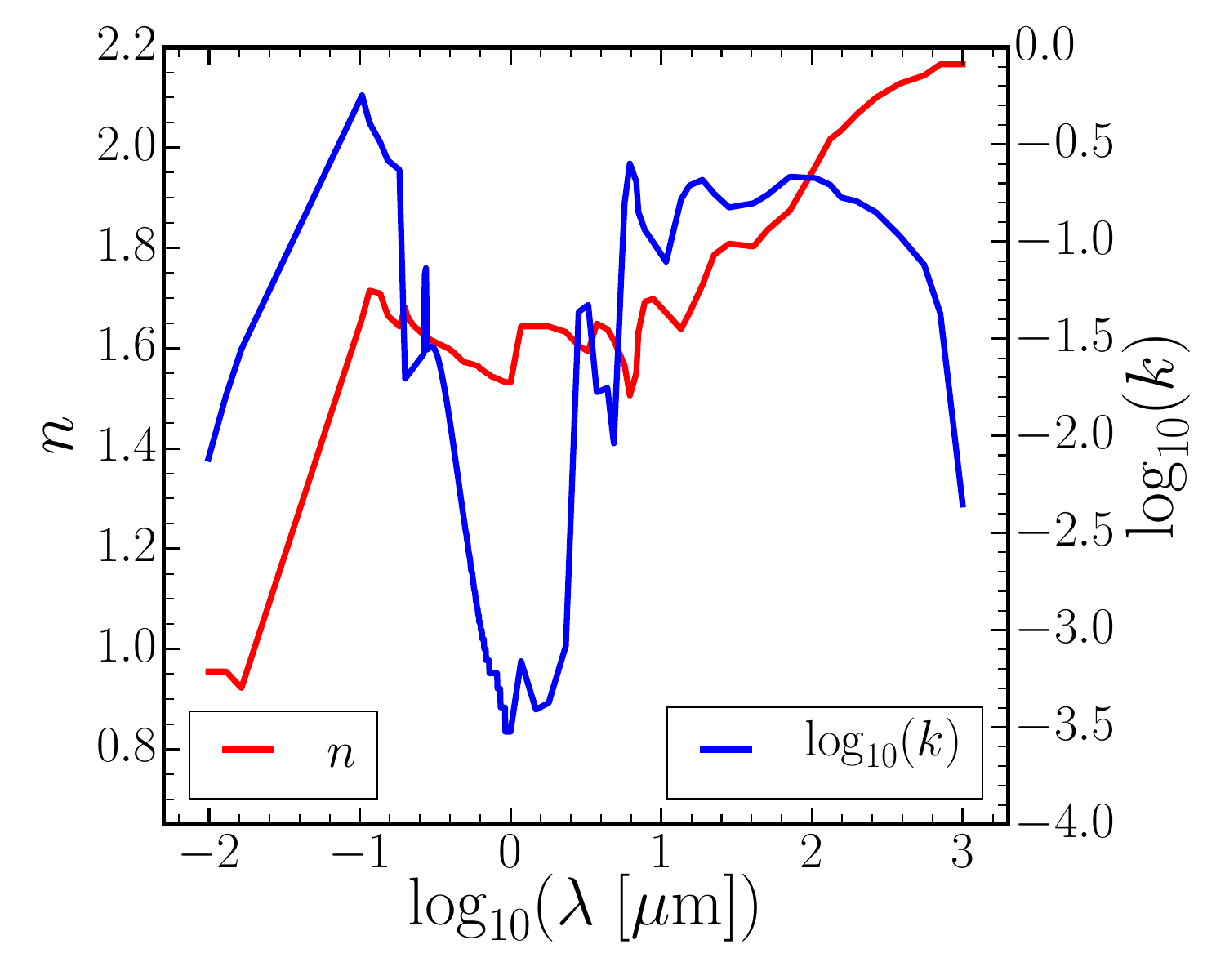}
    \caption{Refractive index data for Tholins.% as used in \cite{Wakeford2015}. 
    Data between $0.2-1\mu$m is from \cite{Ramirez2002} otherwise from \cite{Khare1984}.}
    \label{fig:pgastau1zoomed}
\end{figure}

 {  Cloud particles tend to dominate the optical depth of an atmosphere such that a planet would appear with a larger radius compared to a purely gaseous atmosphere, as shown for example in Fig. 20 in \cite{2019arXiv190608127H} for the ultra-hot Jupiter HAT-P-7b. In addition to mineral cloud particles,  hydrocarbon-haze particles may form in higher atmospheric regions driven by the high-energy fraction of the host star's radiation field. Figure~\ref{fig:a_nd_comp} shows that they may appear with a relatively high number density in the upper atmosphere. How would these two cloud particle populations affect the optical depth of an atmosphere?
 
 To provide a first insight about possible opacity effects of  the two cloud particles ensembles, we determine the local gas pressure where the atmosphere reaches an optical depth of 1 along a vertical line of sight from the top of the atmosphere, i.e. $p_{\rm gas}(\tau(\lambda)=1)$. We apply effective medium and Mie theory for the mixed mineral cloud particles of a mean particle size $\langle a\rangle_{\rm A}$, number density $n_{\rm d, A}=(\rho L_3)/(4\pi \langle a\rangle_{\rm A}^3$/3 (see Eq.~\ref{eq:ameanA2_1}),
 and refractive indices as in \cite{2019arXiv190608127H}. The hydrocarbon-haze opacities are calculated using the volume averaged haze particle size $\langle a\rangle_{\rm V}$~($= s_\mathrm{vol}$ in Fig.~\ref{fig:haze}) and the respective volume average particle number density $n_{\rm V}$~($= n_\mathrm{vol}$ in Fig.~\ref{fig:haze}),
 %(see Fig.~\ref{fig:haze}), 
 with refractive indices for tholins 
 %as in \cite{Wakeford2015}, taken
 from \cite{Khare1984,Ramirez2002} (Fig.~\ref{fig:pgastau1zoomed}).  Appendix~\ref{aA} demonstrates that uncertainties  arise from  how the actual mean particle size and the particle number density are calculated, in particular if considering agglomerate particles like in the the case of hydrocarbons hazes (see Fig.~\ref{fig:a_nd_comp2}).

 For completeness and for linking to previous calculations, two opacity cases are studied for hydrocarbon hazes: a) compact haze particles (dashed lines in Fig~\ref{fig:pgastau1}), and b) non-compact haze particles by incorporating particle shape effects into the opacity calculations through a distribution of hollow spheres as described in \cite{Samra20} (dotted lines in Fig~\ref{fig:pgastau1}). The optical depth is integrated from the top of the respective atmospheres for the hazes ($10^{-10}\ {\rm [bar]}$) and clouds ($10^{-6}\ {\rm [bar]}$) to the atmospheric gas pressure where the wavelength-dependent optical depth reaches unity. 
 Please note that for the haze particles the integrated columns  shown here never reach the $\tau=1$.  These optical depths are calculated vertically, for a slant geometry the effects can increase (see \citealt{Fortney2005}).
 
 Figure~\ref{fig:pgastau1} suggests that hydrocarbon haze particles  would 
 appear optically thin in the layers above the mineral cloud particles, i.e. for $p_{\rm gas}<10^{-4.5}$ bar for $\lambda\approx 0.1\mu$m but up to higher gas pressures of $\approx$ 1 bar for the IR wavelength in the terminator profiles. The optical depth of the hydrocarbon hazes is considerably lower than that of the mineral particle clouds such that a wavelength-dependent radius measurement of WASP-43b would be determined by the mineral clouds particles but not by hazes. 
 
 If hydrocarbon hazes are optically thin on WASP-43b as suggested by our present study, JWST is likely to  observe spectral features of the mineral cloud particles in the IR wavelength of $\approx 15-20 \mu$m at the morning terminator ($\phi=270^o$). We would expect more hydrocarbon haze particles if the monomer production rate is higher than what we have assumed in this paper; in the case such that some other species also act as the haze precursors in addition to those considered in this study, \ce{CH4}, \ce{HCN}, and \ce{C2H2}.
 % mass conversation rate, which is guided by Titan measurements,  of the photochemicaly triggered  hydrocarbon precursors \ce{CH4}, \ce{HCN}, and \ce{C2H2}  to the haze monomers is more efficient compared to what we have assumed in this paper. 
}

\section{Conclusion}

\begin{itemize}
    \item If WASP-43b has an oxygen-rich atmosphere of approximately solar element composition, then WASP-43b can be expected to be covered with mineral clouds on the day- and on the nightside, similar to  HD\,189733b and HD\,209438b.
    
    \item Photochemically driven hydrocarbon hazes form on the dayside where the stellar radiation field has the largest affect on the chemistry of the originally oxygen-rich atmospheric gas (C/O<1) on WASP-43b.   Also the terminator regions are affected. The effect of UV-driven hydrocarbon haze particles is strongest on the day-side, and negligible on the night-side if the horizontal transport %of haze particles 
    is negligible. %, the effect of which is of our future study, is negligible}.
    
    \item WASP-43b can therefore be expected to have clouds made of two distinct ensembles of particles,  mineral particles and hydrocarbon haze particles on the dayside. A gradual  change to a mineral-particle only cloud occurs across the terminator regions into the nightside.
    
    \item Mineral cloud particles reach size  $\approx 100\mu$m  in the inner, denser atmospheric regions. Hydrocarbon haze particles may remain smaller, of $\approx 0.1\mu$m in inner, warmer parts of a cloud.
 
 \item The cloud and haze particle sizes are subject to precise definition of the term 'particles size'. The use of different definitions enables to estimate the systematic errors of this cloud property. 
 
     \item We suggest to search for the spectral features of mineral and of hydrocarbon clusters in the optically thin, low-density regimes of exoplanet atmospheres. We note that similar clusters are studied in AGB star outflows.
     
     \item A wavelength-dependent radius measurement of WASP-43b would be determined by the mineral clouds  but not by hazes.
     
     \item We suggest further to search for mineral absorption features in the 15-20$\mu$m spectral range.
\end{itemize}

\begin{acknowledgements}
We thank Peter Woitke for insightful discussions on  hydrocarbon hazes. G.V. acknowledges the hospitality of the School of Physics \& Astronomy at the University of St Andrews and the summer student funding from the Royal Astronomical Society.
Y.K. and K.L.C are supported by the European Union’s Horizon 2020 Research and Innovation Programme under Grant Agreement 776403.
This work has made use of the MUSCLES Treasury Survey High-Level Science Products; doi:10.17909/T9DG6F.
\end{acknowledgements}

\bibliographystyle{aa}
\bibliography{reference.bib}

\appendix

\section{Different ways of calculating mean particle sizes}\label{aA}
Defining particle sizes and number densities is easy as long as the particle considered is spherically symmetric. Defining an average particle size becomes challenging if the particles appear with a non-analytic size distribution, such as in clouds. The next step up in complexity for defining particles sizes is if agglomerates form that are made of many monomers, for example, as result of coagulation processes. Such agglomerates could have all sorts of geometries.
The definitions of the different representations of the mean particles size $\langle a\rangle$ [cm] are therefore given here.
Equations~\ref{eq:amean2}, ~\ref{eq:ameanA2} follow from Eqs.~\ref{eq:amean1} and ~\ref{eq:ameanA1}, respectively, by applying $f(a)da = f(V)dV$ and the definition of the dust moments $L_{\rm j}$ [cm$^{-3}$g$^{-1}$] ($\rho L_{\rm j}$ [cm$^{\rm j-3}$]
\begin{equation}
\label{eq:Lj}
    \rho L_{\rm j}= \int V^{j/3} f(V)\,dV,
\end{equation}
where $V=4\pi a^3/3$ is the volume of the cloud particle of size $a$.

\begin{figure}
    \centering
    \includegraphics[width=\linewidth]{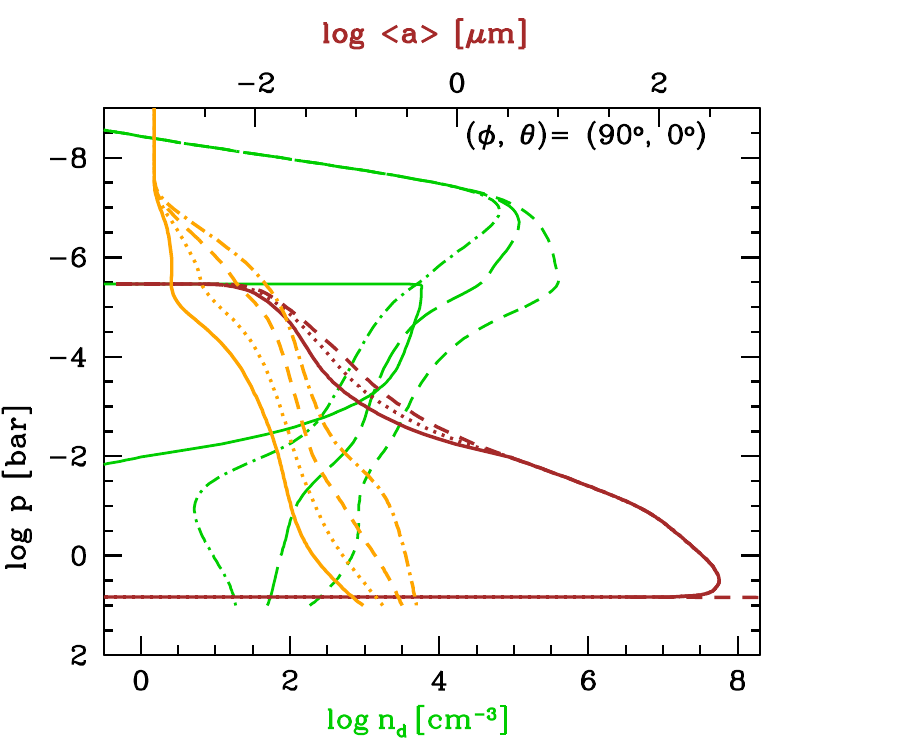}\\
    \includegraphics[width=\linewidth]{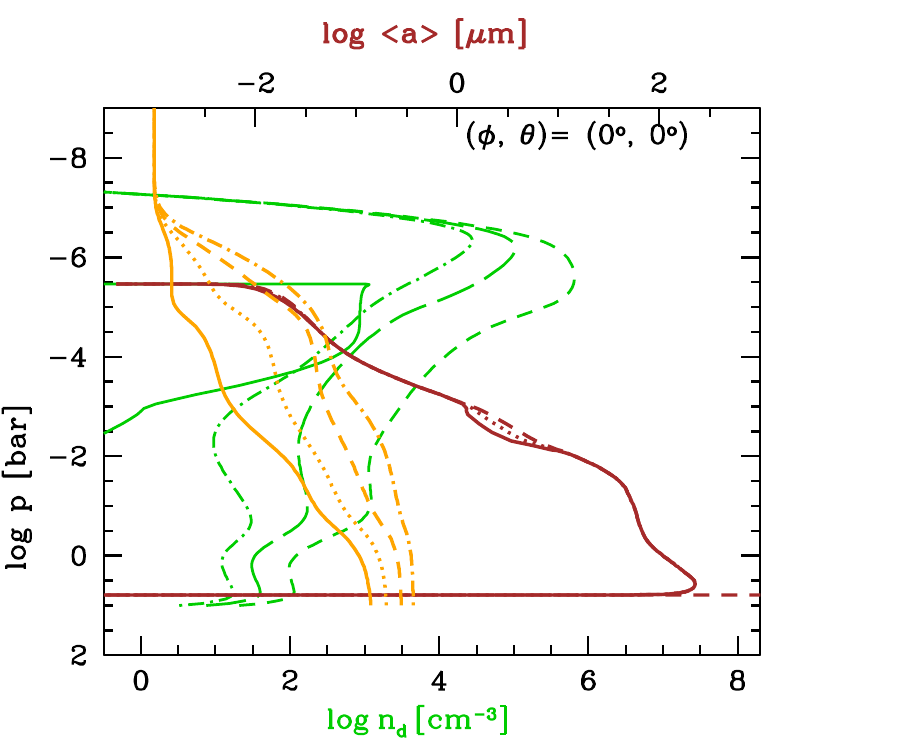}\\
   \includegraphics[width=\linewidth]{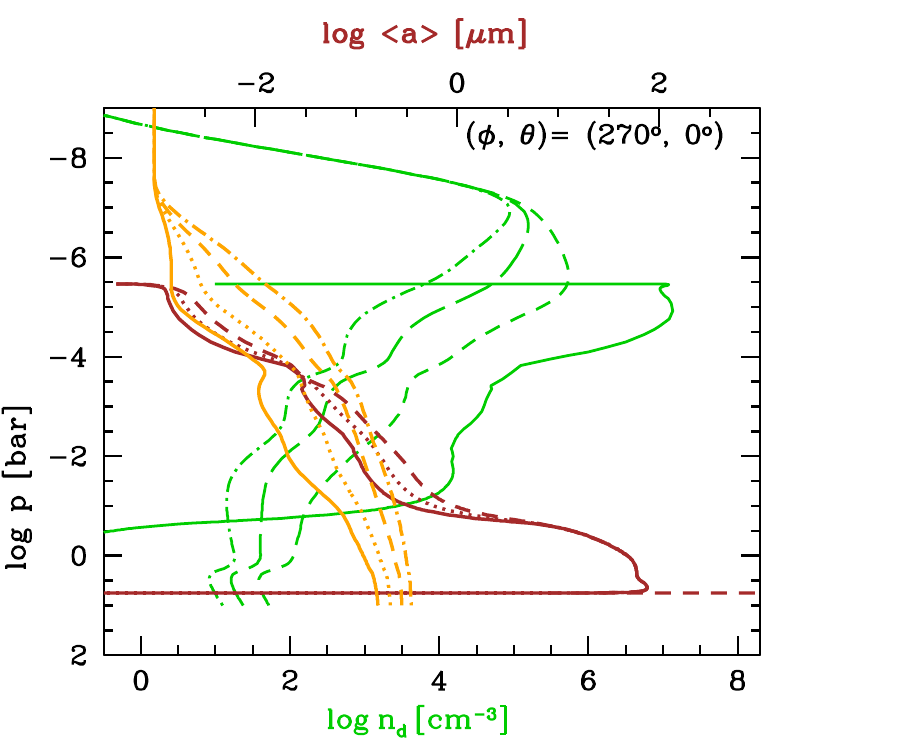}
    \caption{{\bf Particle number density} (n$_{\rm d}$ [cm$^{-3}$], bottom axis):  green solid  -- mineral particles, green dashed --  hydrocarbon haze (short dashed --  total number density $\sum n(a_i)$ with $a_i$ the size of the individual haze particle; long dashed --  surface averaged $n_{\rm surf}$  as in Eq.~65 in, dash-dot -- volume averaged  $n_{\rm Vol}$ surface as in Eq.~66, both in \cite{2018ApJ...853....7K}). 
    {\bf Mean particles size} ($\langle a\rangle$ [$\mu$m], top axis): mineral particles -- brown; hydrocarbon hazes -- orange ($\langle a\rangle$ (Eq.~\ref{eq:amean2}) -- solid lines; $\langle a\rangle_{\rm A}=s_{\rm surf}$ (Eq.~\ref{eq:ameanA2}) -- dashed lines;  $\sqrt[3]{\langle a^3\rangle}$ (Eq.~\ref{eq:amean32}) -- dotted lines, $\langle a\rangle_{\rm V}=s_{\rm vol}$ as in Eq.~38 in \cite{2018ApJ...853....7K} -- dash-dot). }
    \label{fig:a_nd_comp2}
\end{figure}

The weighted averaged particle size $\langle a\rangle$ [cm] (solid brown and orange lines in Figs.~\ref{fig:a_nd_comp} and \ref{fig:a_nd_comp2}) for a particle size distribution $f(a)$ [cm$^{-3}$cm$^{-1}$]
\begin{eqnarray}
\label{eq:amean1}
\langle a\rangle &=& \frac{\int a f(a)\,da}{\int f(a)\,da}\\ 
\label{eq:amean2}
                 &=&\sqrt[3]{\frac{3}{4\pi}}\, \frac{L_1}{L_0}.
\end{eqnarray}
%\sout{Equation~\ref{eq:amean1} is similar to Eq. (64) in \cite{2018ApJ...853....7K} when equating $f(a)=n(s_{\rm i})$ with $n(s_{\rm i})$ being the number density of the cloud particles coagulates of size $s_{\rm i}$.  The smallest size is here the monomer $s_{\rm 0}=1$nm.}

One can also utilize the moments  $L_3$ and $L_2$ based on the definition in Eq.~\ref{eq:Lj} to define a  surface averaged mean particle size $\langle a\rangle_{\rm A}$ (brown and orange dashed lines 
%\sout{in Fig.~\ref{fig:a_nd_comp} and} 
in Fig.~\ref{fig:a_nd_comp2}) with $A(a) = 4\pi a^2$:
\begin{eqnarray}
\label{eq:ameanA1}
\langle a\rangle_{\rm A} &=& 3 \frac{\int V(a)f(a)\,da}{\int A(a) f(a)\,da}\\
&=& \frac{\int a^3 f(a)\,da}{\int a^2 f(a)\,da}\\
\label{eq:ameanA2}
&=& \sqrt[3]{\frac{3}{4\pi}}\, \frac{L_3}{L_2}.
%&=& \frac{1}{\sqrt[3]{(36\pi)}}\, \frac{L_3}{L_2}.
\label{eq:ameanA2_1}
\end{eqnarray}

Note that $s_{\rm surf}$ presented in Sect.~\ref{ss:haze} is exactly the same as $\langle a\rangle_{\rm A}$, which can be recognized by equating $n(s_{\rm i}) = f(a)\,da$ in Eq. (64) of \cite{2018ApJ...853....7K} with $n(s_{\rm i})$ being the number density of the cloud particles of size $s_{\rm i}$. Note that Eq.~\ref{eq:Lj} results in $\rho L_2 = (4\pi/3)^{2/3} \langle a\rangle^2 n_{\rm d}$,   $\rho L_3 = (4\pi/3) \langle a\rangle^3 n_{\rm d}$ etc.  such that the total number of cloud particles is $\int f(V)dV = n_{\rm d}$ for a delta-peak-like f(V). From $\rho L_3 = (4\pi/3) \langle a\rangle_{\rm A}^3 n_{\rm d}$ follows then with $\langle a\rangle = \langle a\rangle_{\rm A}$ that $n_{\rm d, A} = (\rho L_3)/(4\pi  \langle a\rangle_{\rm A}^3/3)$.

We use another expression for the mean particle size (brown and orange dotted lines
in Fig.~\ref{fig:a_nd_comp2}) as
\begin{eqnarray}
\label{eq:amean3}
\sqrt[3]{\langle a^3\rangle} &=& \sqrt[3]{\frac{\int a^3 f(a)\,da}{\int  f(a)\,da}}\\
\label{eq:amean32}
&=& \sqrt[3]{\frac{3}{4\pi}\frac{L_3}{L_0}}
\end{eqnarray}

In Figure~\ref{fig:a_nd_comp2}, we demonstrate the results for these different definitions of particles sizes (brown and orange lines) as result of two different modelling methods. The hydrocarbon haze model of \cite{2018ApJ...853....7K} which was applied in this paper uses a binning method in order to follow the growth of the agglomerates.  The method of \cite{Woitke2003,Woitke2004,Helling2006,2008A&A...485..547H} that we used to calculate the mineral cloud particles solves differential moment method where the moments are weighted integrals over the particle size distribution as defined in Eq.~\ref{eq:Lj}. 
Applying the same definitions (Eq.~\ref{eq:amean1} - Eq.~\ref{eq:amean32}) to the results from both methods allows for a more scientific comparison between the resulting values of the mean particles sizes (or any other property). It enables us further to provide information about what uncertainties cloud particle sizes should be expected. 
%\ch{The largest uncertainties occur in atmospheric regions where cloud particles sizes change only mildly with pressure. 
Comparing the mean particles sizes derived by different formula further demonstrate that the definition of {\it the} particles size becomes increasingly more challenging if particle-particle processes are modelled as in coagulation simulations. This is also the case when defining particle number densities (green dashed lines in   Fig.~\ref{fig:a_nd_comp2}): The volume averaged particles number density, which is of relevance for opacity calculations, can differ by two orders of magnitudes from the number density that sums up all existing particles as $\sum_i n(a_i)$ with $a_i(V_{\rm i})$ the size of the individual haze particle of volume $V_{\rm i}$.

%\yui{
%\begin{eqnarray}
%s_{\rm surf} &=& \frac{\int a^3f(a)\,da}{\int a^2 f(a)\,da} = 3 \langle a\rangle_{\rm A}
%\end{eqnarray}
%}

\section{Supplementary plots. }
%____________________________________________________________________

Here we provide the detailed results that have been used to produce the more condensed results of the previous sections.

\begin{figure}[h]
    \includegraphics[width=1.0\linewidth]{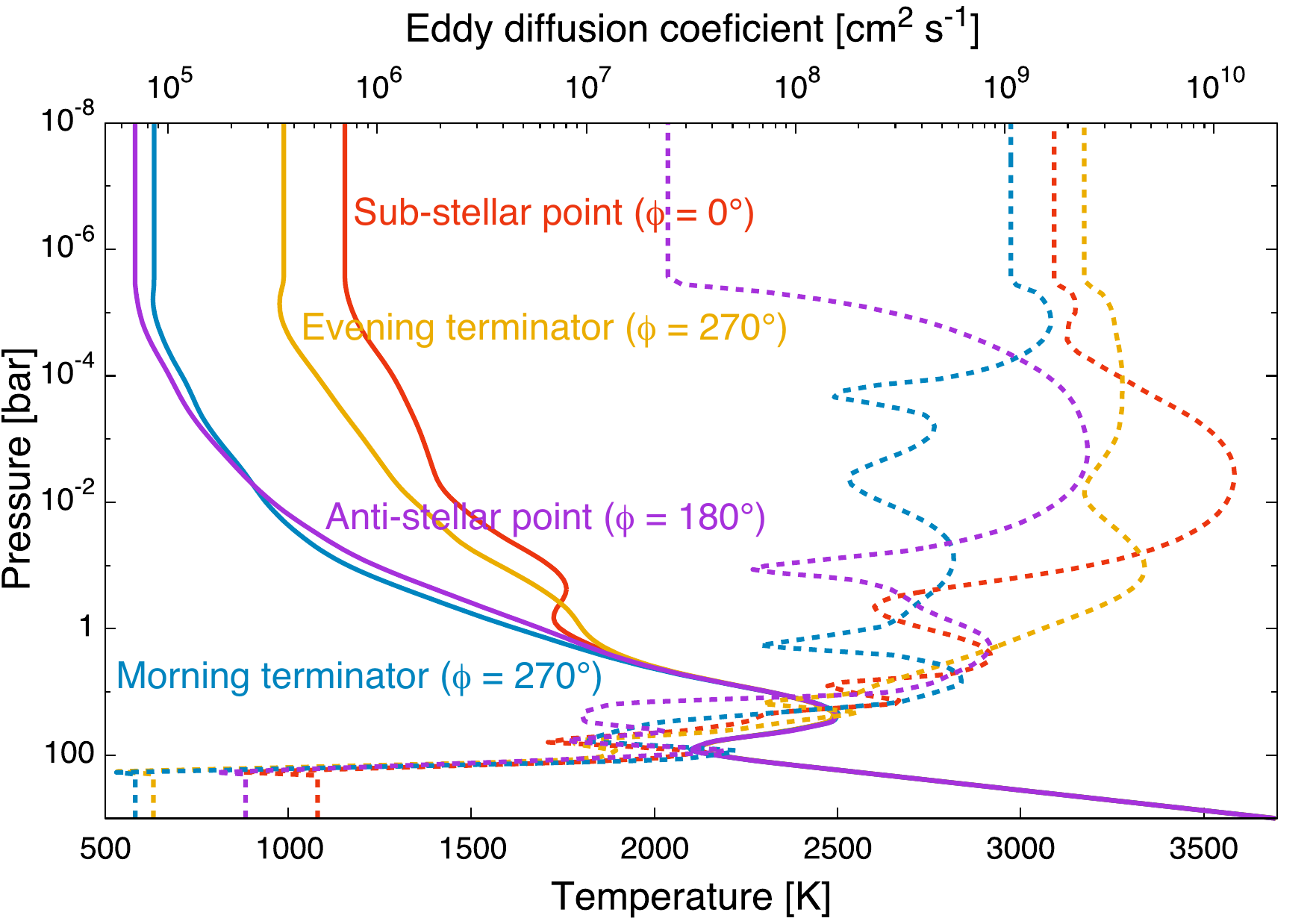}\\
    \caption{The 1D profiles (T$_{\rm gas}$ - solid lines, K$_{\rm zz}$ --  dotted lines) used to study the formation of hydrocarbon hazes at three distinct locations in the atmosphere of WASP-43b. These profiles are isothermaly extended into higher altitudes than those used in Fig.~\ref{TpNuc}. }
    \label{fig:Kzz}
\end{figure}

\smallskip
{\it Input for photochemical haze calculation for WASP-43b:} Figure~\ref{fig:Kzz} provides the 1D (T$_{\rm gas}$, p$_{\rm gas}$, K$_{\rm zz}$)-profiles that have been used for the photochemical gas and haze calculations in Sect~\ref{s:ncheq}.

\begin{figure*}
    \includegraphics[width=0.5\linewidth]{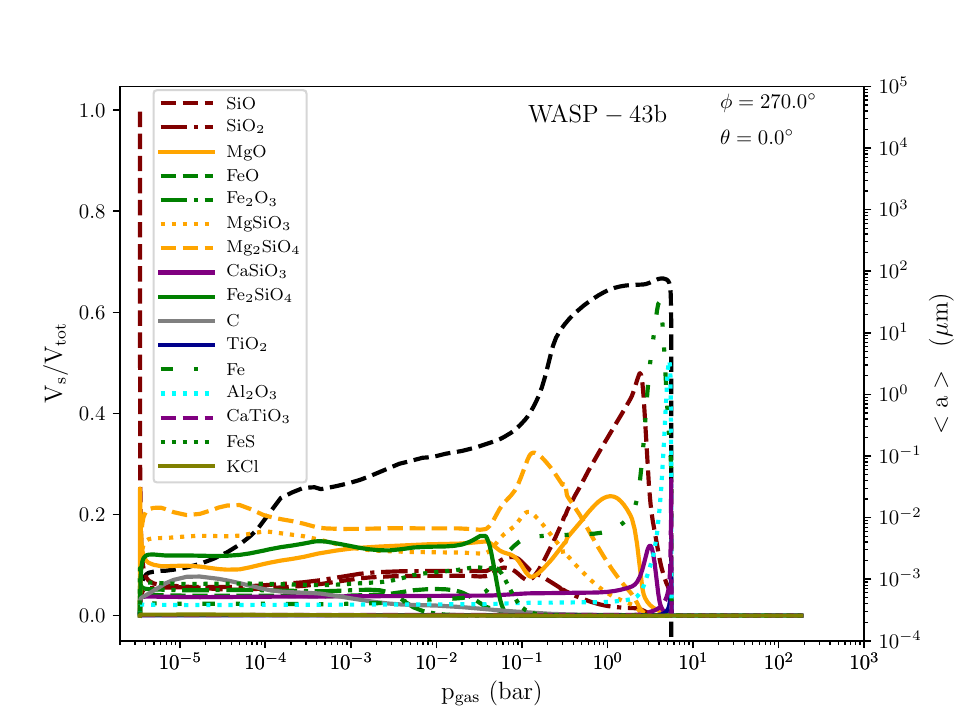}
        \includegraphics[width=0.5\linewidth]{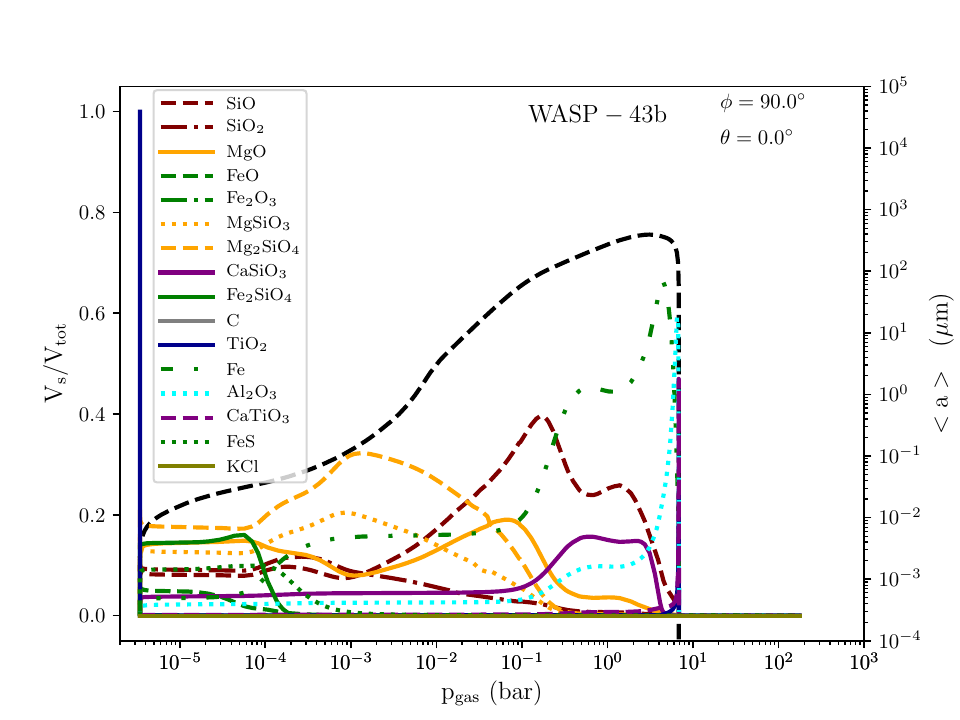}\\
    \includegraphics[width=0.5\linewidth]{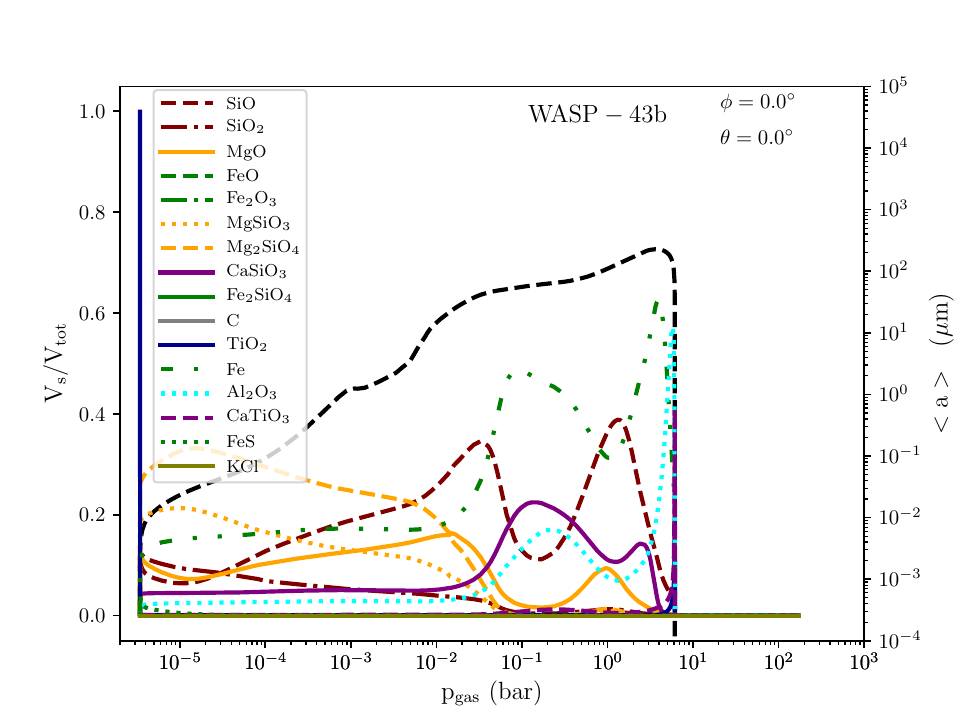}
    \includegraphics[width=0.5\linewidth]{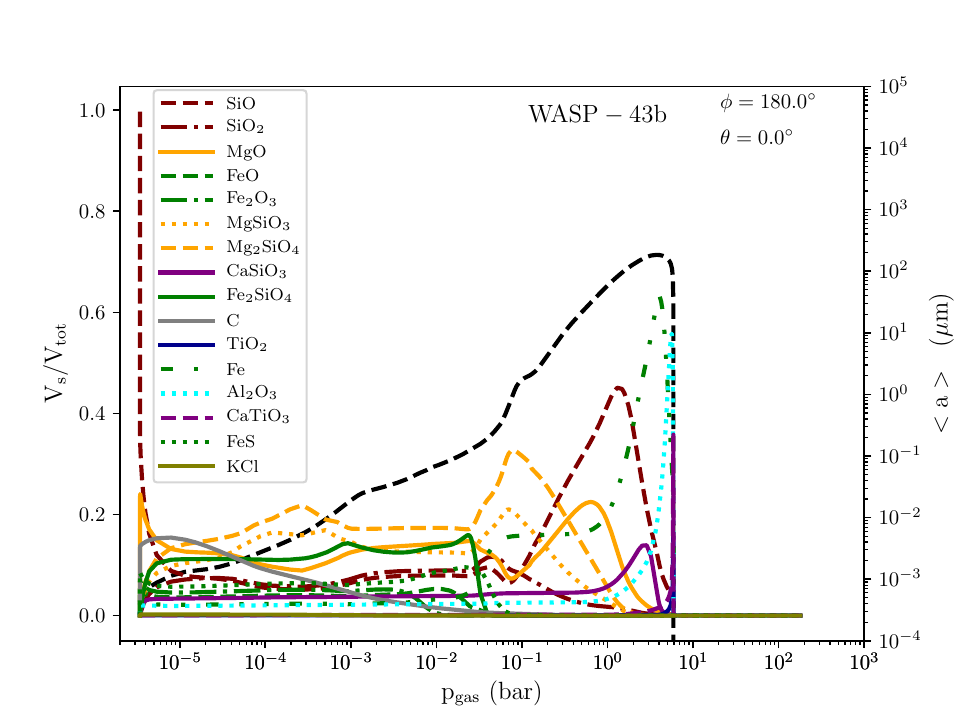}
    \caption{Detailed material volume fraction for the  morning terminator (top left), evening terminator (top right), sub-stellar point (bottom left), anti-stellar point (bottom right).}%\\*[-2cm]
    \label{fig:matvolfrac_ASP}
\end{figure*}

{\it Details for mineral cloud particle composition :} Figure~\ref{fig:matvolfrac_ASP} details how the material composition of the  mineral cloud particles changes vertically for the four example 1D (T$_{\rm gas}$, p$_{\rm gas}$)-profiles. The details in Fig.~\ref{fig:matvolfrac_ASP} are the base for the 5 material categories in Fig.~\ref{fig:5line_MT} grouped as follows:\\
-- silicates (\ce{MgSiO3}[s], \ce{Mg2SiO4}[s], \ce{Fe2SiO4}[s], \ce{CaSiO3}[s])\\
-- metal oxides (MgO[s], SiO[s], \ce{SiO2}[s], FeO[s], \ce{Fe2O3}[s], \ce{CaTiO3}[s])\\
-- high temperature condensates (\ce{TiO2}[s],  Fe[s], \ce{Al2O3}[s])\\
-- carbon (C[s]), and\\
-- salts (here KCl[s]).

%All lines:

\begin{figure*}
    \includegraphics[width=0.5\linewidth,page=3]{images/WASP_43b_Gas_Phase.pdf}
        \includegraphics[width=0.5\linewidth,page=7]{images/WASP_43b_Gas_Phase.pdf}\\*[-0.5cm]
    \includegraphics[width=0.5\linewidth,page=1]{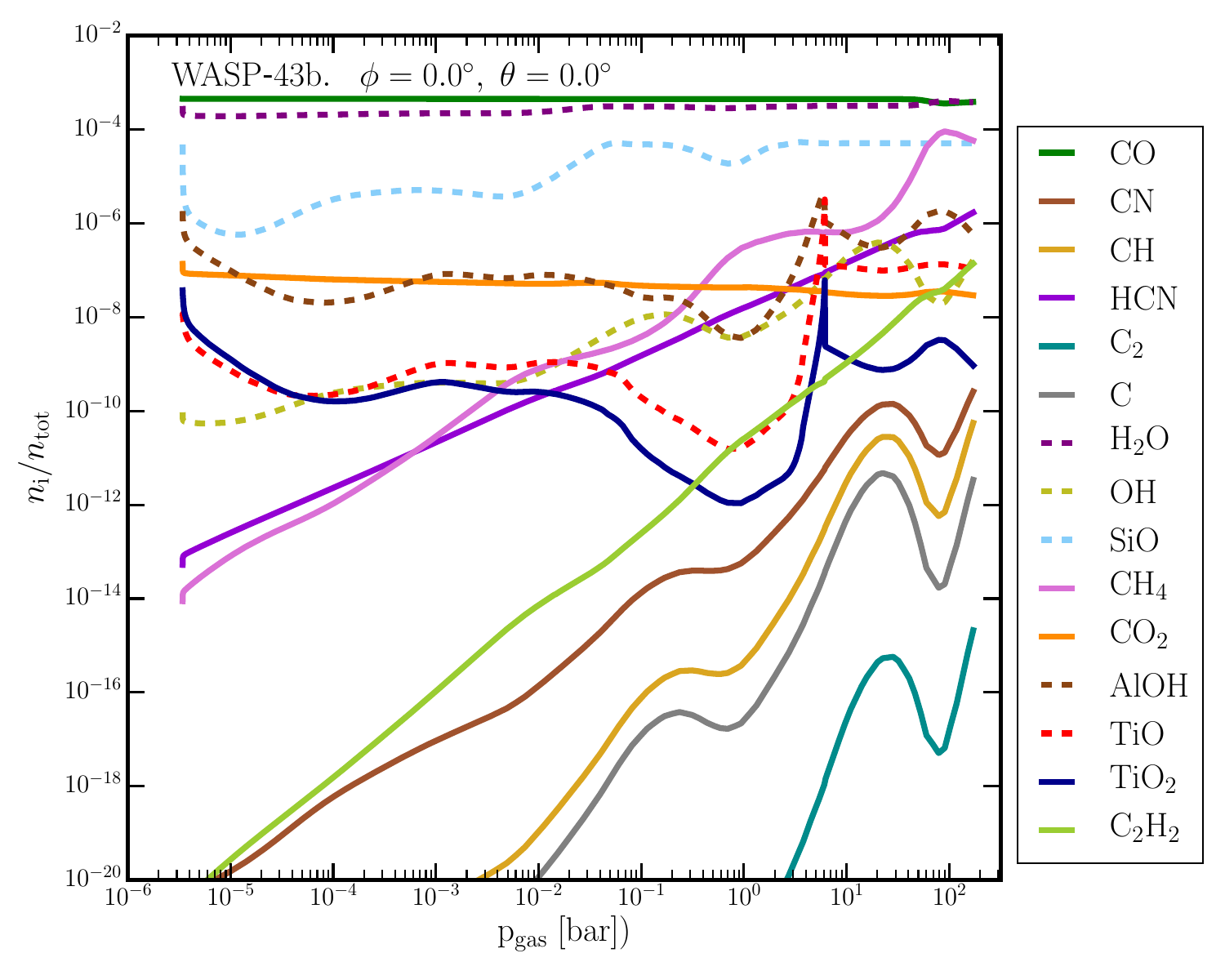}
    \includegraphics[width=0.5\linewidth,page=5]{images/WASP_43b_Gas_Phase.pdf}
    \caption{Concentrations, $n_{\rm i}/n_{\rm tot}$,   for the 15 most abundant molecules (after \ce{H2/H}), in chemical equilibrium based on the cloud-depleted element abundances on WASP-43b. {\bf Top left:} morning terminator, {\bf Top right:} evening terminator, {\bf Bottom left:} sub-stellar point, {\bf Bottom right:} anti-stellar point.}
    \label{fig:Mol_ASP}
\end{figure*}

\begin{figure*}
    \includegraphics[width=0.5\linewidth,page=4]{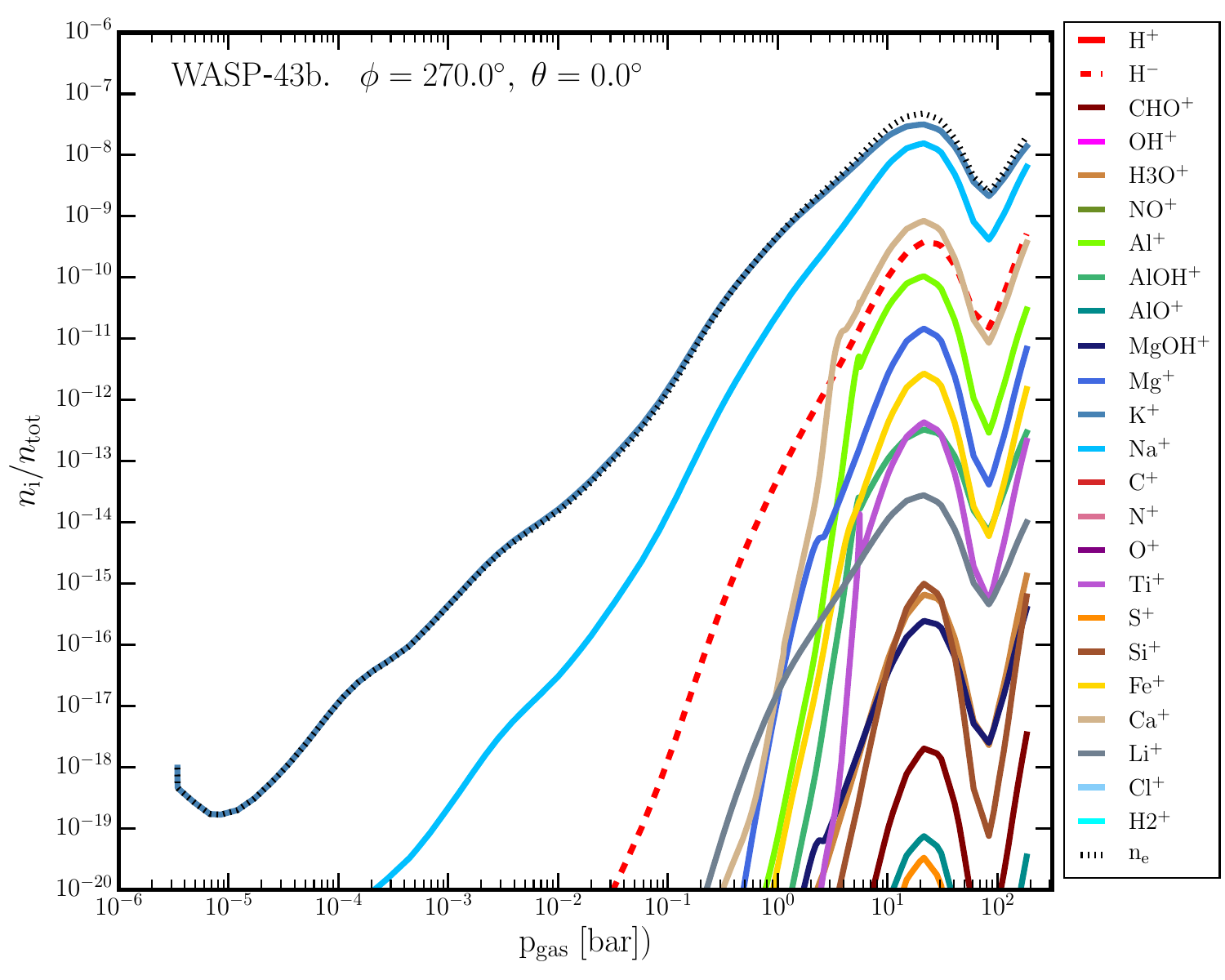}
        \includegraphics[width=0.5\linewidth,page=8]{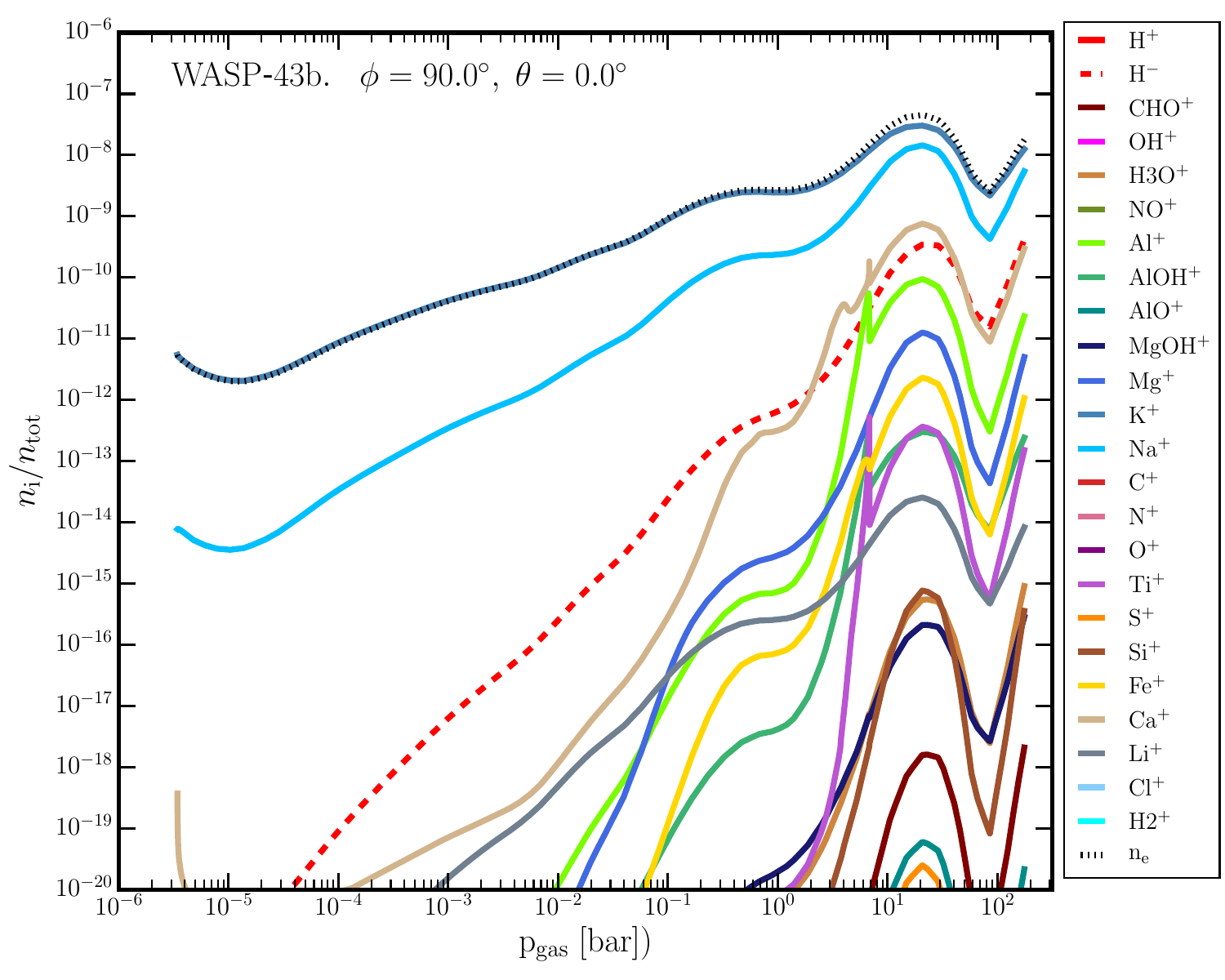}\\*[-0.5cm]
    \includegraphics[width=0.5\linewidth,page=2]{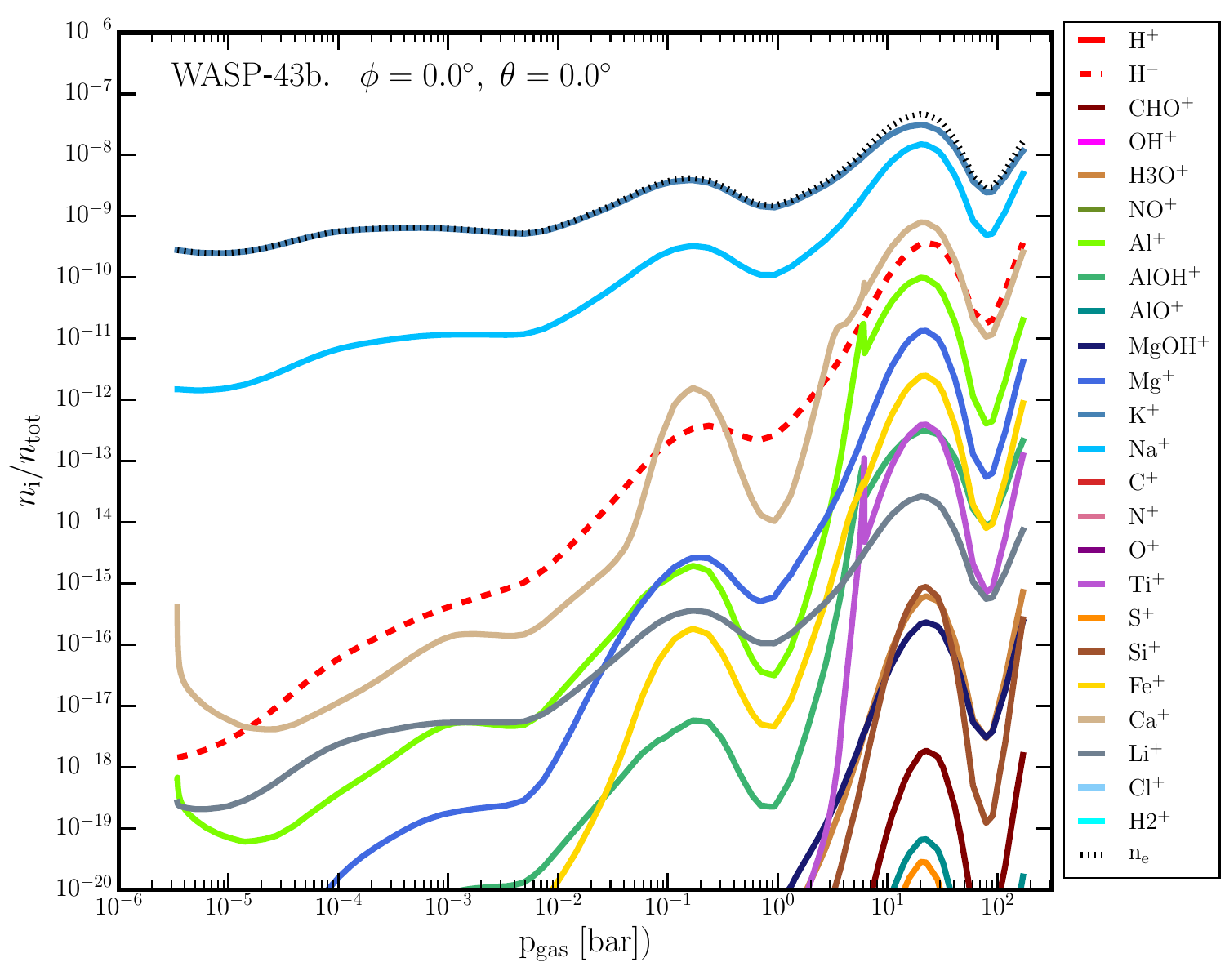}
    \includegraphics[width=0.5\linewidth,page=6]{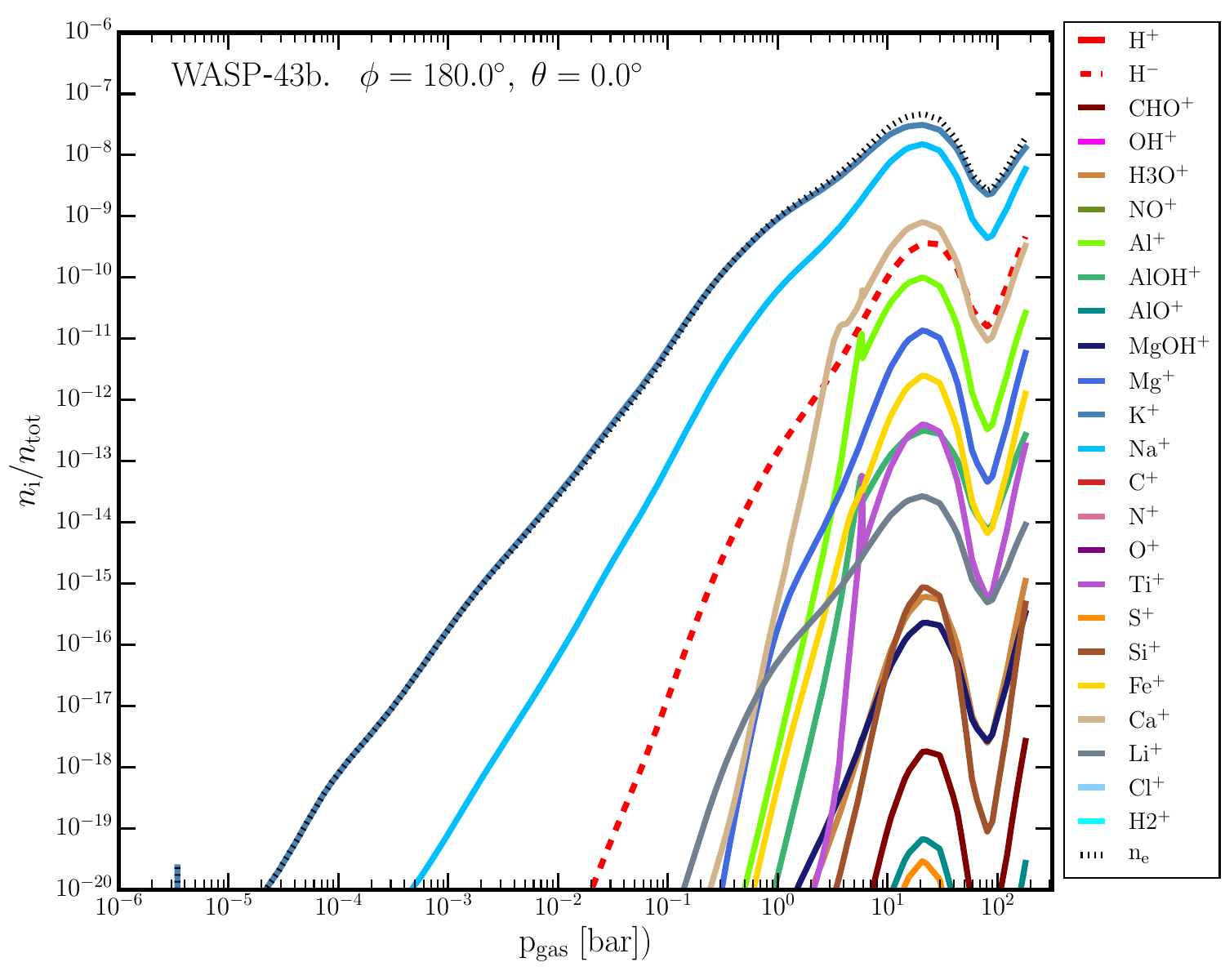}
    \caption{The ion concentrations, $n_{\rm i}/n_{\rm tot}$,  in thermal equilibrium for the four selected 1D profiles based on the cloud-depleted element abundances: {\bf Top left:} morning terminator, {\bf Top right:} evening terminator,  {\bf bottom left:} sub-stellar point, {\bf bottom right:} anti-stellar point. The electron number density is shown as black dash line.}
    \label{fig:Ions_ASP}
\end{figure*}

\smallskip
{\it Details of chemical equilibrium gas-phase composition for WASP-43b:} Figure~\ref{fig:Mol_ASP} summarised the 15 most abundant molecules for the WASP-43b substellar and anti-stellar point, and the two equatorial terminators. The most abundant molecules in chemical equilibrium are CO and \ce{H2O}, after \ce{H2}, and He. \ce{H2}, CO and \ce{H2O} will be affected by photochemical reactions as demonstrated in Fig.~\ref{fig:photo}. The next abundant molecule is SiO on the dayside but \ce{CH4} on the nightside in chemical equilibrium. Both are not affected by photochemistry, SiO because it is not considered in the photochemical code and \ce{CH4} because no stellar photons reach the nightside in our present calculations.  Molecules like AlOH, \ce{CO2} and TiO are considerably less abundant. The abundance of AlOH and TiO is further affected by element depletion through mineral cloud formation.

Figure~\ref{fig:Ions_ASP} gives an account of the atomic and molecular ion species for the WASP-43b substellar and anti-stellar point, and the two equatorial terminators. The most abundant ions are \ce{K+}, \ce{Na+} and \ce{Ca+} followed by \ce{H-}. Photochemistry does not affect \ce{H-} in our simulations.

\end{document}